\documentclass{aastex631}

\newcommand{\CI}{\ion{C}{1}}
\newcommand{\CIfull}{\CI{} $^3$P$_1$--$^3$P$_0$}

\newcommand{\CII}{\ion{C}{2}}
\newcommand{\CIIfull}{\CII{} $^2$P$_{3/2}$--$^2$P$_{1/2}$}

\graphicspath{{./}{figures/}}

\usepackage{natbib}
\usepackage{amsfonts}

\begin{document}

\title{Primordial or Secondary? Testing models of debris disk gas with ALMA\footnote{{\it Herschel} is an ESA space observatory with science instruments provided by European-led Principal Investigator consortia and with important participation from NASA.}}

\correspondingauthor{Gianni Cataldi}
\email{cataldi.gia@gmail.com}

\author[0000-0002-2700-9676]{Gianni Cataldi}
\affil{National Astronomical Observatory of Japan, Osawa 2-21-1, Mitaka, Tokyo 181-8588, Japan}
\affil{Department of Astronomy, Graduate School of Science, The University of Tokyo, Tokyo 113-0033, Japan}

\author[0000-0003-3283-6884]{Yuri Aikawa}
\affil{Department of Astronomy, Graduate School of Science, The University of Tokyo, Tokyo 113-0033, Japan}

\author[0000-0002-2707-7548]{Kazunari Iwasaki}
\affil{National Astronomical Observatory of Japan, Osawa 2-21-1, Mitaka, Tokyo 181-8588, Japan}

\author[0000-0002-5352-2924]{Sebastian Marino}
\affil{School of Physics and Astronomy, University of Exeter, Stocker Road, Exeter, EX4 4QL, United Kingdom}

\author{Alexis Brandeker}
\affil{Department of Astronomy, Stockholm University, AlbaNova University Center, 106\,91 Stockholm, Sweden}

\author{Antonio Hales}
\affil{Joint ALMA Observatory, Avenida Alonso de C\'ordova 3107, Vitacura 7630355, Santiago, Chile}
\affil{National Radio Astronomy Observatory, 520 Edgemont Road, Charlottesville, VA 22903-2475, United States of America}

\author{Thomas Henning}
\affil{MPI for Astronomy, K\"onigstuhl 17, 69117 Heidelberg, Germany}

\author[0000-0002-9221-2910]{Aya E. Higuchi}
\affil{Division of Science, School of Science and Engineering, Tokyo Denki University, Ishizaka, Hatoyama-machi, Hiki-gun, Saitama 350-0394, Japan}

\author[0000-0002-4803-6200]{A. Meredith Hughes}
\affil{Astronomy Department and Van Vleck Observatory, Wesleyan University, 96 Foss Hill Drive, Middletown, CT 06459, USA}

\author{Markus Janson}
\affil{Department of Astronomy, Stockholm University, AlbaNova University Center, 106\,91 Stockholm, Sweden}

%\author{Inga Kamp}
%\affil{}

\author{Quentin Kral}
\affil{LESIA, Observatoire de Paris, Université PSL, CNRS, Sorbonne Université, Univ. Paris Diderot, Sorbonne Paris Cité, 5 place Jules Janssen, 92195 Meudon, France}

\author[0000-0003-4705-3188]{Luca Matrà}
\affil{School of Physics, Trinity College Dublin, the University of Dublin, College Green, Dublin 2, Ireland}

\author{Attila Moór}
\affil{Konkoly Observatory, Research Centre for Astronomy and Earth Sciences, E\"otv\"os Lor\'and Research Network (ELKH), Konkoly-Thege Mikl\'os \'ut 15-17, H-1121 Budapest, Hungary.}

%\author{Nagayoshi Ohashi}
%\affil{}

\author{Göran Olofsson}
\affil{Department of Astronomy, Stockholm University, AlbaNova University Center, 106\,91 Stockholm, Sweden}

\author[0000-0003-3786-3486]{Seth Redfield}
\affil{Astronomy Department and Van Vleck Observatory, Wesleyan University, 96 Foss Hill Drive, Middletown, CT 06459, USA}

\author{Aki Roberge}
\affil{Astrophysics Science Division, NASA Goddard Space Flight Center, Greenbelt MD, 20771, USA}

%\author{Alycia Weinberger}
%\affil{}

%\author{Yanqin Wu}
%\affil{}



%The AAS Journals, the Astrophysical Journal (ApJ), the Astrophysical Journal Letters (ApJL), the Astronomical Journal (AJ), and the Planetary Science Journal (PSJ) all have a 250 word limit for the  abstract.
\begin{abstract}
%Context
The origin and evolution of gas in debris disks is still not well understood. Secondary gas production from cometary material or a primordial origin have been proposed. So far, observations have mostly concentrated on CO, with only few C observations available.
%Aims
We create an overview of the C and CO content of debris disk gas and use it test state-of-the-art models.
%Methods
We use new and archival ALMA observations of CO and \CI{} emission, complemented by \CII{} data from Herschel, for a sample of 14 debris disks. This expands the number of disks with ALMA measurements of both CO and \CI{} by ten disks. We present new detections of \CI{} emission towards three disks: HD~21997, HD~121191 and HD~121617. We use a simple disk model to derive gas masses and column densities.
%Results
We find that current state-of-the-art models of secondary gas production overpredict the C$^0$ content of debris disk gas. This does not rule out a secondary origin, but might indicate that the models require an additional C removal process. Alternatively, the gas might be produced in transient events rather than a steady-state collisional cascade. We also test a primordial gas origin by comparing our results to a simplified thermo-chemical model. This yields promising results, but more detailed work is required before a conclusion can be reached.
%Conclusions
Our work demonstrates that the combination of C and CO data is a powerful tool to advance our understanding of debris disk gas.
\end{abstract}

%% https://astrothesaurus.org`


\section{Introduction}\label{sec:intro}
The extrasolar analogues of the solar system's asteroid and Kuiper belts are known as \textit{debris disks}. The study of debris disks yields important insights into the formation and evolution of planetary systems \citep[e.g.][]{Wyatt08,Hughes18}. It is believed that debris disks derive their dust from the continuous collisional grinding of asteroidal or cometary bodies (planetesimals). However, it has now become clear that some debris disks also harbour detectable amounts of gas. In particular, sensitive observations of CO emission with the Atacama Large Millimeter/submillimeter Array (ALMA) greatly increased the number of known gaseous debris disks \citep[e.g.][]{Lieman-Sifry16,Moor17}. Today, more than 20 debris disks with gas are known. So far, gas has been mostly found around young (age $\lesssim$50\,Myr) A-type stars, but debris disks with gas also exist around much older \citep[e.g.\ the 1--2\,Gyr old $\eta$~Corvi,][]{Marino17} and cooler \citep[e.g.\ the M-type star TWA~7,][]{Matra19_luminosity_correlation} stars. The presence of gas in debris disks might have important implications. For example, the gas might affect the dust dynamics and therefore our interpretation of continuum or scattered light observations \citep[e.g.][]{Takeuchi01,Thebault05,Krivov09,Lyra13,Lin19,Marino22,Olofsson22}. However, the origin and evolution of the gas is still poorly understood.

The photodissociation timescale of a CO molecule exposed to the interstellar radiation field (ISRF) is only about 120 years \citep{Visser09}. Thus, unless CO is shielded, it needs to be replenished continuously from a mechanism that liberates CO (or CO$_2$) from the ice phase, for example by collisions of comets \citep[e.g.][]{Moor11,Zuckerman12} or photodesorption \citep[e.g.][]{Grigorieva07,Oberg09}. Such mechanisms would be considered a \textit{secondary} gas origin. Gas production from comets can readily explain disks with low CO masses ($M_\mathrm{CO}\lesssim10^{-5}$\,M$_\oplus$) where the CO production rate required to replenish CO is in reasonable agreement with comet destruction rates estimated from continuum observations. Examples include the disks around $\beta$~Pic \citep{Dent14,Matra17_betaPic}, HD~181327 \citep{Marino16} and Fomalhaut \citep{Matra17_Fomalhaut}. Secondary gas allows us to indirectly study the chemical composition of the solids from which the gas is derived \citep[e.g.][]{Matra17_Fomalhaut,Matra18_SMA}.

However, a number of debris disks around young A-stars host significantly higher CO masses ($M_\mathrm{CO}\gtrsim10^{-3}$\,M$_\oplus$, hereafter `CO-rich disks'). Examples include the disks around HD~21997 \citep{Kospal13}, 49~Ceti and HD~32297 \citep[e.g.][]{Moor19}. While CO self-shielding prolongs the CO lifetime in these systems, it is still much shorter than the age of the system \citep{Kospal13}, again suggesting that CO production is needed. But the large CO masses of CO-rich disks require a steady-state CO production rate that greatly exceeds what can be expected from a reasonable comet population \citep{Moor11,Kospal13,Kral17}. Instead, it was proposed that these disks host \textit{primordial} gas, that is, gas leftover from the protoplanetary phase. Compared to secondary gas, primordial gas would have a much lower metallicity, meaning there would be abundant H$_2$ that could shield CO. \citet{Nakatani21} proposed that inefficient photo-evaporation of protoplanetary disks around A-stars could prolong the lifetime of protoplanetary gas sufficiently to produce CO-rich debris disks. The primordial scenario would have important implications for gas disk dispersal mechanism and thus giant planet formation, because the typical ages of CO-rich disks (tens of Myr) significantly exceed the canonical lifetime of a protoplanetary disk as estimated from infrared excess studies \citep[${\lesssim}10$\,Myr, e.g.][]{Ribas14}.

On the other hand, a different class of models attempts to apply the secondary gas scenario even to the CO-rich debris disks \citep{Kral19,Moor19}. Here, another shielding agent comes into play: neutral carbon (C$^0$), which is continuously produced through CO photo-dissociation. Once a sufficient amount of C$^0$ is present, the CO destruction rate drops significantly. The reduced destruction rate allows a large CO mass to accumulate without the need for unrealistically high CO production rates. \citet{Marino20} used this model to produce a population synthesis of gaseous debris disks. They found generally favorable agreement with the observations, although the data available to test models were limited, especially in terms of \CI{} observations. If the gas in CO-rich disks is indeed secondary, it could have important consequences for planet formation. Indeed, \citet{Kral20_terrestrial_atmospheres} showed that terrestrial planets embedded in a disk of secondary gas can accrete this high metallicity gas, thereby altering their atmospheric composition. Eventually, this scenario might be tested by detecting C bearing species and determining their abundances in exoplanetary atmospheres \citep[e.g.][]{Rustamkulov23}.

Clearly, observations of \CI{} are needed to clarify its role as shielding agent. In addition, \CI{} data can help to better understand the chemistry of debris disk gas \citep{Higuchi17}. In particular, it is unclear how efficiently C produced from CO photo-dissociation can reform CO. There is therefore a strong motivation to study the atomic carbon component of debris disks gas. But observations have so far mostly focussed on CO because it is easier to observe. In this paper, we present new ALMA observations of \CI{} emission. Combined with archival observations of CO, \CI{} and \CII{}, we study the C and CO content of a sample of 14 debris disks with the aim to constrain current models. Our paper is structured as follows. In section \ref{sec:observations} we describe the observations and data reduction. In section \ref{sec:data_analysis} we describe how the data were analysed. Section \ref{sec:modelling} presents the methodology used to derive disk-integrated C and CO masses and column densities. We present our results in section \ref{sec:results}. In section \ref{sec:discussion} we discuss our results by comparing to models of secondary gas production as well as a thermo-chemical model. We summarise our results in section \ref{sec:summary}.

\section{Observations}\label{sec:observations}
\subsection{Disk sample}\label{sec:disk_sample}
The goal of the present study is to get an overview of the C and CO gas in debris disks. Therefore, our sample includes any disks listed in the ``Debris'' category of the Catalog of Circumstellar Disks\footnote{\url{https://www.circumstellardisks.org}, accessed September 2021} that had observations of the \CIfull{} line and at least one CO line in the ALMA archive as of September 2021. We used the \texttt{ALminer} software to query the ALMA archive. We identified 14 disks meeting our criteria. Among these, ten disks have no previously published ALMA observations of \CI{}. Table~\ref{tab:disk_parameters} gives an overview of our sample and lists the stellar and disk parameters.

For each disk, we define an inner radius $r_\mathrm{in}$ and an outer radius $r_\mathrm{out}$ based on a dust or gas model presented in the literature, as indicated in the last row of Table~\ref{tab:disk_parameters}. If the literature model is a Gaussian ring (HD~121191, HD~121617, HD~32297, HD~181327, HD~48370), we set $r_\mathrm{in}=r_0-\mathrm{FWHM}/2$ and $r_\mathrm{out}=r_0+\mathrm{FWHM}/2$, where $r_0$ is the center of the Gaussian and FWHM is its full width at half maximum. For the disk models of 49~Ceti by \citet{Hughes17} and HD~110058 by \citet{Hales22}, no outer radius is defined. Thus, we adopt the radius where the model surface density profile is at 50\% of the peak value as $r_\mathrm{out}$. The same procedure is employed for HD~61005. While the power law disk model by \citet{MacGregor18} for HD~61005 does specify an outer cutoff radius (at 188\,au), the power law index is very steep, meaning that our adopted $r_\mathrm{out}=78$\,au better represents the location of the bulk disk material. In all other cases, $r_\mathrm{in}$ and $r_\mathrm{out}$ are adopted directly from the literature. Where necessary, $r_\mathrm{in}$ and $r_\mathrm{out}$ (as well as the stellar luminosity) were scaled to take into account updated distance measurements.

\movetabledown=70mm
\begin{rotatetable}
\begin{deluxetable}{lccccccccccccccl}
\tablecaption{Stellar and disk parameters used in this study. References are given in parenthesis.\label{tab:disk_parameters}}
\tablewidth{0pt} %0pt means natural width
\tablehead{
\colhead{star} & \colhead{SpT} & \colhead{d} & \colhead{$v_\mathrm{sys}$} & \colhead{$M_*$} & \colhead{$T_\mathrm{eff}$} & \colhead{$\log g$} & \colhead{[M/H]} & \colhead{$L_*$} & \colhead{age} & \colhead{$f$} & \colhead{$i$} & \colhead{PA} & \colhead{$r_\mathrm{in}$} & \colhead{$r_\mathrm{out}$} & \colhead{$r$ reference}\\
& & \colhead{[pc]} & \colhead{[km\,s$^{-1}$]} & \colhead{[$M_\odot$]} & \colhead{[K]} & \colhead{[log(cm\,s$^{-2}$)]} & \colhead{[-]} & \colhead{[$L_\odot$]} & \colhead{[Myr]} & \colhead{[$\times10^{-3}$]} & \colhead{[deg]} & \colhead{[deg]} & \colhead{[au]} & \colhead{[au]} &
}
\colnumbers
\startdata
49 Ceti & A1V (1) & 57.1 & 11.92 (2) & 2.1 (2) & 9120 (1) & 4.32 (1) & 0 (1) & 17.2 (3) & 40 (4) & 1.1 (5) & 79.5 (2) & -72 (2) & 19.3$^\ddagger$ & 153.8$^\ddagger$ & CO (2)\\
$\beta$ Pictoris & A6V (6) & 19.4 & 20.5 (7) & 1.75 (8) & 8052 (6) & 4.15 (6) & 0.05 (6) & 8.1 (9) & 23 (10) & 2.1 (9) & 86.6 (11) &  & 50 & 160 & CO (12)\\
HD 21997 & A3IV/V (1) & 69.6 & 17.92 (13) & 1.8 (13) & 8520 (1) & 4.27 (1) & 0 (1) & 9.29 (9)$^\ddagger$ & 42 (14) & 0.6 (9) & 32.6 (13) & 202.6 (13) & 25.1$^\ddagger$ & 133.5$^\ddagger$ & CO (13)\\
HD 32297 & A6V (15) & 132.8 & 20.67 (16) & 1.59 (16) & 7980 (1) & 3.77 (1) & -0.5 (1) & 8.46 (3)$^\ddagger$ & $<30$ (17) & 4.4 (18) & 77.9 (16) &  & 67.6$^\ddagger$ & 151.2$^\ddagger$ & \ion{C}{1} (16)\\
HD 48370 & G8V (19) & 36.1 & 23.6 (20,21) & 0.94 (22) & 5600 (22) & 4.5 (22) & -0.03 (23) & 0.76 (22)$^\ddagger$ & 42 (14,22) & 0.6 (22) & 69.8 (24) & 67.3 (24) & 73.6 & 145.7 & mm dust (24)\\
HD 61005 & G8Vk (6) & 36.5 & 22.5 (25) & 0.9 (9) & 5500 (25) & 4.5 (25) & 0.01 (25) & 0.75 (9)$^\ddagger$ & 40 (9) & 2.3 (9) & 85.6 (26) & 70.3 (26) & 41.9 & 78 & mm dust (26)\\
HD 95086 & A8III (27) & 86.4 & 17 (28) & 1.6 (29) & 7750 (28) & 4 (28) & -0.25 (28) & 6.4 (30)$^\ddagger$ & 13 (31) & 1.7 (5) & 31 (32) & 278 (32) & 110.4$^\ddagger$ & 337.3$^\ddagger$ & mm dust (32)\\
HD 110058 & A6/7V (33) & 130.1 & 12.26 (33) & 1.84 (33) & 7839 (34) & 4 (34) & -0.5 (1) & 8.65 (30)$^\ddagger$ & 15 (35,1) & 1.4 (18) & 85.5 (33) & 335.1 (33) & 7.4 & 21.2 & CO (33)\\
HD 121191 & A5IV/V (1) & 132.1 & 10.1 (36) & 1.6 (36) & 7970 (1) & 4.38 (1) & 0 (1) & 7.75 (30)$^\ddagger$ & 16 (35,1) & 4.7 (37) & 28 (36) & 25 (36) & 14 & 30 & CO (36)\\
HD 121617 & A1V (1) & 116.9 & 7.8 (38) & 1.9 (9) & 9160 (1) & 4.13 (1) & 0 (1) & 14.13 (30)$^\ddagger$ & 16 (35,1) & 4.8 (39) & 37 (30) & 43 (30) & 49.2$^\ddagger$ & 101.2$^\ddagger$ & mm dust (30)\\
HD 131835 & A2IV (1) & 133.7 & 3.45 (40) & 1.77 (41) & 8610 (1) & 4.24 (1) & 0 (1) & 11 (40) & 16 (35,1) & 3 (41) & 68.6 (40) & 57.7 (40) & 22.8$^\ddagger$ & 174.9$^\ddagger$ & mm dust (42)\\
HD 146897 & F2/F3V (1) & 131.5 & -3.1 (1) & 1.5 (9) & 6700 (1) & 4.3 (1) & 0 (1) & 3.25 (9)$^\ddagger$ & 10 (9) & 8.2 (9) & 84.6 (43) & 114.6 (44) & 75.9$^\ddagger$ & 97.3$^\ddagger$ & mm dust (42)\\
HD 181327 & F5/F6V (1) & 48.2 & 0.1 (45) & 1.36 (46) & 6360 (1) & 4.09 (1) & -0.05 (1) & 2.87 (47) & 23 (10,1) & 4.1 (47) & 30 (45) & 98.8 (45) & 69.2$^\ddagger$ & 90.8$^\ddagger$ & mm dust (45)\\
HR 4796 & A0V (48) & 71.9 & 7.5 (1) & 2.18 (49) & 10060 (1) & 4.44 (1) & -0.5 (1) & 22.83 (50)$^\ddagger$ & 8 (18) & 3 (51) & 76.6 (52) & 26.7 (52) & 70.5$^\ddagger$ & 84.4$^\ddagger$ & mm dust (52)\\
\enddata
\tablerefs{
(1) \citet{Rebollido20}, (2) \citet{Hughes17}, (3) \citet{Moor19}, (4) \citet{Zuckerman12}, (5) \citet{Moor15_stirring}, (6) \citet{Gray06}, (7) \citet{Brandeker11}, (8) \citet{Crifo97}, (9) \citet{Matra18_radius_luminosity_relation}, (10) \citet{Mamajek14}, (11) \citet{Matra19}, (12) \citet{Dent14}, (13) \citet{Kospal13}, (14) \citet{Bell15}, (15) \citet{Rodigas14}, (16) \citet{Cataldi20}, (17) \citet{Kalas05}, (18) \citet{Chen14}, (19) \citet{Torres08}, (20) \citet{GAIA_mission}, (21) \citet{GAIA_DR2}, (22) \citet{Moor16}, (23) \citet{Gaspar16}, (24) \citet{Moor_inprep}, (25) \citet{Desidera11}, (26) \citet{MacGregor18}, (27) \citet{Houk75}, (28) \citet{Moor13_HD95086}, (29) \citet{Booth19}, (30) \citet{Moor17}, (31) \citet{Booth21}, (32) \citet{Su17}, (33) \citet{Hales22}, (34) \citet{Saffe21}, (35) \citet{Pecaut16}, (36) \citet{Kral20_HD129590}, (37) \citet{Vican16}, (38) \citet{Rebollido18}, (39) \citet{Moor11}, (40) \citet{Hales19}, (41) \citet{Moor15_APEX}, (42) \citet{Lieman-Sifry16}, (43) \citet{Engler17}, (44) \citet{Goebel18}, (45) \citet{Marino16}, (46) \citet{Lebreton12}, (47) \citet{Pawellek21}, (48) \citet{Houk82}, (49) \citet{Gerbaldi99}, (50) \citet{Kral17}, (51) \citet{Meeus12}, (52) \citet{Kennedy18}
}
\tablecomments{(1) star name; (2) spectral type; (3) distance from \citet{GAIA_DR2}, except for $\beta$~Pictoris where we use the more precise value by \citet{vanLeeuwen07}; (4) systemic velocity in the barycentric frame; (5) stellar mass; (6) stellar effective temperature; (7) surface gravity; (8) stellar metallicity; (9) stellar luminosity; (10) system age; (11) disk fractional luminosity; (12) disk inclination; (13) disk position angle (missing values indicate that it was not necessary to adopt a PA because no flux measurements were carried out for that target); (14) adopted inner edge of the disk; (15) adopted outer edge of the disk; (16) type of observations that were used to define $r_\mathrm{in}$ and $r_\mathrm{out}$. $^\ddagger$ indicates scaling to updated GAIA distance measures.}
\end{deluxetable}
\end{rotatetable}

Besides CO and \CI{} observations by ALMA, we also used \CIIfull{} observations from Herschel/PACS where available. Table~\ref{tab:disk_fluxes} provides an overview of the emission line data we used for each disk. Note the following data sets available in the ALMA archive that we ignored. For \CI{} toward HD~121191, we found that when combining the ALMA data of program 2019.1.01175.S with the ACA data of program 2018.1.00633.S, the SNR of the disk-integrated flux worsens. Thus, we ignored the noisier ACA data. For $^{12}$CO~3--2 toward HD~181327, we ignored data from program 2013.1.00025.S that had a total integration time of only 60\,s. For $^{12}$CO~2--1 toward HD~61005, we follow \citet{Olofsson16} and only use the data acquired on March 20, 2014 for which the amount of water vapor in the atmosphere was smallest.

\startlongtable
\begin{deluxetable}{llccc}
\tablecaption{Overview of the data analysed in this study.\label{tab:disk_fluxes}}
\tablewidth{0pt} %0pt means natural width
\tablehead{ \colhead{star} & \colhead{emission line} &\colhead{flux} & \colhead{flux reference} & \colhead{observation ID}\\ &  &  \colhead{[10$^{-20}$ W/m$^2$]} &  & }
\colnumbers
\startdata
49 Ceti & \CIfull{} & $25.1\pm 1.6$  & 1 &\\
 & $^{13}$\CIfull{} & $2.1\pm 0.5$ \tablenotemark{a} & 2 &\\
 & \CIIfull{} & $370\pm 80$  & 3 &\\
 & $^{12}$CO 2--1 & $2.8\pm 0.3$  & & 2016.2.00200.S, 2018.1.01222.S\\
 & $^{12}$CO 3--2 & $7.4\pm 0.6$ \tablenotemark{b} & 4, 5 &\\
 & $^{13}$CO 2--1 & $0.92\pm 0.10$  & & 2016.2.00200.S, 2018.1.01222.S\\
 & C$^{18}$O 2--1 & $0.069\pm 0.014$  & & 2016.2.00200.S, 2018.1.01222.S\\
\hline
$\beta$ Pictoris & \CIfull{} & $16\pm 3$ \tablenotemark{c} & 6 &\\
 & \CIIfull{} & $(3.3\pm0.5)\times 10^{3}$  & & 1342198171\\
 & $^{12}$CO 2--1 & $3.5\pm 0.4$  & 7 &\\
 & $^{12}$CO 3--2 & $6.7\pm 0.7$  & 7 &\\
\hline
HD 21997 & \CIfull{} & $7.1\pm 0.8$  & & 2018.1.00633.S, 2019.1.01175.S\\
 & \CIIfull{} & $6\pm 20$ (${<}66$)  & & 1342247736\\
 & $^{12}$CO 2--1 & $1.58\pm 0.16$  & & 2011.0.00780.S\tablenotemark{d}\\
 & $^{12}$CO 3--2 & $2.9\pm 0.3$  & & 2011.0.00780.S\tablenotemark{d}\\
 & $^{13}$CO 2--1 & $0.67\pm 0.07$  & & 2011.0.00780.S\tablenotemark{d}\\
 & $^{13}$CO 3--2 & $1.64\pm 0.12$ \tablenotemark{e} & & 2011.0.00780.S, 2017.1.01575.S\tablenotemark{d}\\
 & C$^{18}$O 2--1 & $0.39\pm 0.05$  & & 2011.0.00780.S\tablenotemark{d}\\
 & C$^{18}$O 3--2 & $0.70\pm 0.08$  & & 2017.1.01575.S\tablenotemark{d}\\
\hline
HD 32297 & \CIfull{} & $4.0\pm 0.4$  & 8 &\\
 & \CIIfull{} & $270\pm 70$  & 9 &\\
 & $^{12}$CO 2--1 & $0.80\pm 0.06$ \tablenotemark{f} & 10, 11 &\\
 & $^{13}$CO 2--1 & $0.37\pm 0.05$  & 11 &\\
 & C$^{18}$O 2--1 & $0.20\pm 0.04$  & 11 &\\
\hline
HD 48370 & \CIfull{} & $0.0\pm 0.6$ (${<}1.8$)  & & 2019.2.00208.S\\
 & $^{12}$CO 2--1 & $0.04\pm 0.07$ (${<}0.26$)  & & 2016.2.00200.S\\
 & $^{13}$CO 2--1 & $0.04\pm 0.08$ (${<}0.27$)  & & 2016.2.00200.S\\
 & C$^{18}$O 2--1 & $0.00\pm 0.04$ (${<}0.11$)  & & 2016.2.00200.S\\
\hline
HD 61005 & \CIfull{} & $0.05\pm 0.15$ (${<}0.49$)  & & 2019.1.01603.S\\
 & $^{12}$CO 2--1 & $0.022\pm 0.015$ (${<}0.066$)  & & 2012.1.00437.S\\
\hline
HD 95086 & \CIfull{} & $0.0\pm 0.4$ (${<}1.1$)  & & 2019.1.01175.S\\
 & $^{12}$CO 1--0 & $(9.5\pm6)\times 10^{-3}$ (${<}0.028$)  & & 2016.A.00021.T\\
 & $^{12}$CO 2--1 & $(0\pm5)\times 10^{-3}$ (${<}0.016$)  & & 2013.1.00612.S, 2013.1.00773.S\\
 & $^{12}$CO 3--2 & $0.06\pm 0.14$ (${<}0.48$)  & & 2016.A.00021.T\\
\hline
HD 110058 & \CIfull{} & $0.15\pm 0.11$ (${<}0.46$)  & & 2019.1.01175.S\\
 & $^{12}$CO 2--1 & $0.064\pm 0.007$  & & 2012.1.00688.S, 2018.1.00500.S\\
 & $^{12}$CO 3--2 & $0.22\pm 0.02$  & & 2018.1.00500.S\\
 & $^{13}$CO 2--1 & $0.037\pm 0.005$  & & 2018.1.00500.S\\
 & $^{13}$CO 3--2 & $0.136\pm 0.016$  & & 2018.1.00500.S\\
\hline
HD 121191 & \CIfull{} & $0.32\pm 0.08$  & & 2019.1.01175.S\\
 & $^{12}$CO 2--1 & $0.161\pm 0.015$  & 12 &\\
 & $^{13}$CO 2--1 & $0.052\pm 0.016$  & 13 &\\
 & C$^{18}$O 2--1 & $0.000\pm 0.012$ (${<}0.037$)  & 13 &\\
\hline
HD 121617 & \CIfull{} & $6.1\pm 0.6$  & & 2019.1.01175.S\\
 & $^{12}$CO 2--1 & $0.98\pm 0.10$  & 13 &\\
 & $^{13}$CO 2--1 & $0.39\pm 0.05$  & 13 &\\
 & C$^{18}$O 2--1 & $0.058\pm 0.017$  & 13 &\\
\hline
HD 131835 & \CIfull{} & $4.6\pm 0.7$  & 14 &\\
 & \CIIfull{} & $0\pm 18$ (${<}53$)  & 15 &\\
 & $^{12}$CO 2--1 & $0.61\pm 0.03$  & 16 &\\
 & $^{12}$CO 3--2 & $1.29\pm 0.14$  & & 2013.1.01166.S\\
 & $^{13}$CO 3--2 & $0.45\pm 0.05$  & & 2013.1.01166.S\\
 & C$^{18}$O 3--2 & $0.20\pm 0.04$  & & 2013.1.01166.S\\
\hline
HD 146897 & \CIfull{} & $0.03\pm 0.15$ (${<}0.47$)  & & 2018.1.00633.S\\
 & $^{12}$CO 2--1 & $0.046\pm 0.012$  & 16 &\\
\hline
HD 181327 & \CIfull{} & $0.11\pm 0.16$ (${<}0.60$)  & & 2016.1.01253.S\\
 & \CIIfull{} & $0\pm 300$ (${<}760$)  & 17 &\\
 & $^{12}$CO 2--1 & $0.023\pm 0.004$  & 18 &\\
 & $^{12}$CO 3--2 & $0.02\pm 0.07$ (${<}0.23$)  & & 2015.1.00032.S\\
 & $^{13}$CO 2--1 & $0.089\pm 0.06$ (${<}0.26$)  & & 2013.1.01147.S\\
\hline
HR 4796 & \CIfull{} & $-0.1\pm 0.4$ (${<}1.0$)  & & 2017.A.00024.S\\
 & \CIIfull{} & $0\pm 180$ (${<}540$)  & 19 &\\
 & $^{12}$CO 2--1 & $0.00\pm 0.03$ (${<}0.077$)  & 12 &\\
 & $^{12}$CO 3--2 & $0.000\pm 0.010$ (${<}0.029$)  & 20 &\\
 & $^{12}$CO 6--5 & $0\pm 3$ (${<}9.2$)  & & 2016.A.00010.S\\
\enddata
\tablerefs{((1) \citet{Higuchi19}, (2) \citet{Higuchi19_13C}, (3) \citet{Roberge13}, (4) \citet{Hughes17}, (5) \citet{Nhung17}, (6) \citet{Cataldi18}, (7) \citet{Matra17_betaPic}, (8) \citet{Cataldi20}, (9) \citet{Donaldson13}, (10) \citet{MacGregor18}, (11) \citet{Moor19}, (12) \citet{Kral20_HD129590}, (13) \citet{Moor17}, (14) \citet{Kral19}, (15) \citet{Moor15_APEX}, (16) \citet{Lieman-Sifry16}, (17) \citet{Riviere-Marichalar14}, (18) \citet{Marino16}, (19) \citet{Meeus12}, (20) \citet{Kennedy18}}
\tablenotetext{a}{\citet{Higuchi19_13C} derive a \CI{}/$^{13}$\CI{} peak intensity ratio of $12\pm3$. We assume that the same ratio applies to the disk-integrated fluxes. To estimate the error on the disk-integrated $^{13}$\CI{} flux, we conservatively assume it has the same SNR as the peak ratio: 4.}
\tablenotetext{b}{The weighted average of the fluxes (see equation \ref{eq:error_on_weighted_mean}) reported by \citet{Hughes17} and \citet{Nhung17} is adopted, where we added a 10\% calibration error in quadrature to each of the reported error bars.}
\tablenotetext{c}{We added a 10\% calibration error to the flux error reported by \citet{Cataldi18}.}
\tablenotetext{d}{Data cubes used to derive the fluxes will be presented in a forthcoming publication (Matr\`a et al.\ in prep.).}
\tablenotetext{e}{Weighted average (see equation \ref{eq:error_on_weighted_mean}) of the disk-integrated fluxes measured from data sets 2011.0.00780.S and 2017.1.01575.S.}
\tablenotetext{f}{The two flux measurements were combined as described in table note \textit{b}, except that the calibration error was already included in the flux error reported by \citet{Moor19}.}
\tablecomments{(3) Values in parenthesis indicate 3$\sigma$ upper limits. In cases where only an upper limit was given in the literature, we adopt $0\pm K/n$ as the measured flux, with $K$ the reported upper limit and $n$ its significance in units of $\sigma$.}
\end{deluxetable}

Table~\ref{tab:disk_fluxes} also lists the disk-integrated line fluxes we used. For emission lines without published fluxes, we derived disk-integrated fluxes as described in Section \ref{sec:data_analysis}. Otherwise, we used fluxes from the literature, with the following exceptions. For HD~21997, fluxes of $^{12}$CO~2--1 and 3--2, $^{13}$CO~2--1 and 3--2, and C$^{18}$O~2--1 from program 2011.0.00780.S were published by \citet{Kospal13}. However, these data were recently re-analysed\footnote{We used visibility Measurement Sets provided directly by the European ALMA ARC in 2017, as those publicly available in the archive had missing or corrupted data. At the time, the ALMA Helpdesk reported that these data sets, originally published in \citet{Kospal13}, had several issues including frequency labelling. Those issues were fixed in the data sets provided by the ARC.} and supplemented by new observations of the 3--2 transition of $^{13}$CO and C$^{18}$O (program 2017.1.01575.S). We use naturally weighted data cubes derived from the re-analysed 2011.0.00780.S data and the 2017.1.01575.S data. These will be presented in more detail in a forthcoming publication (Matr{\`a} et al.\ in prep.). For $^{13}$CO~3--2, which was observed by both programs, we measure the disk-integrated fluxes from both data cubes (as described in section \ref{sec:flux_measurements_ALMA}) and adopt the weighted average as follows. The weights are $w_i=\sigma_i^{-2}$, with $\sigma_i$ the errors on the fluxes. The error on the weighted mean is then calculated as
\begin{equation}\label{eq:error_on_weighted_mean}
    \sigma = \sqrt{\left(\sum_{i=1}^{2} \sigma_i^{-2}\right)^{-1}}
\end{equation}
Our results are consistent with \citet{Kospal13}, except for C$^{18}$O~2--1 where we find a significantly higher flux.

For HD~95086, \citet{Booth19} presented fluxes of $^{12}$CO~1--0, 2--1 and 3--2, where the 2--1 transition was tentatively detected \citep[see also][]{Zapata18}. However, \citet{Booth21} recently proposed that HD~95086 belongs to the Carina young association, implying a different systemic velocity than assumed by \citet{Booth19}. Thus, the $^{12}$CO~2--1 signal is likely spurious and we re-derived the fluxes using the updated systemic velocity.

For HD~110058, \citet{Hales22} published CO fluxes after we already had conducted our flux measurements. Furthermore, for $^{12}$CO~2--1, they used data from program 2018.1.00500.S only, while we included additional data from program 2012.1.00688.S. We stick with the CO fluxes derived in this work, which are consistent with the results by \citet{Hales22}.

Finally, we also re-derived the \CIIfull{} flux toward $\beta$~Pictoris, as described in Section \ref{sec:flux_measurements_Herschel}. The same data were already analysed by \citet{Brandeker16}.

We note that after September 2021, more disks that meet our selection criteria as well as more line observations for the already selected disks have become available in the ALMA archive. Analysing an extended sample including these new data will be the subject of a future publication.

\subsection{ALMA data}
Here we describe the calibration and imaging of the ALMA (and ACA) data that were used to derive disk-integrated fluxes. We start with the pipeline-calibrated visibility data downloaded from the ALMA archive that we process and image as described below. The one exception is the CO data of HD~21997 where we use image cubes that will be presented in more detail in a forthcoming publication (Matr{\`a} et al.\ in prep., see Section \ref{sec:disk_sample}).

\subsubsection{Calibration}
We used modular \texttt{CASA} version 6.3 for processing of the pipeline-calibrated visibilities. In a first step, we produced a single MS file for each disk and emission line using the \texttt{concat} command. For the emission lines that were observed several times at different epochs, we verified that the proper motion of the target is sufficiently small compared to the beam size such that an alignment before concatenating of the different data sets is not necessary. The exception is the $^{12}$CO~2--1 line of HD~110058, observed in December 2013 (beam size ${\sim}0.8\arcsec$) and in April 2019 (beam size ${\sim}0.5\arcsec$). Between the two observations, the target moved by $0.18\arcsec$. For simplicity, we still chose to not align the two data sets. We verified that the disk-integrated flux does not change significantly when using the concatenated data set, or the higher SNR 2019 data only, suggesting that alignment would not change our results. We also note that we did not concatenate the HD~110058 $^{12}$CO~2--1 data sets initially, but waited with concatenation until after the continuum subtraction (see below). This is because \texttt{CASA}, for an unknown reason, was not able to subtract the continuum from the concatenated data.

The second step consisted of transforming the data to the barycentric reference frame using the \texttt{mstransform} command. We also removed the first and last ten channels of each data set, because edge channels often showed high noise. Next, for the emission lines where we used data from more than one observation program, we used the task \texttt{statwt} to reweight the visibilities according to their scatter\footnote{\url{https://casa.nrao.edu/docs/taskref/statwt-task.html}, accessed 2022-07-16.}. Reweighting assures that the relative weights between the different data sets are correct. The CLEAN algorithm we used to image the visibilities indeed depends on correct relative weights between the visibilities, although it is insensitive to an overall scaling of the weights. We then proceed to subtract the continuum in the $uv$-plane using the task \texttt{uvcontsub} with \texttt{fitorder=1}. For both the reweighting and the continuum subtraction, we exclude a range of $\pm20$\,km\,s$^{-1}$ around the wavelength of emission lines. We also exclude a small number of channels with abnormally high amplitudes. Finally, we use the task \texttt{split} to extract a range of $\pm30$\,km\,s$^{-1}$ around each emission line that we use for imaging.

We note that HD~48370 shows prominent CO~2--1 (and $^{13}$CO~2--1) emission at $v_\mathrm{bary}\approx41$\,km\,s$^{-1}$, potentially due to cloud contamination. The emission is sufficiently separated from the systemic velocity (23.6\,km\,s$^{-1}$) to not affect our analysis of CO emission from the disk. The contaminating CO emission was excluded for the continuum subtraction and continuum imaging.

\subsubsection{Imaging}\label{sec:imaging}
We use the \texttt{tclean} command with \texttt{deconvolver='multiscale'} and \texttt{scales=[0,5,15,25]} to produce image cubes. Briggs weighting with \texttt{robust=0.5} is employed, as we found it to yield similar SNR compared to natural weighting, but overall better beam properties that allow us to circumvent the JvM correction (see below). We use an elliptical CLEAN mask constructed based on the disk inclination, position angle, outer radius as well as the systemic velocity of the star. The mask is made conservatively large to ensure that all locations with potential disk emission are included. The threshold for the CLEAN algorithm is set to four times the rms measured from the dirty image. We also produce continuum images with the same procedure, using the non-continuum-subtracted visibilities and excluding data within $\pm20$\,km\,s$^{-1}$ of emission lines.

We produced additional data cubes by employing the JvM correction \citep{Jorsater95} as described by \citet[][but see also \citet{Casassus22} for a critique of the JvM correction]{Czekala21}. Here we provide a brief summary of the JvM correction. The final output image of the CLEAN algorithm is the sum of the residual map and the CLEAN model, where the latter is convolved with the CLEAN beam (a Gaussian fit to the dirty beam). However, the two maps have inconsistent units (Jy/(dirty beam) vs Jy/(CLEAN beam)), which can lead to wrong flux measurements. The JvM correction attempts to correct for this by scaling the residual map before adding it to the CLEAN model. The scaling factor is given by the ratio of the clean beam and dirty beam areas. This correction becomes important in situations where the areas differ strongly, i.e.\ where the dirty beam is strongly non-Gaussian. Scaling of the residual map implies that the noise in the image is scaled as well (because the noise is determined by the residual map). On the other hand, the total flux is only affected if a significant portion of it resides in the residual map (i.e.\ below the CLEANing threshold). Ultimately we decided to use the non-JvM-corrected images since the disk-integrated fluxes we derive are mostly unaffected by the correction (see section \ref{sec:flux_measurements_ALMA}).

\subsection{Herschel data}\label{sec:Herschel_data}
When available, we use published \CIIfull{} fluxes measured with Herschel (Table \ref{tab:disk_fluxes}). However, for HD~21997 the Herschel data have not been published yet. We also decided to re-analyse the data for $\beta$~Pic. For both targets, we used observations by Herschel/PACS \citep{Pilbratt10,Poglitsch10}. The PACS spectrometer array consists of $5\times5$ spatial pixels (spaxels) covering a $47\arcsec\times47\arcsec$ field of view. We directly use the data downloaded from the Herschel Science Archive.

HD~21997 was observed in 'Mapping' observing mode, meaning that a raster map was constructed using several pointings. We use the spatially resampled and mosaicked 'projected cube' (HPS3DPR). The beam FWHM is approximately $11.5\arcsec$\footnote{PACS Observer's Manual v.\ 2.5.1, Fig.\ 4.12, \url{https://www.cosmos.esa.int/documents/12133/996891/PACS+Observers\%27+Manual}}, thus we do not expect the target to be resolved. We extract a spectrum by spatially integrating the data cube over a circular aperture with a diameter equal to the beam FWHM, centred on the proper motion-corrected position of the star. We only consider the region between 157.1 and 158.5\,$\mu$m because the noise increases towards the edges of the spectral window. The FWHM of an unresolved line is $\Delta \lambda=0.126$\,$\mu$m (239\,km\,s$^{-1}$)\footnote{PACS Observer's Manual v.\ 2.5.1, Table~4.1}, implying that the \CII{} emission line will not be resolved. We subtract the continuum by fitting a linear polynomial to the spectrum, excluding the region $\lambda_0\pm(1.5\Delta\lambda)$ where $\lambda_0$ is the rest wavelength of the line.

$\beta$~Pic was observed in 'Pointed' observing mode, meaning that only a single pointing was carried out. Thus, instead of a raster map as in the case of HD~21997, we only get a single spectrum for each of the 25 spaxels of PACS. We use the spectrally rebinned cube (HPS3DRR). Given the beam size of $11.5\arcsec$, the spaxel size of $9.4\arcsec\times9.4\arcsec$ and a disk diameter of roughly $10\arcsec$, we can expect that emission is restricted to the central nine spaxels \citep[see also Fig.\ 3 of][]{Brandeker16}. We extract a single spectrum by summing over the central nine spaxels. We then subtract the continuum in the same way as for HD~21997.

\section{Data analysis}\label{sec:data_analysis}
In this section we describe how we measured disk-integrated fluxes.

\subsection{CO and \texorpdfstring{\CI{}}{CI} fluxes from ALMA data}

\subsubsection{Moment 0 maps}
\begin{figure*}
\plotone{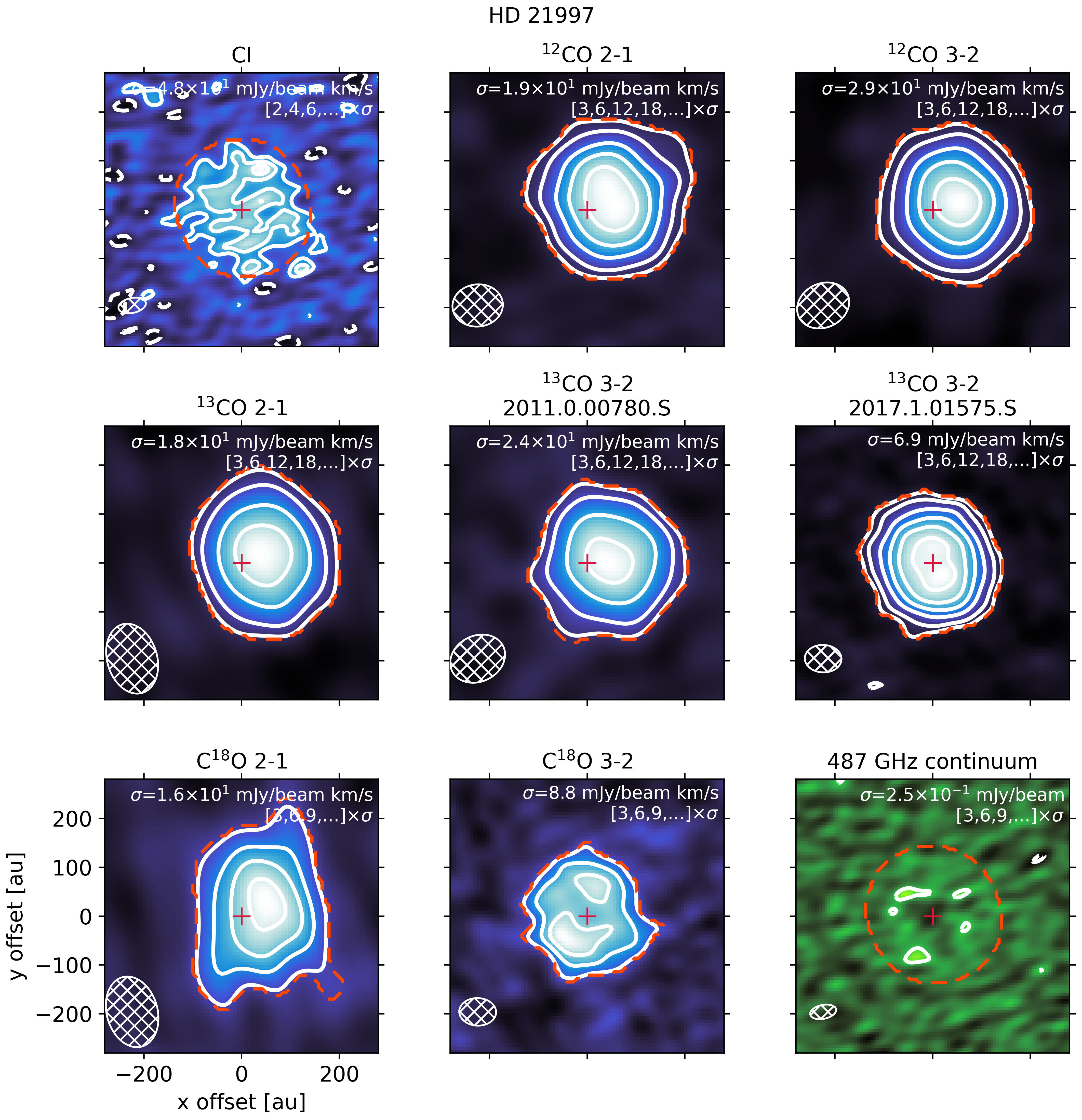}
\caption{Moment 0 and continuum maps used for flux measurements towards HD~21997. The noise and contour levels are indicated in the upper right of each image. The beam size is indicated by the hatched ellipse in the bottom left. The stellar position is marked by the red cross. The orange contours mark the aperture used to measure the flux. We note significant offsets between the expected stellar position and the disk center for the lines observed during cycle 0 ($^{12}$CO 2--1 and 3--2, $^{13}$CO 2--1 and 3--2 (from program 2011.0.00780.S for the latter) and C$^{18}$O 2--1). The moment 0 maps of the targets not shown in the main text are available in Appendix \ref{appendix:moment0_maps}.\label{fig:mom0_HD21997}}
\end{figure*}

\begin{figure*}
\plotone{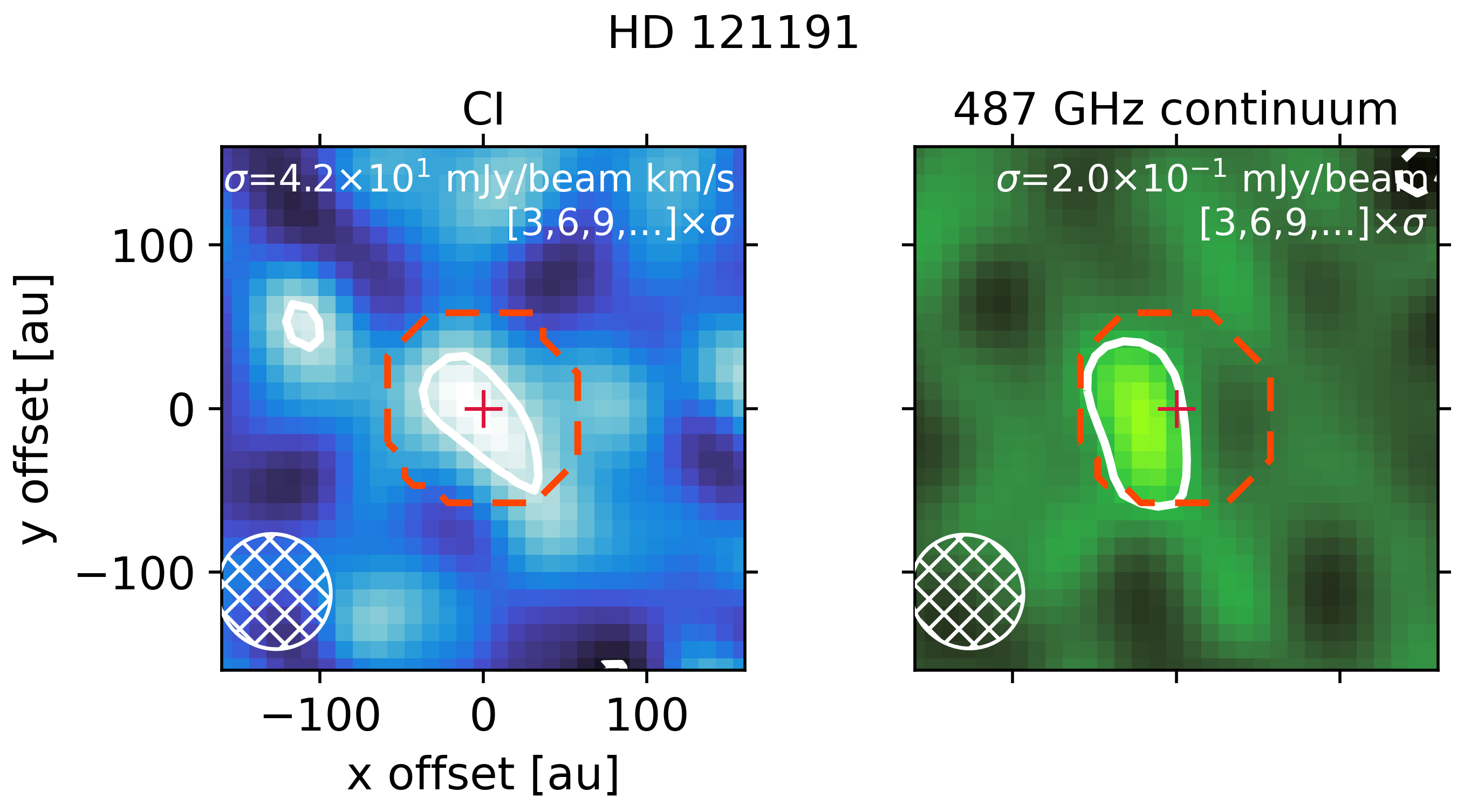}
\caption{Same as Fig.\ \ref{fig:mom0_HD21997}, but for HD~1211191. \label{fig:mom0_HD121191}}
\end{figure*}

\begin{figure*}
\plotone{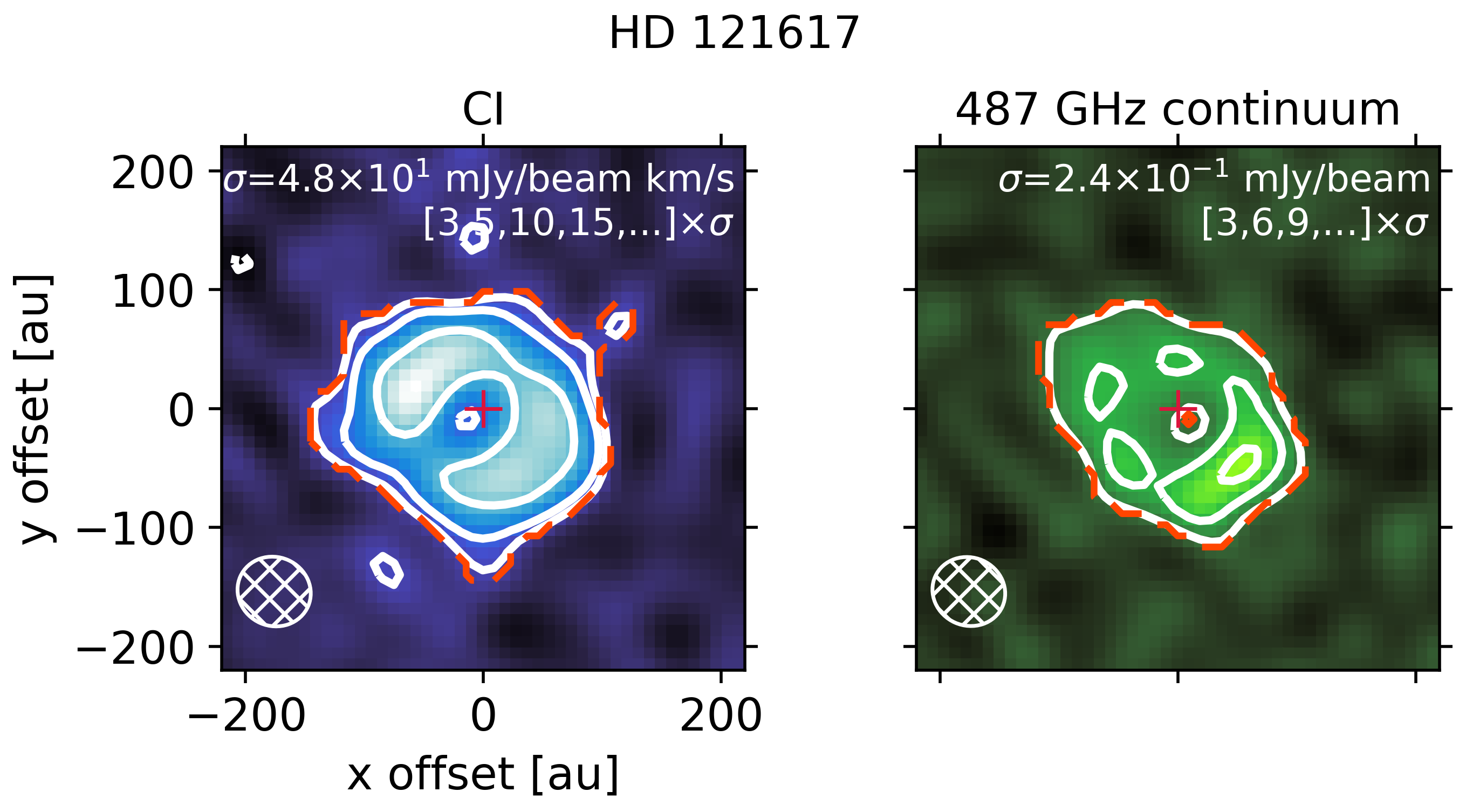}
\caption{Same as Fig.\ \ref{fig:mom0_HD21997}, but for HD~121617. \label{fig:mom0_HD121617}}
\end{figure*}

To measure CO and \CI{} fluxes, we first produce moment 0 maps. For each emission line, we first define a velocity range $v_\mathrm{sys}\pm\Delta v$ where $v_\mathrm{sys}$ is the systemic velocity and $\Delta v$ is the projected Keplerian velocity at the inner edge of the disk $r_\mathrm{in}$ (see Table~\ref{tab:disk_parameters} for the adopted values of $r_\mathrm{in}$). Thus,
\begin{equation}
    \Delta v = \sqrt{\frac{GM_*}{r_\mathrm{in}}}\sin i
\end{equation}
with the stellar mass $M_*$ and the disk inclination $i$ as listed in Table~\ref{tab:disk_parameters}. For low inclinations, this tends to zero, but our sample does not include very low inclination disks and thus the smallest $\Delta v$ is 1.9\,km\,s$^{-1}$. The adopted velocity range always spans several channels. We then extract a spectrum by spatially integrating the image cube over an elliptical region defined by the disk inclination, position angle and disk outer edge $r_\mathrm{out}$. We use this spectrum to verify by eye that our velocity range encompasses all emission (obviously, this is only possible in the cases where emission is detected). This is indeed the case, with one exception: for HD~121617, we needed to manually increase the velocity range from 6.7\,km\,s$^{-1}$ to 10\,km\,s$^{-1}$. Since for this disk, $r_\mathrm{in}$ is based on continuum observations, this could indicate that the gas extends further inwards than the dust. We then create a moment 0 map using the \texttt{bettermoments} package by integrating the image cube over the adopted velocity range. The moment 0 maps for the disks with new \CI{} detections are shown in Fig.\ \ref{fig:mom0_HD21997} (HD~21997), \ref{fig:mom0_HD121191} (HD~121191) and \ref{fig:mom0_HD121617} (HD~121617). For each disk, we show all moment 0 maps and continuum maps we created. The remaining moment 0 and continuum maps we produced and used for flux measurements are shown in Appendix \ref{appendix:moment0_maps}.

\subsubsection{Measurement of disk-integrated fluxes}\label{sec:flux_measurements_ALMA}

We now measure disk-integrated CO and \CI{} fluxes from the moment 0 maps by using apertures. We will consider two kinds of apertures: 1) apertures constructed using the known geometrical parameters of the disk ('geometrical aperture'), and 2) empirical apertures based on the observed emission in the moment 0 map ('empirical aperture'). The latter can only be applied if the SNR is sufficiently high.

Let us first consider the geometrical apertures. We consider a disk model with inner and outer cutoffs given by $r_\mathrm{in}$ and $r_\mathrm{out}$ (Table~\ref{tab:disk_parameters}) and constant scale height
\begin{equation}\label{eq:scale_height}
H=\sqrt{\frac{kTr_\mathrm{mean}^3}{\mu m_pGM_*}}
\end{equation}
with $k$ the Boltzmann constant, $T$ the temperature (fixed to 50\,K for simplicity), $\mu$ the mean molecular weight of the gas \citep[for simplicity fixed to 14,][]{Kral17} and $m_p$ the proton mass. We define the mean radius $r_\mathrm{mean}$ as
\begin{equation}
    r_\mathrm{mean}^2 = \frac{r_\mathrm{in}^2+r_\mathrm{out}^2}{2}
\end{equation}
such that the areas (and thus gas masses, given the constant scale height) inside and outside of $r_\mathrm{mean}$ are equal. The disk is placed in a 3D grid taking into account its inclination and position angle. We set any point within our grid to 1 if it satisfies $r_\mathrm{in}\leq r\leq r_\mathrm{out}$ (with $r$ the cylindrical coordinate) and is within $\pm2H$ of the disk midplane. All other points are set to 0. We then sum the grid along the line of sight. In the resulting 2D map, any point larger than 0 is set to 1. Finally, we convolve the map with the beam of the observations and normalise it to a peak of 1. Our aperture then consists of all points larger than 0.2.

In some cases (most significantly for the CO lines of HD~21997, see Fig.\ \ref{fig:mom0_HD21997}), there was a visible offset between the expected and observed disk center. We then shifted the geometrical aperture by eye so that all flux is included.

We then measure the noise $\xi$ in a region outside of the disk. Note that we use the non-primary beam corrected images to assure that the noise is uniform throughout the image. This is justified because the disks are small compared to the primary beam. Indeed, considering the normalized primary beam response (peak 1), no aperture extends further out than the 0.73 level of the primary beam. In 95\% of the cases, the aperture stays within the 0.86 level of the primary beam. We also verified that disk-integrated fluxes measured from primary beam corrected images are not significantly different. To this end, we measured fluxes from the primary beam corrected images using the same apertures as for the uncorrected images. The difference was always smaller than 0.4$\sigma$, and for the vast majority of the images, smaller than 0.1$\sigma$, where $\sigma$ denotes the error on the disk-integrated flux.

Next, we measure the peak SNR within the geometrical aperture. If the peak SNR within the geometrical aperture is below 8, the flux obtained by integrating over this aperture is used as our final flux measurement. If the SNR is above 8, we measured the flux from an empirical aperture instead. The empirical aperture is constructed as follows. We consider all pixels of the image with a flux $F>2.5\xi$, where $\xi$ is the noise in the moment 0 map. This results in a number of disjoint regions. The region corresponding to the disk can be easily identified as the one containing the highest flux. This region is adopted as our empirical aperture. We found that a cutoff at $2.5\xi$ gives a good compromise at including most of the real emission and excluding noise.

The moment 0 and continuum maps of Figures \ref{fig:mom0_HD21997}, \ref{fig:mom0_HD121191} and \ref{fig:mom0_HD121617} and in Appendix \ref{appendix:moment0_maps} show the apertures we used to measure the fluxes.

To estimate the error $\sigma$ on the disk-integrated flux, we place the (geometrical or empirical) aperture at various positions outside the disk (without overlap) and take the standard deviation of the collected flux samples. However, sometimes it is not possible to collect a sufficient number of non-overlapping flux samples because the aperture is too big and/or the image is too small. Thus, if less than 20 flux samples can be collected, we instead calculate $\sigma$ analytically as
\begin{equation}
    \sigma = \xi\Omega_p\sqrt{N_p}f = \xi\Omega_\mathrm{b}\sqrt{N_b}
\end{equation}
where $\Omega_p$ and $\Omega_b$ are the solid angles (in units of [sr]) of a pixel and the beam respectively, $N_p$ is the number of pixels in the aperture, $f=\sqrt{\Omega_b/\Omega_p}$ is the noise correlation ratio \citep[e.g.][]{Booth17} and $N_b$ is the number of beams in the aperture. Here the noise $\xi$ is in units of [W\,m$^{-2}$\,sr$^{-1}$].

For the lines for which enough flux samples can be collected, we compared the two error estimation methods. We find that the ratio $\sigma_\mathrm{sample}/\sigma_\mathrm{analytical}=0.7\pm0.1$, i.e.\ the analytical method tends to give larger errors, but the two methods are in reasonable agreement. Finally, we conservatively add a 10\% calibration error in quadrature\footnote{ALMA Cycle 9 Technical Handbook ver.\ 1.0, \url{https://almascience.nao.ac.jp/documents-and-tools/cycle9/alma-technical-handbook}}. Our results are listed in Table~\ref{tab:disk_fluxes}. Our work results in three new detections of \CI{} towards HD~21997, HD~121191 and HD~121617 (see Figures \ref{fig:mom0_HD21997}, \ref{fig:mom0_HD121191} and \ref{fig:mom0_HD121617}).

We also measured continuum fluxes with the same procedure as for line emission. For HD~95086 we excluded the background source identified by \citet{Zapata18} from our aperture, which implies that our measurements will tend to underestimate the total disk flux. Our continuum measurements are presented in Table~\ref{tab:continuum_fluxes} of appendix \ref{appendix:continuum_fluxes}.

As discussed in section \ref{sec:imaging}, we used image cubes without the JvM correction applied. As a test, we computed disk-integrated line fluxes using JvM-corrected image cubes. We find that for all disks and lines, the fluxes agree with the values given in Table~\ref{tab:disk_fluxes} within $0.7\sigma$. For the continuum, 84\% of the flux measurements are within $1\sigma$, with the maximum difference of $2.2\sigma$ occurring for HD~181327. These differences are partially due to the fact that the aperture chosen to measure the flux depends on the noise in the image, and the JvM correction changes the noise. We also find that the derived flux errors agree well. Overall, this indicates that our images have a dirty beam that is well approximated by a Gaussian. For naturally weighted images, there is a larger proportion of cases where the corrected and uncorrected line fluxes differ: 84\% are within $1\sigma$, with the maximum difference of $3.0\sigma$ being identified for $^{13}$CO~3--2 towards HD~110058.

\subsection{\texorpdfstring{\CII{}}{CII} fluxes from Herschel data}\label{sec:flux_measurements_Herschel}
We measure fluxes from the \CII{} spectra extracted from the Herschel/PACS data (see section \ref{sec:Herschel_data}). First, we measure the standard deviation $\varsigma$ of the flux data points, excluding the region $\lambda_0\pm(1.5\Delta\lambda)$ (where again $\lambda_0$ and $\Delta\lambda$ are the rest wavelength and FWHM of the line, respectively). We define a model for the emission line as
\begin{equation}
    F(\lambda,F_0)=\frac{F_0}{\sqrt{2\pi}\sigma}\exp\left(-\frac{(\lambda-\lambda_0)^2}{2\sigma^2}\right)
\end{equation}
where $\sigma=\Delta\lambda/(2\sqrt{2\ln2})$ and $F_0$ is the total line flux. We then consider the likelihood
\begin{equation}
    \mathcal{L}(F_0) = \prod_i\frac{1}{\sqrt{2\pi}f\varsigma}\exp\left(-\frac{(F(\lambda_i,F_0)-D_i)^2}{2(f\varsigma)^2}\right)
\end{equation}
where the product goes over all data points $i$ of the spectrum, $D_i$ are the measured flux values, and $f=\sqrt{\Delta\lambda/\delta\lambda}$ (with $\delta\lambda$ the channel width) is the noise correlation ratio \citep[e.g.][]{Booth17}. We verified that the likelihood is very well approximated by a Gaussian. Using $\mathcal{L}$ we can then determine the maximum likelihood estimate \citep[e.g.][]{Barlow89}, and error bars corresponding to the fluxes where $\ln\mathcal{L}$ has decreased by 0.5 with respect to its peak. This yields \CII{} fluxes of $(0.6\pm2.0)\times10^{-19}$\,W\,m$^{-2}$ for HD~21997 and $(3.3\pm0.5)\times10^{-17}$\,W\,m$^{-2}$ for $\beta$~Pic. For HD~21997, a $3\sigma$ upper limit of $6.6\times10^{-19}$\,W\,m$^{-2}$ is derived by determining the flux where $\ln\mathcal{L}$ has decreased by 4.5 with respect to its peak. These maximum likelihood estimates of the \CII{} fluxes are reported in Table~\ref{tab:disk_fluxes} and used in our MCMC modelling. However, we may go one step further and derive a posterior probability distribution by multiplying the likelihood with a prior that is zero for $F_0<0$, and flat for $F_0\geq0$. For HD~21997, this yields $1.6^{+1.6}_{-1.1}\times10^{-19}$\,W\,m$^{-2}$ (50th percentile with 15.9th and 84.1th percentile as error bars), or a 99\% upper limit of $5.6\times10^{-19}$\,W\,m$^{-2}$. For $\beta$~Pic, the maximum likelihood and the posterior distribution yield identical results because the line is clearly detected.

\citet{Brandeker16} also analysed the \CII{} data of $\beta$~Pic, but they only report the fluxes from the four spaxels for which they detect emission. Summing their values (with errors summed in quadrature) gives $(3.2\pm0.3)\times10^{-17}$\,W\,m$^{-2}$, consistent with our result. Note that \CII{} towards $\beta$~Pic was also observed by Herschel/HIFI, but the HIFI beam barely covers the disk \citep{Cataldi14}. Therefore, the data from Herschel/PACS are better suited to measure the total flux.

\section{Modelling}\label{sec:modelling}
Our main goal is to derive the masses and disk-averaged column densities of carbon and CO for each of our targets. We employ a simple disk model that is fitted to the disk-integrated fluxes that we derived in the previous section.

\subsection{LTE models}\label{sec:LTE_models}
We first present a model where local thermodynamic equilibrium (LTE) is assumed. The free parameters of the model are the total CO mass $M(\mathrm{CO})$, the total C$^0$ mass $M(\mathrm{C^0})$, the total C$^+$ mass $M(\mathrm{C^+})$ (if \CII{} was observed), the CO temperature $T_\mathrm{CO}$, and the carbon temperature $T_\mathrm{C}$ (see Table \ref{tab:fit_parameters}). Thus, we assume that C$^0$ and C$^+$ share the same temperature. This assumption is made because half of our disks do not have \CII{} data. In Appendix \ref{appendix:test_TC0_TCplus}, for the disks with \CII{} data, we run additional fits where the temperatures of C$^0$ and C$^+$ are separated. No significant changes in the derived gas masses are observed when separating the C$^0$ and C$^+$ temperatures.

The mass ratios of C$^0$/$^{13}$C$^0$, $^{12}$CO/$^{13}$CO and $^{12}$CO/C$^{18}$O are fixed to ISM abundances \citep[$^{12}\mathrm{C}/^{13}\mathrm{C}=69$, $^{16}\mathrm{O}/^{18}\mathrm{O}=557$,][]{Wilson99}. With this assumption we explicitly ignore the possibility of CO isotopologue-selective photo-dissociation which would tend to increase the $^{12}$CO/$^{13}$CO and $^{12}$CO/C$^{18}$O ratios \citep[e.g.][]{Visser09}. However, the assumption is necessary to constrain the CO mass in the cases where the $^{12}$CO emission is optically thick, although it might lead to an underestimation of the CO mass if isotopologue-selective photodissociation is active.

We assume a disk with a uniform density for $r_\mathrm{in}\leq r \leq r_\mathrm{out}$ (see Table~\ref{tab:disk_parameters} for the adopted values), and within $\pm1.5H$ of the disk midplane, where $H$ is the scale height given by equation \ref{eq:scale_height}. For simplicity, the scale height is fixed, that is, it does not vary with radius nor with the temperature: $r=r_\mathrm{mean}$ and $T=50$\,K are fixed to calculate $H$. Choosing $T=50$\,K is arbitrary, but given the weak dependence of $H$ on temperature, we do not expect this choice to affect our results. Indeed, compared to $H(T=50\mathrm{K})$, $H$ would vary by a factor ${\sim}2$ at most for the temperature range (10--200\,K) we consider. We also note that in reality, the vertical structure is unlikely to be the same for each species as it depends on a number of factors and the interplay between photochemistry and vertical diffusion \citep[e.g.][]{Marino22}.

For a given set of parameter values, we can then compute the total line emission by ray-tracing the emission along the line of sight of the 3D disk model \citep[e.g.][]{Cataldi14}, assuming a Keplerian velocity field that is considered independent of $z$ (the direction perpendicular to the midplane). However, for low gas densities, the gas might not be in Keplerian rotation, but rather be blown out as a wind \citep{Kral23}. As discussed in Sect.\ \ref{sec:wind}, this caveat in unlikely to affect our results. Since we assume LTE, the level populations are computed from the Boltzmann equation. The atomic data is taken from the LAMDA database\footnote{\url{https://home.strw.leidenuniv.nl/~moldata/}}. The local intrinsic emission was assumed as a square line profile with a fixed width of 0.5\,km\,s$^{-1}$. In Appendix \ref{appendix:line_width_tests} we discuss how our fits change when the line width is changed.

We initially explored an even simpler model that ignored the velocity field of the disk, but found that such a model can easily overestimate the optical depth of the lines, resulting in highly uncertain column density estimates, particularly for highly inclined disks.

The parameter space is explored using the MCMC method implemented by the \texttt{emcee} package \citep{Foreman-Mackey13}. The log likelihood is given by
\begin{equation}
    \ln \mathcal{L}(\mathcal{P}) = -\sum_i \frac{(F_i(\mathcal{P})-D_i)^2}{2\sigma_i^2}
\end{equation}
where the sum runs over all observed emission lines, $\mathcal{P}$ represents the parameter values, $F_i$ is the model flux, and $D_i$ and $\sigma_i$ are the observed flux and error as given in Table~\ref{tab:disk_fluxes}. The posterior probability is found by multiplying the likelihood $\mathcal{L}$ with a flat prior for each parameter, with limits detailed in Table~\ref{tab:fit_parameters}. The influence of the chosen temperature priors on our results is discussed in Appendix \ref{appendix:T_prior}.

An additional prior is applied that ensures $T_\mathrm{CO}\leq T_\mathrm{C}$. This represents our expectation that C is either well-mixed with CO ($T_\mathrm{CO}=T_\mathrm{C}$) or preferentially occupies regions where CO is efficiently photodissociated such as the disk atmosphere \citep{Marino22} or the inner disk regions where the temperature is expected to be higher \citep[e.g.][]{Kral19}. Note though that our disk model implicitly assumes that the gas is well-mixed by considering an identical spatial distribution for each species.

\begin{deluxetable}{CcCC}
\tablecaption{Parameters used in our model fitting.\label{tab:fit_parameters}}
\tablewidth{0pt}
\tablehead{ \colhead{parameter} & \colhead{units} & \colhead{min}& \colhead{max}}
\colnumbers
\startdata
\log M(\mathrm{CO}) & $\log M_\oplus$ & -8 & 1\\
\log M(\mathrm{C^0}) & $\log M_\oplus$ & -8 & 1\\
\log M(\mathrm{C^+})\tablenotemark{a} & $\log M_\oplus$ & -8 & 1\\
T_\mathrm{CO} & K & 10 & 200\\
T_\mathrm{C} & K & 10 & 200\\
\log n(\mathrm{H_2})\tablenotemark{b} & $\log$ cm$^{-3}$ & 1 & 7\\
\log n(\mathrm{e}^-)\tablenotemark{b} & $\log$ cm$^{-3}$ & -1 & 5\\
\enddata
\tablenotetext{a}{Used only if \CII{} data were available.}
\tablenotetext{b}{Used for the non-LTE models only.}
\tablecomments{(3) Value below which the flat prior is zero. (4) Value above which the flat prior is zero.}
\end{deluxetable}

For computational efficiency, we pre-compute the model fluxes on a grid of parameter values and interpolate on that grid during the MCMC run. The grid has a resolution of 5\,K (for $T<30$\,K) respectively 25\,K (for $T>30$\,K) for temperature, at least 0.5 dex for gas masses, and 0.5 dex for collider number densities (used for the non-LTE models, see below). We tested the grid by comparing interpolated and directly calculated fluxes and found excellent agreement. For the MCMC, we employ 1000 walkers, each taking 20'000 steps. We discard the first 15'000 steps as burn-in.

\subsection{Non-LTE models}\label{sec:nonLTE_models}
We explore additional models where the LTE assumption is dropped. We introduce an additional free parameter: either the H$_2$ number density $n(\mathrm{H}_2)$, or the electron number density $n(\mathrm{e}^-)$. Note that in the latter case, we do not assume $n(\mathrm{C}^+)=n(\mathrm{e}^-)$, that is, $n(\mathrm{e}^-)$ is an independent parameter. The level populations are determined by using a statistical equilibrium calculation in the radiative transfer code \texttt{pythonradex}\footnote{\url{https://github.com/gica3618/pythonradex}}, where the input for the \texttt{pythonradex} code is the column density along the line of sight $N_\mathrm{los}$ corresponding to a given gas mass. The radiative transfer is then calculated in the same way as for the LTE models.

\section{Results}\label{sec:results}
\subsection{LTE models}\label{sec:results_LTE}
In Table~\ref{tab:fit_results_LTE} we present summary statistics of the posterior probability distribution we derived for each parameter in the LTE case. We also transformed the derived masses to column densities. We derive the averaged column density in the $z$ direction $N_\perp$ by dividing the mass by the face-on projection of the disk $\pi(r_\mathrm{out}^2-r_\mathrm{in}^2)$ and the molecular mass. We also derive the averaged column density along the line of sight $N_\mathrm{los}$ by considering the area of the disk projected onto the sky plane. To determine the latter, we project our 3D disk model along the line of sight and then sum the area of all non-zero pixels.

In Table~\ref{tab:fit_results_LTE}, we list upper limits (99th percentile) for the mass and column density whenever the corresponding disk-integrated fluxes have a $\mathrm{SNR}<3$. The table also lists lower limits whenever a parameter has a substantial probability to be close to the upper bound of the prior. This is formalised as
\begin{equation}\label{eq:lower_limit_condition}
    \int_{p_\mathrm{max}-0.2W}^{p_\mathrm{max}} P(p)\mathrm{d}p > 5\%
\end{equation}
where $p_\mathrm{max}$ is the maximum parameter value allowed by the prior, $W$ is a proxy for the width of the posterior distribution calculated as the difference between the 99th and the 1st percentile, and $P$ is the posterior probability density.

If the \CII{} line was not observed, we estimate the amount of ionised carbon with a simple ionisation equilibrium calculation. For a given C$^0$ mass, we calculate the ionisation balance by assuming uniform density and that the electron density equals the C$^+$ density. We use ionisation cross sections \citep{Nahar91} and recombination coefficients \citep{Nahar95,Nahar96} from the NORAD-Atomic-Data database\footnote{\url{https://norad.astronomy.osu.edu}}. For the stellar flux, we use ATLAS9 models \citep{Castelli03} with the parameters listed in Table~\ref{tab:disk_parameters} and scaled by $\exp(-\alpha_\mathrm{C} n(\mathrm{C}^0)(r_\mathrm{mean}-r_\mathrm{in}))$ to take into account extinction by C$^0$ ionisation. Here, $\alpha_\mathrm{C}$ is the ionisation cross section and $n(\mathrm{C}^0)$ the uniform C$^0$ number density. For the interstellar radiation, we use the Draine field \citep{Draine78,Lee84} scaled by $\exp(-\alpha_\mathrm{C} N_\perp(\mathrm{C}^0)/2)$. This calculation implicitly assumes that C$^0$ and C$^+$ are well-mixed. \citet{Marino22} found that C$^+$ tends to form an upper layer in the disk independently of the strength of vertical mixing. Therefore, our simple model may underestimate the total C$^+$ mass because it only accounts for the C$^+$ co-located with C$^0$.

\movetabledown=60mm
\begin{rotatetable}
\begin{deluxetable}{lLLLLLLLLLLL}
\tablecaption{Median as well as 16th and 84th percentiles of the posterior probability distribution derived with our MCMC runs assuming LTE.\label{tab:fit_results_LTE}}
\tablewidth{0pt}
\tablehead{ \colhead{star} & \colhead{$\log M(\mathrm{CO})$} & \colhead{$\log M(\mathrm{C^0})$} & \colhead{$\log M(\mathrm{C^+})$} & \colhead{$T_\mathrm{CO}$} & \colhead{$T_\mathrm{C}$} & \colhead{$\log N_\perp(\mathrm{CO})$} & \colhead{$\log N_\perp(\mathrm{C^0})$} & \colhead{$\log N_\perp(\mathrm{C^+})$} & \colhead{$\log N_\mathrm{los}(\mathrm{CO})$} & \colhead{$\log N_\mathrm{los}(\mathrm{C^0})$} & \colhead{$\log N_\mathrm{los}(\mathrm{C^+})$}\\
& \colhead{[$\log M_\oplus$]} & \colhead{[$\log M_\oplus$]} & \colhead{[$\log M_\oplus$]}& \colhead{[K]} & \colhead{[K]}& \colhead{[$\log \mathrm{cm}^{-2}$]} & \colhead{[$\log \mathrm{cm}^{-2}$]} & \colhead{[$\log \mathrm{cm}^{-2}$]} & \colhead{[$\log \mathrm{cm}^{-2}$]} & \colhead{[$\log \mathrm{cm}^{-2}$]} & \colhead{[$\log \mathrm{cm}^{-2}$]}}
\colnumbers
\startdata
49 Ceti& -2.15_{-0.07}^{+0.07}& -2.05_{-0.17}^{+0.2}& -1.1_{-1.0}^{+0.8}& 15.2_{-0.6}^{+0.6}& 24_{-2}^{+3}& 16.74_{-0.07}^{+0.07}& 17.21_{-0.17}^{+0.2}& 18.1_{-1.0}^{+0.8}& 17.41_{-0.07}^{+0.07}& 17.88_{-0.17}^{+0.2}& 18.8_{-1.0}^{+0.8}\\
 &  &  &  {>}-3.6 &  &  &  &  &  {>}15.7 &  &  &  {>}16.3\\
$\beta$ Pictoris& -4.32_{-0.15}^{+0.10}& -3.75_{-0.09}^{+0.08}& -3.45_{-0.16}^{+1.5}& 11.8_{-1.1}^{+2}& 92_{-60}^{+70}& 14.58_{-0.15}^{+0.10}& 15.52_{-0.09}^{+0.08}& 15.82_{-0.16}^{+1.5}& 15.59_{-0.15}^{+0.10}& 16.52_{-0.09}^{+0.08}& 16.82_{-0.16}^{+1.5}\\
 &  &  &  {>}-3.7 &  &  &  &  &  {>}15.5 &  &  &  {>}16.5\\
HD 21997& -1.32_{-0.09}^{+0.07}& -2.90_{-0.09}^{+1.0} & {<}-0.33& 10.25_{-0.18}^{+0.3}& 67_{-50}^{+90}& 17.71_{-0.09}^{+0.07}& 16.50_{-0.09}^{+1.0} & {<}19.1& 17.77_{-0.09}^{+0.07}& 16.55_{-0.09}^{+1.0} & {<}19.1\\
 &  &  {>}-3.1 &  &  &  &  &  {>}16.3 &  &  &  {>}16.3 & \\
HD 32297& -1.28_{-0.15}^{+0.12}& -2.53_{-0.05}^{+0.08}& -2.4_{-0.5}^{+1.6}& 25.3_{-1.8}^{+1.8}& 56_{-16}^{+90}& 17.72_{-0.15}^{+0.12}& 16.84_{-0.05}^{+0.08}& 16.9_{-0.5}^{+1.6}& 18.26_{-0.15}^{+0.12}& 17.37_{-0.05}^{+0.08}& 17.5_{-0.5}^{+1.6}\\
 &  &  &  {>}-3.8 &  &  &  &  &  {>}15.6 &  &  &  {>}16.1\\
HD 48370 & {<}-4.9 & {<}-4.4 & {<}-3.9\tablenotemark{a}& 69_{-40}^{+60}& 150_{-60}^{+40} & {<}14.2 & {<}15.1 & {<}15.5\tablenotemark{a} & {<}14.5 & {<}15.4 & {<}15.9\tablenotemark{a}\\
HD 61005 & {<}-5.4 & {<}-4.9 & {<}-4.6\tablenotemark{a}& 69_{-40}^{+60}& 150_{-60}^{+40} & {<}14.2 & {<}15.1 & {<}15.4\tablenotemark{a} & {<}15.1 & {<}15.9 & {<}16.3\tablenotemark{a}\\
HD 95086 & {<}-5.4 & {<}-3.9 & {<}-3.0\tablenotemark{a}& 70_{-40}^{+60}& 150_{-60}^{+40} & {<}12.9 & {<}14.8 & {<}15.6\tablenotemark{a} & {<}12.9 & {<}14.8 & {<}15.6\tablenotemark{a}\\
HD 110058& -2.03_{-0.12}^{+0.14} & {<}-3.5 & {<}-4.1\tablenotemark{a}& 102_{-8}^{+8}& 150_{-30}^{+30}& 18.63_{-0.12}^{+0.14} & {<}17.6 & {<}16.9\tablenotemark{a}& 19.62_{-0.12}^{+0.14} & {<}18.5 & {<}17.9\tablenotemark{a}\\
HD 121191& -2.25_{-0.3}^{+0.20}& -3.68_{-0.13}^{+0.11}& -5.01_{-0.04}^{+0.03}\tablenotemark{a}& 91_{-8}^{+9}& 140_{-40}^{+40}& 18.16_{-0.3}^{+0.20}& 17.10_{-0.13}^{+0.11}& 15.78_{-0.04}^{+0.03}\tablenotemark{a}& 18.21_{-0.3}^{+0.20}& 17.15_{-0.13}^{+0.11}& 15.82_{-0.04}^{+0.03}\tablenotemark{a}\\
HD 121617& -1.59_{-0.07}^{+0.07}& -2.42_{-0.06}^{+0.12}& -3.49_{-0.09}^{+0.07}\tablenotemark{a}& 42_{-4}^{+4}& 100_{-50}^{+70}& 17.78_{-0.07}^{+0.07}& 17.32_{-0.06}^{+0.12}& 16.24_{-0.09}^{+0.07}\tablenotemark{a}& 17.85_{-0.07}^{+0.07}& 17.39_{-0.06}^{+0.12}& 16.31_{-0.09}^{+0.07}\tablenotemark{a}\\
HD 131835& -1.76_{-0.08}^{+0.06}& -2.48_{-0.11}^{+0.8} & {<}-0.45& 12.0_{-0.3}^{+0.3}& 74_{-60}^{+90}& 17.02_{-0.08}^{+0.06}& 16.67_{-0.11}^{+0.8} & {<}18.7& 17.41_{-0.08}^{+0.06}& 17.05_{-0.11}^{+0.8} & {<}19.1\\
HD 146897& -4.4_{-0.2}^{+1.2} & {<}-3.9 & {<}-4.2\tablenotemark{a}& 43_{-30}^{+70}& 140_{-60}^{+50}& 15.3_{-0.2}^{+1.2} & {<}16.2 & {<}15.8\tablenotemark{a}& 15.8_{-0.2}^{+1.2} & {<}16.7 & {<}16.4\tablenotemark{a}\\
HD 181327& -5.72_{-0.12}^{+0.3} & {<}-4.5 & {<}-1.4& 39_{-20}^{+60}& 130_{-70}^{+50}& 14.00_{-0.12}^{+0.3} & {<}15.6 & {<}18.7& 13.99_{-0.12}^{+0.3} & {<}15.6 & {<}18.7\\
HR 4796 & {<}-6.0 & {<}-4.0 & {<}-0.75& 65_{-40}^{+60}& 140_{-70}^{+40} & {<}13.9 & {<}16.3 & {<}19.5 & {<}14.2 & {<}16.6 & {<}19.8\\
\enddata
\tablenotetext{a}{Estimated from an ionisation calculation, because no \CII{} data were available. See text for details.}
\tablecomments{Upper limits corresponding to the 99th percentile are given if the disk-integrated flux has $\mathrm{SNR}<3$. Lower limits corresponding to the 1st percentile are given if equation \ref{eq:lower_limit_condition} is satisfied. (7), (8), (9): column density in the $z$ direction (perpendicular to the midplane). (10), (11), (12): column density along the line of sight.}
\end{deluxetable}
\end{rotatetable}

In Table~\ref{tab:fitted_tau} we present the optical depth we derived for each emission line.

\startlongtable
\begin{deluxetable}{llL}
\tablecaption{Median as well as 16th and 84th percentile of the optical depth along the line of sight, derived from MCMC runs assuming LTE.\label{tab:fitted_tau}}
\tablehead{ \colhead{star} & \colhead{emission line} & \colhead{$\tau(\nu_0)$} }
\tablewidth{0pt}
\startdata
49 Ceti & \CIfull{} & 11_{-5}^{+8}\\
 & $^{13}$\CIfull{} & 0.15_{-0.06}^{+0.09}\\
 & \CIIfull{} & 110_{-100}^{+600}\;({>}0.084)\\
 & $^{12}$CO 2--1 & 121_{-14}^{+16}\\
 & $^{12}$CO 3--2 & 120_{-20}^{+20}\\
 & $^{13}$CO 2--1 & 1.53_{-0.18}^{+0.20}\\
 & C$^{18}$O 2--1 & 0.20_{-0.02}^{+0.03}\\
\hline
$\beta$ Pictoris & \CIfull{} & 0.080_{-0.04}^{+0.3}\\
 & \CIIfull{} & 0.45_{-0.3}^{+30}\;({>}0.12)\\
 & $^{12}$CO 2--1 & 2.5_{-0.9}^{+1.3}\\
 & $^{12}$CO 3--2 & 1.8_{-0.5}^{+0.3}\\
\hline
HD 21997 & \CIfull{} & 0.10_{-0.07}^{+10}\;({>}0.025)\\
 & \CIIfull{} & {<}260\\
 & $^{12}$CO 2--1 & 450_{-90}^{+60}\\
 & $^{12}$CO 3--2 & 280_{-70}^{+60}\\
 & $^{13}$CO 2--1 & 6.0_{-1.4}^{+1.0}\\
 & $^{13}$CO 3--2 & 3.7_{-1.0}^{+0.8}\\
 & C$^{18}$O 2--1 & 0.64_{-0.09}^{+0.18}\\
 & C$^{18}$O 3--2 & 0.46_{-0.07}^{+0.05}\\
\hline
HD 32297 & \CIfull{} & 1.2_{-0.9}^{+1.6}\\
 & \CIIfull{} & 3.9_{-3}^{+200}\;({>}0.087)\\
 & $^{12}$CO 2--1 & 580_{-200}^{+200}\\
 & $^{13}$CO 2--1 & 7.1_{-2}^{+2}\\
 & C$^{18}$O 2--1 & 0.86_{-0.3}^{+0.5}\\
\hline
HD 48370 & \CIfull{} & {<}6.1\times 10^{-3}\\
 & $^{12}$CO 2--1 & {<}0.042\\
 & $^{13}$CO 2--1 & {<}7.0\times 10^{-4}\\
 & C$^{18}$O 2--1 & {<}3.3\times 10^{-4}\\
\hline
HD 61005 & \CIfull{} & {<}0.022\\
 & $^{12}$CO 2--1 & {<}0.18\\
\hline
HD 95086 & \CIfull{} & {<}1.2\times 10^{-3}\\
 & $^{12}$CO 1--0 & {<}4.0\times 10^{-4}\\
 & $^{12}$CO 2--1 & {<}9.2\times 10^{-4}\\
 & $^{12}$CO 3--2 & {<}9.6\times 10^{-4}\\
\hline
HD 110058 & \CIfull{} & {<}5.9\\
 & $^{12}$CO 2--1 & 940_{-200}^{+500}\\
 & $^{12}$CO 3--2 & (1.8_{-0.4}^{+1.2})\times 10^{3}\\
 & $^{13}$CO 2--1 & 12_{-4}^{+7}\\
 & $^{13}$CO 3--2 & 24_{-8}^{+12}\\
\hline
HD 121191 & \CIfull{} & 0.16_{-0.06}^{+0.09}\\
 & $^{12}$CO 2--1 & 39_{-18}^{+20}\\
 & $^{13}$CO 2--1 & 0.48_{-0.2}^{+0.3}\\
 & C$^{18}$O 2--1 & 0.060_{-0.03}^{+0.04}\\
\hline
HD 121617 & \CIfull{} & 0.42_{-0.2}^{+1.0}\\
 & $^{12}$CO 2--1 & 83_{-15}^{+16}\\
 & $^{13}$CO 2--1 & 1.1_{-0.2}^{+0.2}\\
 & C$^{18}$O 2--1 & 0.13_{-0.03}^{+0.03}\\
\hline
HD 131835 & \CIfull{} & 0.30_{-0.18}^{+19}\\
 & \CIIfull{} & {<}240\\
 & $^{12}$CO 2--1 & 170_{-40}^{+40}\\
 & $^{12}$CO 3--2 & 114_{-13}^{+11}\\
 & $^{13}$CO 3--2 & 1.52_{-0.17}^{+0.2}\\
 & C$^{18}$O 3--2 & 0.21_{-0.05}^{+0.04}\\
\hline
HD 146897 & \CIfull{} & {<}0.37\\
 & $^{12}$CO 2--1 & 0.98_{-0.7}^{+130}\\
\hline
HD 181327 & \CIfull{} & {<}0.016\\
 & \CIIfull{} & {<}78\\
 & $^{12}$CO 2--1 & 0.012_{-0.008}^{+0.02}\\
 & $^{12}$CO 3--2 & 0.017_{-0.010}^{+0.015}\\
 & $^{13}$CO 2--1 & (2.7_{-2}^{+6})\times 10^{-4}\\
\hline
HR 4796 & \CIfull{} & {<}0.20\\
 & \CIIfull{} & {<}1.8\times 10^{3}\\
 & $^{12}$CO 2--1 & {<}0.11\\
 & $^{12}$CO 3--2 & {<}0.094\\
 & $^{12}$CO 6--5 & {<}0.025\\
\enddata
\tablecomments{Upper limits correspond to the 99th percentile and are given if the SNR of the disk-integrated flux is less than 3. Lower limits corresponding to the 1st percentile are given in parenthesis if equation \ref{eq:lower_limit_condition} is satisfied.}
\end{deluxetable}

\begin{figure*}
\plotone{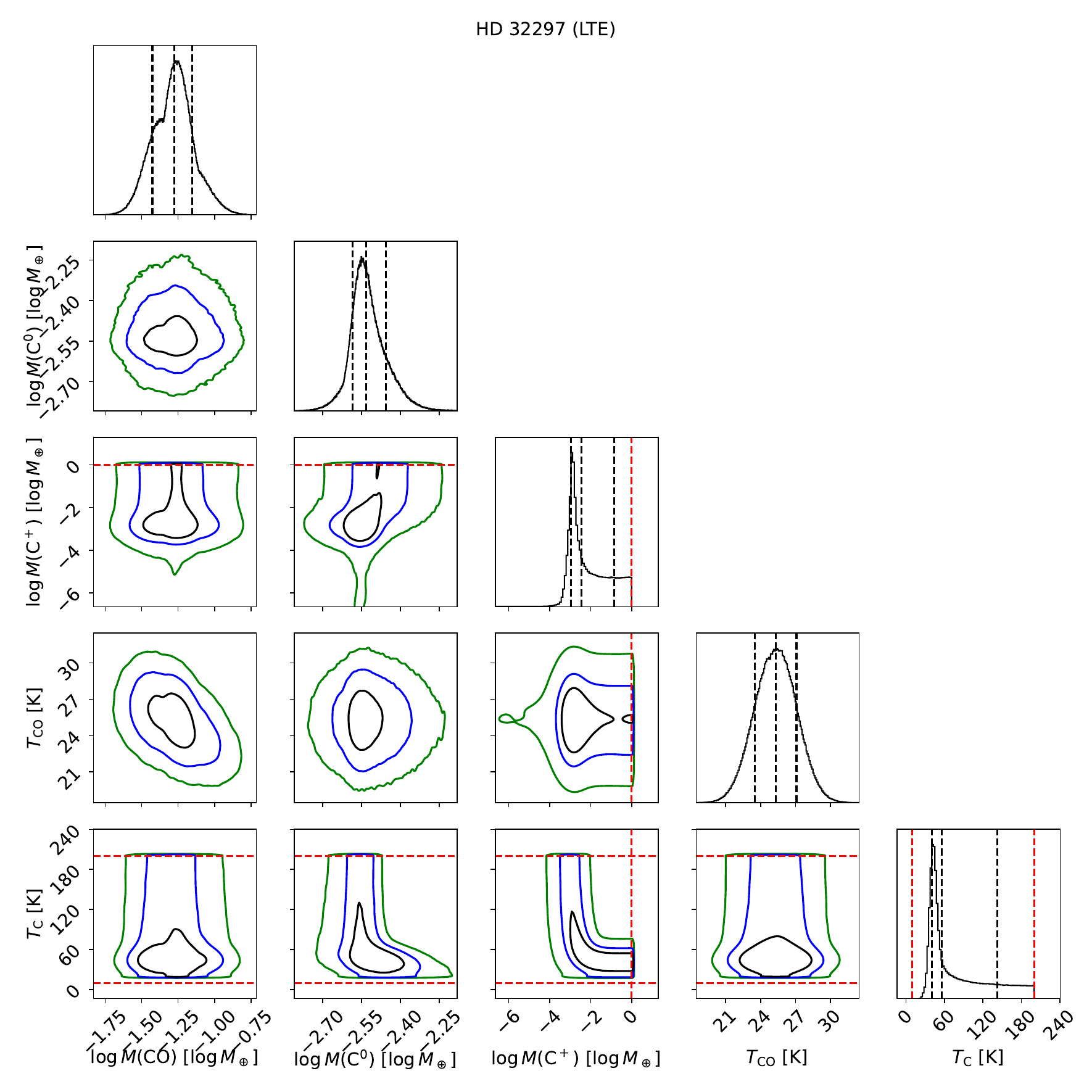}
\caption{Corner plot showing posterior probability distributions derived assuming LTE, for the disk around HD~32297. In the 1D histograms, the black vertical dashed lines indicate the 16th, 50th and 84th percentile. In the 2D histograms, the black, blue and green contours mark the 50th, 90th and 99th percentile. Red dashed lines mark the upper or lower bound of the prior distribution. The corner plots from the LTE runs not shown in the main text can be found in Appendix \ref{appendix:corner_plots_LTE}.}\label{fig:corner_HD32297}
\end{figure*}

\begin{figure*}
\plotone{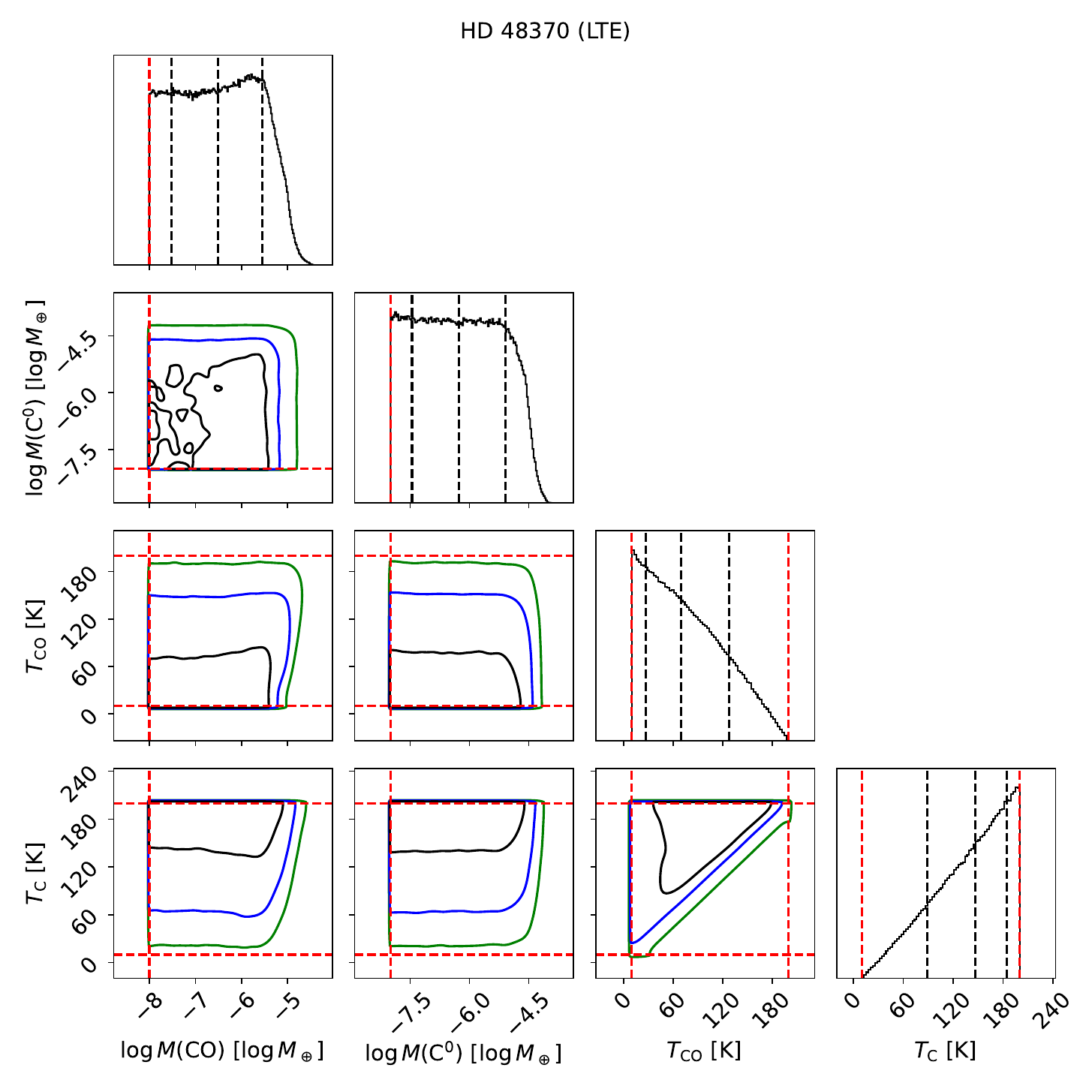}
\caption{Same as Fig.\ \ref{fig:corner_HD32297}, but for the disk around HD~48370.\label{fig:corner_HD48370}}
\end{figure*}

Figures \ref{fig:corner_HD32297} and \ref{fig:corner_HD48370} show corner plots (1D and 2D histograms of the posterior probability distribution) for two example disks: the CO-rich disk around HD~32297 and the disk around HD~48370 where no gas emission was detected. For HD~32297, we see that the CO and C$^0$ masses are well constrained, while the C$^+$ mass could be very large if the temperature $T_\mathrm{C}$ is low. Indeed, for low temperatures, the \CII{} emission becomes optically thick and the mass cannot be meaningfully constrained anymore. High optical depth at low temperature and consequently large uncertainties on the underlying CO, C$^0$ or C$^+$ masses are an issue for several other disks. The CO temperature in the HD~32997 disk is well-constrained. The posterior of the carbon temperature has a distinct peak, but also broad wings.

For the HD~48370 disk, the posterior distributions of the masses clearly allow us to derive upper limits, while the temperatures are essentially unconstrained. Note that the shape of the posterior distributions of $T_\mathrm{CO}$ and $T_\mathrm{C}$ reflects our prior condition that $T_\mathrm{CO}<T_\mathrm{C}$.

The remaining corner plots of the LTE fits are shown in Appendix \ref{appendix:corner_plots_LTE}. We emphasize that the values presented in Table~\ref{tab:fit_results_LTE} are merely summary statistics of the posterior distributions. The reader interested in a particular parameter or disk is invited to look at the corner plots to fully understand the posterior distributions and their mutual correlations.

\begin{figure*}
\plotone{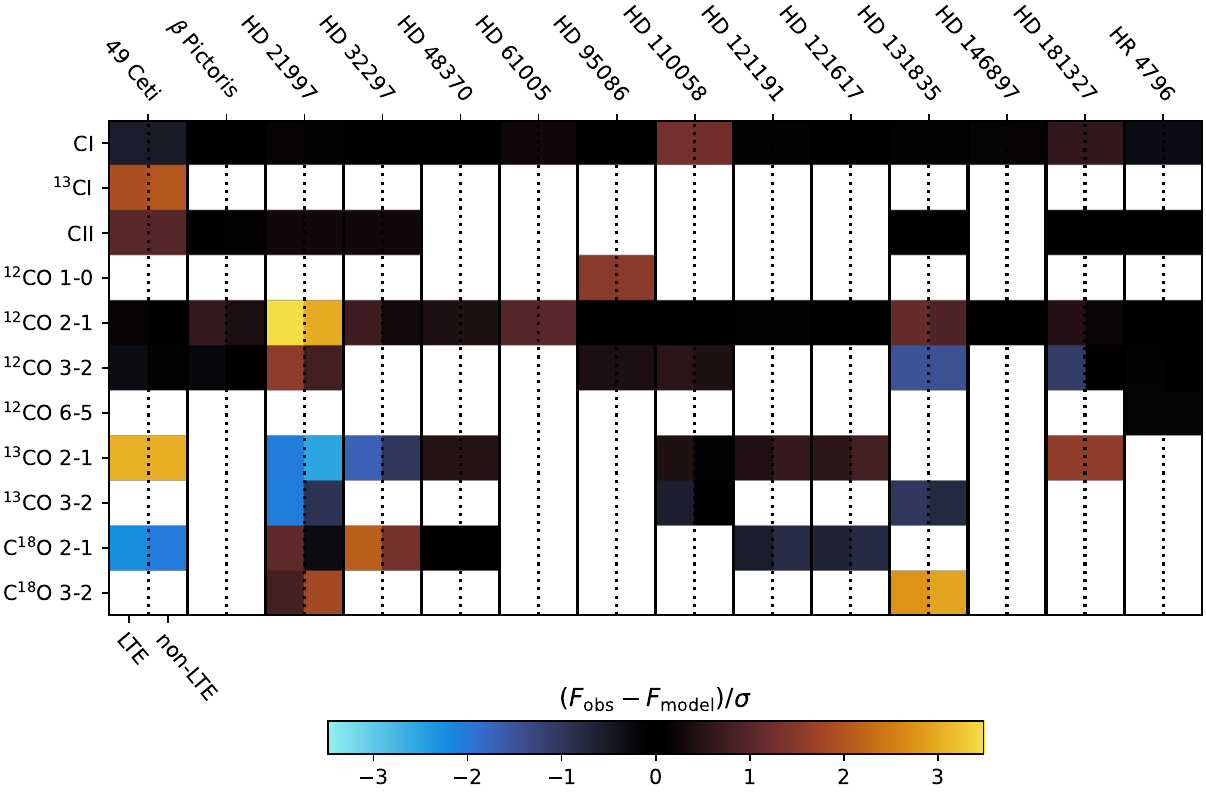}
\caption{Overview of fit residuals. For each disk, the first column corresponds to the LTE fits and the second column to the non-LTE fits (collisions with H$_2$). The color indicates the difference between the observed flux and the median of the model fluxes, normalised by the observational error. White indicates absence of data.}\label{fig:fit_residuals}
\end{figure*}

Figure \ref{fig:fit_residuals} compares the observed line fluxes $F_\mathrm{obs}$ to the median of the model line fluxes $F_\mathrm{model}$, normalised by the observational error. We see that our model, in general, successfully reproduces the observed line emission. There are some exceptions, such as HD~21997 where the model underpredicts the $^{12}$CO~2--1 flux by $3.5\sigma$ and overpredicts the 2--1 and 3--2 fluxes of $^{13}$CO by $2\sigma$.

\subsection{Non-LTE models}
The derived masses, temperatures and column densities for the non-LTE cases are presented in Tables \ref{tab:fit_results_nonLTE_H2} and \ref{tab:fit_results_nonLTE_e} for collisions with H$_2$ and e$^-$ respectively. The two cases deliver consistent gas masses, although we note that the case with e$^-$ collisions generally gives less constraining results for the C$^0$ masses (e.g.\ higher upper limits).

\movetabledown=60mm
\begin{rotatetable}
\begin{deluxetable}{lLLLLLLLLLLL}
\tablecaption{Same as Table~\ref{tab:fit_results_LTE}, but for the non-LTE case assuming collisions with H$_2$.\label{tab:fit_results_nonLTE_H2}}
\tablewidth{0pt}
\tablehead{ \colhead{star} & \colhead{$\log M(\mathrm{CO})$} & \colhead{$\log M(\mathrm{C^0})$} & \colhead{$\log M(\mathrm{C^+})$} & \colhead{$T_\mathrm{CO}$} & \colhead{$T_\mathrm{C}$} & \colhead{$\log N_\perp(\mathrm{CO})$} & \colhead{$\log N_\perp(\mathrm{C^0})$} & \colhead{$\log N_\perp(\mathrm{C^+})$} & \colhead{$\log N_\mathrm{los}(\mathrm{CO})$} & \colhead{$\log N_\mathrm{los}(\mathrm{C^0})$} & \colhead{$\log N_\mathrm{los}(\mathrm{C^+})$}\\
& \colhead{[$\log M_\oplus$]} & \colhead{[$\log M_\oplus$]} & \colhead{[$\log M_\oplus$]}& \colhead{[K]} & \colhead{[K]}& \colhead{[$\log \mathrm{cm}^{-2}$]} & \colhead{[$\log \mathrm{cm}^{-2}$]} & \colhead{[$\log \mathrm{cm}^{-2}$]} & \colhead{[$\log \mathrm{cm}^{-2}$]} & \colhead{[$\log \mathrm{cm}^{-2}$]} & \colhead{[$\log \mathrm{cm}^{-2}$]}}
\colnumbers
\startdata
49 Ceti& -2.04_{-0.15}^{+0.5}& -2.0_{-0.2}^{+0.3}& -1.1_{-1.1}^{+0.7}& 15.7_{-0.8}^{+1.1}& 24_{-2}^{+6}& 16.86_{-0.15}^{+0.5}& 17.2_{-0.2}^{+0.3}& 18.2_{-1.1}^{+0.7}& 17.52_{-0.15}^{+0.5}& 17.9_{-0.2}^{+0.3}& 18.9_{-1.1}^{+0.7}\\
 &  &  &  {>}-3.5 &  &  &  &  &  {>}15.7 &  &  &  {>}16.4\\
$\beta$ Pictoris& -4.36_{-0.14}^{+0.14}& -3.77_{-0.09}^{+0.09}& -3.2_{-0.3}^{+0.5}& 38_{-30}^{+70}& 130_{-80}^{+50}& 14.53_{-0.14}^{+0.14}& 15.50_{-0.09}^{+0.09}& 16.1_{-0.3}^{+0.5}& 15.54_{-0.14}^{+0.14}& 16.50_{-0.09}^{+0.09}& 17.1_{-0.3}^{+0.5}\\
HD 21997& -1.17_{-0.14}^{+0.3}& -2.91_{-0.09}^{+0.9} & {<}-0.33& 10.7_{-0.5}^{+0.6}& 72_{-60}^{+90}& 17.85_{-0.14}^{+0.3}& 16.48_{-0.09}^{+0.9} & {<}19.1& 17.91_{-0.14}^{+0.3}& 16.54_{-0.09}^{+0.9} & {<}19.1\\
HD 32297& -0.5_{-0.6}^{+0.4}& -2.46_{-0.11}^{+0.19}& -1.8_{-0.7}^{+1.0}& 26.3_{-1.8}^{+1.8}& 79_{-30}^{+80}& 18.5_{-0.6}^{+0.4}& 16.91_{-0.11}^{+0.19}& 17.6_{-0.7}^{+1.0}& 19.0_{-0.6}^{+0.4}& 17.44_{-0.11}^{+0.19}& 18.1_{-0.7}^{+1.0}\\
 &  {>}-1.5 &  &  {>}-3.4 &  &  &  {>}17.5 &  &  {>}15.9 &  {>}18.0 &  &  {>}16.5\\
HD 48370 & {<}-4.1 & {<}-4.1 & {<}-3.7\tablenotemark{a}& 66_{-40}^{+60}& 140_{-60}^{+40} & {<}14.9 & {<}15.3 & {<}15.7\tablenotemark{a} & {<}15.2 & {<}15.7 & {<}16.0\tablenotemark{a}\\
HD 61005 & {<}-4.7 & {<}-4.7 & {<}-4.5\tablenotemark{a}& 66_{-40}^{+60}& 140_{-60}^{+40} & {<}15.0 & {<}15.3 & {<}15.5\tablenotemark{a} & {<}15.8 & {<}16.1 & {<}16.4\tablenotemark{a}\\
HD 95086 & {<}-4.5 & {<}-3.5 & {<}-2.9\tablenotemark{a}& 65_{-40}^{+60}& 140_{-60}^{+40} & {<}13.7 & {<}15.1 & {<}15.7\tablenotemark{a} & {<}13.8 & {<}15.1 & {<}15.7\tablenotemark{a}\\
HD 110058& -1.8_{-0.3}^{+1.2} & {<}-3.0 & {<}-4.0\tablenotemark{a}& 104_{-9}^{+9}& 150_{-30}^{+30}& 18.8_{-0.3}^{+1.2} & {<}18.0 & {<}17.0\tablenotemark{a}& 19.8_{-0.3}^{+1.2} & {<}19.0 & {<}18.0\tablenotemark{a}\\
 &  {>}-2.3 &  &  &  &  &  {>}18.4 &  &  &  {>}19.4 &  & \\
HD 121191& -2.1_{-0.4}^{+1.3}& -3.61_{-0.17}^{+0.7}& -5.02_{-0.2}^{+0.05}\tablenotemark{a}& 102_{-14}^{+30}& 160_{-40}^{+30}& 18.3_{-0.4}^{+1.3}& 17.17_{-0.17}^{+0.7}& 15.76_{-0.2}^{+0.05}\tablenotemark{a}& 18.4_{-0.4}^{+1.3}& 17.21_{-0.17}^{+0.7}& 15.80_{-0.2}^{+0.05}\tablenotemark{a}\\
 &  {>}-3.2 &  &  &  &  &  {>}17.2 &  &  &  {>}17.3 &  & \\
HD 121617& -1.62_{-0.09}^{+0.09}& -2.42_{-0.06}^{+0.14}& -3.49_{-0.09}^{+0.07}\tablenotemark{a}& 43_{-4}^{+4}& 100_{-50}^{+60}& 17.75_{-0.09}^{+0.09}& 17.32_{-0.06}^{+0.14}& 16.25_{-0.09}^{+0.07}\tablenotemark{a}& 17.82_{-0.09}^{+0.09}& 17.39_{-0.06}^{+0.14}& 16.32_{-0.09}^{+0.07}\tablenotemark{a}\\
HD 131835& -1.70_{-0.13}^{+0.7}& -2.48_{-0.12}^{+0.8} & {<}-0.53& 12.0_{-0.3}^{+0.5}& 82_{-70}^{+80}& 17.08_{-0.13}^{+0.7}& 16.68_{-0.12}^{+0.8} & {<}18.6& 17.46_{-0.13}^{+0.7}& 17.06_{-0.12}^{+0.8} & {<}19.0\\
 &  {>}-3.3 &  &  &  &  &  {>}15.5 &  &  &  {>}15.9 &  & \\
HD 146897& -4.2_{-0.4}^{+1.5} & {<}-3.6 & {<}-4.1\tablenotemark{a}& 45_{-30}^{+70}& 140_{-60}^{+40}& 15.5_{-0.4}^{+1.5} & {<}16.5 & {<}15.9\tablenotemark{a}& 16.0_{-0.4}^{+1.5} & {<}17.0 & {<}16.5\tablenotemark{a}\\
HD 181327& -5.3_{-0.5}^{+0.9} & {<}-3.9 & {<}-1.2& 54_{-30}^{+60}& 140_{-60}^{+40}& 14.4_{-0.5}^{+0.9} & {<}16.1 & {<}18.9& 14.4_{-0.5}^{+0.9} & {<}16.1 & {<}18.9\\
HR 4796 & {<}-4.1 & {<}-3.5 & {<}-0.66& 59_{-40}^{+60}& 140_{-60}^{+40} & {<}15.8 & {<}16.8 & {<}19.6 & {<}16.1 & {<}17.1 & {<}19.9\\
\enddata
\tablenotetext{a}{Estimated from an ionisation calculation, because no \CII{} data was available. See text for details.}
\end{deluxetable}
\end{rotatetable}

\movetabledown=60mm
\begin{rotatetable}
\begin{deluxetable}{lLLLLLLLLLLL}
\tablecaption{Same as Table~\ref{tab:fit_results_LTE}, but for the non-LTE case assuming collisions with e$^-$.\label{tab:fit_results_nonLTE_e}}
\tablewidth{0pt}
\tablehead{ \colhead{star} & \colhead{$\log M(\mathrm{CO})$} & \colhead{$\log M(\mathrm{C^0})$} & \colhead{$\log M(\mathrm{C^+})$} & \colhead{$T_\mathrm{CO}$} & \colhead{$T_\mathrm{C}$} & \colhead{$\log N_\perp(\mathrm{CO})$} & \colhead{$\log N_\perp(\mathrm{C^0})$} & \colhead{$\log N_\perp(\mathrm{C^+})$} & \colhead{$\log N_\mathrm{los}(\mathrm{CO})$} & \colhead{$\log N_\mathrm{los}(\mathrm{C^0})$} & \colhead{$\log N_\mathrm{los}(\mathrm{C^+})$}\\
& \colhead{[$\log M_\oplus$]} & \colhead{[$\log M_\oplus$]} & \colhead{[$\log M_\oplus$]}& \colhead{[K]} & \colhead{[K]}& \colhead{[$\log \mathrm{cm}^{-2}$]} & \colhead{[$\log \mathrm{cm}^{-2}$]} & \colhead{[$\log \mathrm{cm}^{-2}$]} & \colhead{[$\log \mathrm{cm}^{-2}$]} & \colhead{[$\log \mathrm{cm}^{-2}$]} & \colhead{[$\log \mathrm{cm}^{-2}$]}}
\colnumbers
\startdata
49 Ceti& -1.7_{-0.4}^{+0.2}& -1.0_{-0.9}^{+0.4}& -1.2_{-0.9}^{+0.8}& 17.0_{-1.6}^{+1.6}& 24.4_{-1.9}^{+2}& 17.2_{-0.4}^{+0.2}& 18.2_{-0.9}^{+0.4}& 18.1_{-0.9}^{+0.8}& 17.8_{-0.4}^{+0.2}& 18.9_{-0.9}^{+0.4}& 18.8_{-0.9}^{+0.8}\\
 &  &  {>}-2.5 &  {>}-3.6 &  &  &  &  {>}16.8 &  {>}15.7 &  &  {>}17.4 &  {>}16.4\\
$\beta$ Pictoris& -4.43_{-0.12}^{+0.12}& -3.77_{-0.09}^{+0.09}& -3.52_{-0.11}^{+0.4}& 44_{-30}^{+70}& 130_{-70}^{+50}& 14.47_{-0.12}^{+0.12}& 15.50_{-0.09}^{+0.09}& 15.75_{-0.11}^{+0.4}& 15.48_{-0.12}^{+0.12}& 16.50_{-0.09}^{+0.09}& 16.75_{-0.11}^{+0.4}\\
HD 21997& -1.22_{-0.10}^{+0.14}& -2.91_{-0.10}^{+1.3} & {<}-0.35& 10.7_{-0.4}^{+0.6}& 71_{-60}^{+90}& 17.81_{-0.10}^{+0.14}& 16.48_{-0.10}^{+1.3} & {<}19.0& 17.87_{-0.10}^{+0.14}& 16.54_{-0.10}^{+1.3} & {<}19.1\\
 &  &  {>}-3.1 &  &  &  &  &  {>}16.3 &  &  &  {>}16.3 & \\
HD 32297& -0.5_{-0.5}^{+0.4}& -1.9_{-0.5}^{+0.5}& -2.5_{-0.4}^{+1.5}& 26.7_{-1.9}^{+1.9}& 66_{-20}^{+90}& 18.5_{-0.5}^{+0.4}& 17.5_{-0.5}^{+0.5}& 16.9_{-0.4}^{+1.5}& 19.0_{-0.5}^{+0.4}& 18.0_{-0.5}^{+0.5}& 17.4_{-0.4}^{+1.5}\\
 &  {>}-1.4 &  &  {>}-3.6 &  &  &  {>}17.6 &  &  {>}15.8 &  {>}18.1 &  &  {>}16.3\\
HD 48370 & {<}-3.6 & {<}-2.9 & {<}-3.2\tablenotemark{a}& 65_{-40}^{+60}& 140_{-60}^{+40} & {<}15.5 & {<}16.5 & {<}16.2\tablenotemark{a} & {<}15.8 & {<}16.9 & {<}16.5\tablenotemark{a}\\
HD 61005 & {<}-4.1 & {<}-3.7 & {<}-4.0\tablenotemark{a}& 65_{-40}^{+60}& 140_{-60}^{+40} & {<}15.5 & {<}16.3 & {<}16.0\tablenotemark{a} & {<}16.4 & {<}17.2 & {<}16.8\tablenotemark{a}\\
HD 95086 & {<}-4.4 & {<}-2.2 & {<}-2.3\tablenotemark{a}& 66_{-40}^{+60}& 140_{-60}^{+40} & {<}13.9 & {<}16.4 & {<}16.3\tablenotemark{a} & {<}13.9 & {<}16.4 & {<}16.3\tablenotemark{a}\\
HD 110058& -1.7_{-0.4}^{+1.2} & {<}-2.8 & {<}-4.0\tablenotemark{a}& 105_{-9}^{+9}& 150_{-30}^{+30}& 18.9_{-0.4}^{+1.2} & {<}18.2 & {<}17.0\tablenotemark{a}& 19.9_{-0.4}^{+1.2} & {<}19.2 & {<}18.0\tablenotemark{a}\\
 &  {>}-2.3 &  &  &  &  &  {>}18.4 &  &  &  {>}19.4 &  & \\
HD 121191& -2.1_{-0.5}^{+1.4}& -3.56_{-0.20}^{+1.1}& -5.02_{-0.2}^{+0.05}\tablenotemark{a}& 111_{-20}^{+30}& 160_{-40}^{+30}& 18.4_{-0.5}^{+1.4}& 17.22_{-0.20}^{+1.1}& 15.76_{-0.2}^{+0.05}\tablenotemark{a}& 18.4_{-0.5}^{+1.4}& 17.26_{-0.20}^{+1.1}& 15.80_{-0.2}^{+0.05}\tablenotemark{a}\\
 &  {>}-3.1 &  &  &  &  &  {>}17.3 &  &  &  {>}17.3 &  & \\
HD 121617& -1.64_{-0.09}^{+0.09}& -2.41_{-0.07}^{+0.19}& -3.49_{-0.09}^{+0.07}\tablenotemark{a}& 43_{-4}^{+5}& 100_{-50}^{+60}& 17.72_{-0.09}^{+0.09}& 17.32_{-0.07}^{+0.19}& 16.25_{-0.09}^{+0.07}\tablenotemark{a}& 17.79_{-0.09}^{+0.09}& 17.39_{-0.07}^{+0.19}& 16.32_{-0.09}^{+0.07}\tablenotemark{a}\\
HD 131835& -1.69_{-0.11}^{+0.5}& -2.45_{-0.14}^{+1.1} & {<}-0.46& 12.0_{-0.3}^{+0.4}& 75_{-60}^{+80}& 17.10_{-0.11}^{+0.5}& 16.70_{-0.14}^{+1.1} & {<}18.7& 17.48_{-0.11}^{+0.5}& 17.09_{-0.14}^{+1.1} & {<}19.1\\
 &  {>}-3.4 &  {>}-2.7 &  &  &  &  {>}15.4 &  {>}16.4 &  &  {>}15.8 &  {>}16.8 & \\
HD 146897& -3.6_{-1.0}^{+1.5} & {<}-2.4 & {<}-4.0\tablenotemark{a}& 47_{-30}^{+70}& 140_{-60}^{+40}& 16.1_{-1.0}^{+1.5} & {<}17.7 & {<}16.0\tablenotemark{a}& 16.6_{-1.0}^{+1.5} & {<}18.2 & {<}16.6\tablenotemark{a}\\
HD 181327& -4.4_{-1.3}^{+0.7} & {<}-2.5 & {<}-1.8& 59_{-40}^{+60}& 140_{-60}^{+40}& 15.3_{-1.3}^{+0.7} & {<}17.6 & {<}18.3& 15.3_{-1.3}^{+0.7} & {<}17.6 & {<}18.3\\
HR 4796 & {<}-3.4 & {<}-2.2 & {<}-0.66& 60_{-40}^{+60}& 140_{-70}^{+40} & {<}16.6 & {<}18.1 & {<}19.6 & {<}16.9 & {<}18.4 & {<}19.9\\
\enddata
\tablenotetext{a}{Estimated from an ionisation calculation, because no \CII{} data was available. See text for details.}
\end{deluxetable}
\end{rotatetable}

\begin{figure*}
\plotone{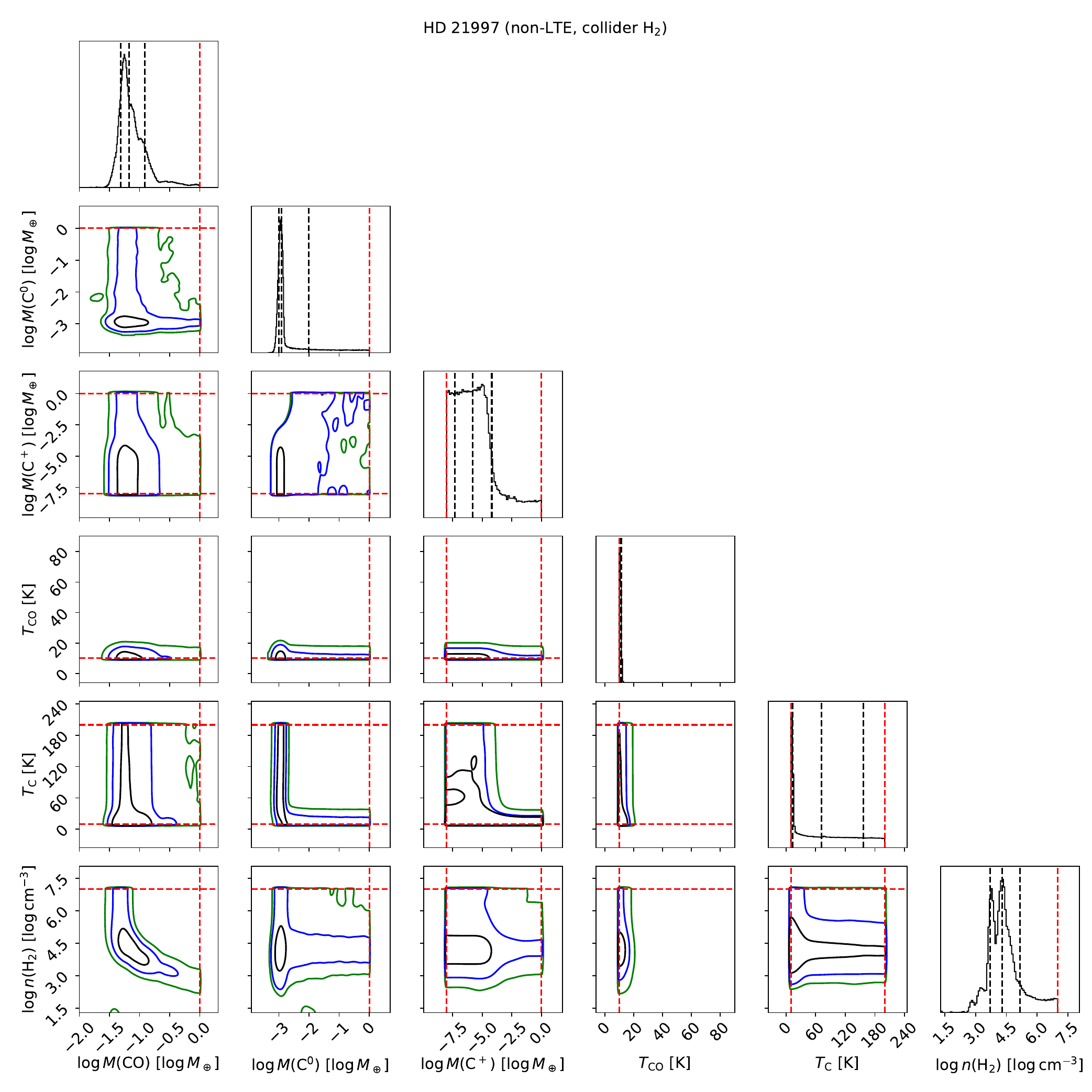}
\caption{Same as Fig.\ \ref{fig:corner_HD32297}, but for the disk around HD~21997 in non-LTE assuming collisions with H$_2$. The remaining corner plots from the non-LTE runs are available in Appendices \ref{appendix:corner_plots_nonLTE_H2} (collisions with H$_2$) and \ref{appendix:corner_plots_nonLTE_e} (collisions with electrons).}\label{fig:corner_HD21997}
\end{figure*}

We find that in general, $n(\mathrm{H}_2)$ and $n(\mathrm{e}^-)$ are not well constrained. By visual inspection of the posterior distributions, we identify the disks where lower limits can be inferred. Although not particularly constraining, they are presented in Table~\ref{tab:nH2_ne_constraints} for completeness. The strongest constraints can be derived for HD~21997, for which we show the corner plot in Fig.\ \ref{fig:corner_HD21997} for the case of collisions with H$_2$ (the corner plots for the other targets for both H$_2$ and e$^-$ collisions can be found in Appendices \ref{appendix:corner_plots_nonLTE_H2} and \ref{appendix:corner_plots_nonLTE_e}, respectively). These collider densities are derived by assuming a single collider species, either H$_2$ or e$^-$. In reality, several colliders might contribute to the excitation. Thus, our approach might overestimate the collider densities of H$_2$ and e$^-$. In Table~\ref{tab:nH2_ne_constraints}, we also list $n(\mathrm{C}^+)$ as estimated from our LTE fits, for comparison with $n(\mathrm{e}^-)$. The values are formally consistent, that is, it could indeed be the case that $n(\mathrm{e}^-)\approx n(\mathrm{C}^+)$, as is often assumed in simple ionisation calculations.

\begin{deluxetable}{lLLL}
\tablecaption{Lower limits on the H$_2$ and e$^-$ density for disks where meaningful constraints can be derived. We note that these values are derived from models that assume a single collider, either H$_2$ or e$^-$. The third column shows the number density of C$^+$ estimated from the LTE fits for comparison.\label{tab:nH2_ne_constraints}}
\tablewidth{0pt}
\tablehead{ \colhead{star} & \colhead{$\log n(\mathrm{H}_2)$} & \colhead{$\log n(\mathrm{e}^-)$} & \colhead{$\log n(\mathrm{C}^+)$}\\
& \colhead{[$\log \mathrm{cm}^{-3}$]} & \colhead{[$\log \mathrm{cm}^{-3}$]} & \colhead{[$\log \mathrm{cm}^{-3}$]} }
\startdata
49 Ceti & >2.1 & >0.3 & >1.4\\
$\beta$ Pictoris &  & >-0.7 & >1.2\\
HD 21997 & >2.8 & >1.7 & <4.8\\
HD 32297 & >1.8 & >0.4 & >1.2\\
HD 110058 &  & >0.4 & <3.8\tablenotemark{a}\\
HD 121191 &  & >-0.1 & 2.43_{-0.04}^{+0.03}\tablenotemark{a}\\
HD 121617 & >1.9 & >0.8 & 2.14_{-0.09}^{+0.07}\tablenotemark{a}\\
HD 131835 &  & >0.6 & <4.3\\
\enddata
\tablenotetext{a}{Estimated from an ionisation calculation, because no \CII{} data were available. See text for details.}
\end{deluxetable}

Figure \ref{fig:fit_residuals} compares the residuals for the LTE and non-LTE fits. The non-LTE fits occasionally reproduce the data slightly better (for example for HD~32297), but overall the residuals are very similar. This suggests that dropping the LTE assumption is not enough to resolve the discrepancy between modelled and observed line fluxes seen for some disks (for example $^{12}$CO~2--1 toward HD~21997). One possible cause could be that our non-LTE model only considers excitation caused by collisions and the CMB background radiation. However, stellar photons (or dust thermal emission) can sometimes be an important contribution to the excitation. For example, \citet{Matra18_SMA} found that pumping to electronic and rovibrational levels by the stellar UV had an important effect on the population of rotational levels of CO in the disk around $\beta$~Pic.

\subsection{The \texorpdfstring{C$^0$}{C0} and CO content of the debris disk sample}\label{sec:CO_vs_C0_overview}
\begin{figure*}
\plotone{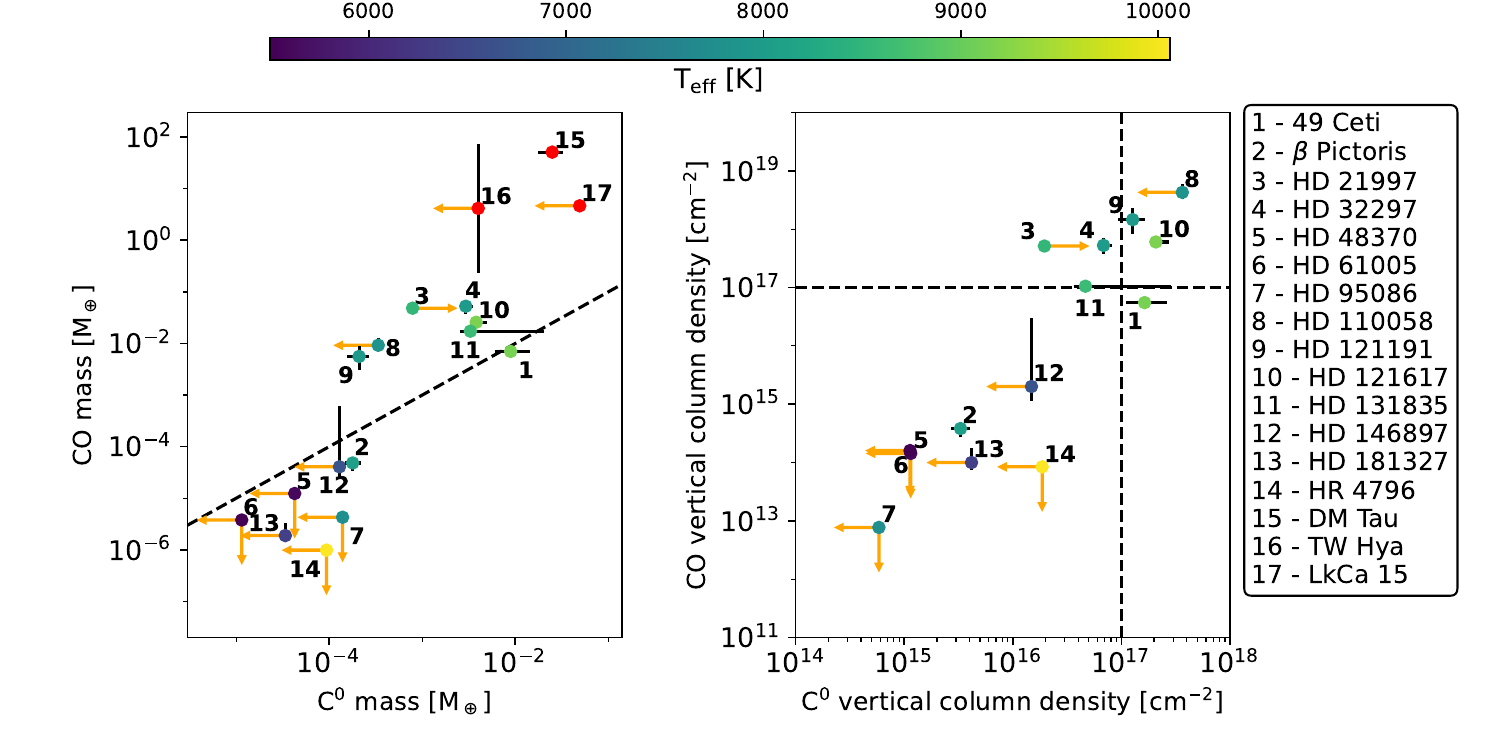}
\caption{The disks of our sample in the plane of C$^0$ vs CO mass (left) and vertical column density (right). Data points are colored by the effective temperature of the host star, except for three protoplanetary disks marked by red points. Upper and lower limits are indicated by orange arrows. The black dashed line on the left corresponds to $M(\mathrm{C}^0)=M(\mathrm{CO})$. The black dashed lines on the right roughly mark the column densities where CO self-shielding and CO shielding by C$^0$ become effective.\label{fig:masses_and_column_densities}}
\end{figure*}

Fig.\ \ref{fig:masses_and_column_densities} shows how the disks of our sample are distributed in the plane of C$^0$ vs CO masses and vertical column densities. Looking at the masses (Fig.\ \ref{fig:masses_and_column_densities} left), we note that for all disks where both masses are well constrained, $M(\mathrm{CO})\gtrsim M(\mathrm{C}^0)$, with one notable exception: $\beta$~Pictoris. Second, we can recognise two groups of disks: CO-rich disks with $M(\mathrm{CO})>10^{-3}$\,M$_\oplus$ and CO-poor disks with $M(\mathrm{CO})<10^{-4}$\,M$_\oplus$. Only one disk (HD~146897) might be located in the CO mass range between $10^{-4}$ and $10^{-3}$\,M$_\oplus$. In a secondary gas scenario, this might indicate that the transition from a CO-poor to a CO-rich disk occurs fast once C shielding becomes significant. We also note that there are no disks with high C$^0$ mass and low CO mass (bottom right of the plot). In the secondary gas production scenario, one might expect such disks to exist. Once the CO production rate declines, the CO mass can decrease rapidly while C stays in the system \citep[Fig.\ 2 in][]{Marino20}. However, our sample is biased and therefore we cannot infer the non-existence of such systems. The analysis of a statistically complete sample will be presented in a forthcoming paper. In section \ref{sec:CO_vs_C0_population_synthesis} we discuss in detail how our results compare to models of secondary gas production.

Fig.\ \ref{fig:masses_and_column_densities} (left) also shows masses for the three protoplanetary disks DM~Tau, LkCa~15 and TW~Hya with red data points. While there are more protoplanetary disks with \CI{} data \citep{Kama16_CI,Kama16_TWHya_HD100546,Alarcon22}, to the best of our knowledge, these are the only disks with published C$^0$ masses \citep{Tsukagoshi15}. The CO masses are from \citet[][their Fig.\ 4]{Zhang19} for DM~Tau and from \citet{Jin19} for LkCa~15. For TW~Hya, we consider two CO mass estimates: the smaller is derived by combining the minimum CO abundance (relative to H$_2$) from \citet{Favre13} with the minimum disk mass from \citet{Bergin13}. The larger CO mass is from \citet[][their Fig.\ 4]{Zhang19}. The error bar for TW~Hya encompasses the range between these two mass estimates. The three protoplanetary disks considered here have a substantially higher CO mass compared to the CO-rich debris disks, although this might not be true in general \citep[e.g.][]{Smirnov-Pinchukov22}. On the other hand, the protoplanetary C$^0$ masses are comparable to (or possibly lower than) the debris disk C$^0$ masses. This probably reflects the typical chemical structure of protoplanetary disks where atomic C is only expected to be present in the disk atmosphere, while most of the mass is located in the molecular layer closer to the midplane \citep[e.g.][]{Kama16_CI}. We also note that all three protoplanetary disks considered here are found around low mass stars, complicating the comparison to our sample that consists mostly of A-type stars.

Looking at the vertical column densities (Fig.\ \ref{fig:masses_and_column_densities} right), we see the same two groups of disks as for the masses. The vertical dashed line marks a C$^0$ column density of $10^{17}$\,cm$^{-2}$, corresponding to a reduction of the CO photodissociation rate by roughly a factor $1/e$ \citep{Heays17}, assuming that C$^0$ forms a distinct layer above and below CO \citep[if they are well-mixed instead, the shielding efficiency is reduced,][]{Marino22}. We see that the CO-rich disks are consistent with being significantly shielded by neutral carbon, as required by the secondary gas scenario. The CO-poor disks on the other hand are not C$^0$-shielded. The horizontal dashed line in Fig. \ref{fig:masses_and_column_densities} (right) marks a CO column density of $10^{17}$\,cm$^{-2}$ where self-shielding reduces the ISRF-induced dissociation rate to 1.6\% of the unshielded rate \citep{Visser09}. It shows that CO-rich disks are self-shielded, while CO-poor disks are not \citep[at a CO column density of $10^{15}$\,cm$^{-2}$, the dissociation rate is still at 40\% of the unshielded value,][]{Visser09}. We emphasize that from a secondary gas perspective, CO self-shielding is not sufficient to explain CO-rich disks. Indeed, unless the CO production rate is very large, CO can never reach a self-shielding column density without the help of C$^0$ shielding \citep[e.g.][]{Cataldi20}.

\subsection{The carbon ionisation fraction}\label{sec:ionisation_fraction}
We test whether the carbon ionisation fraction can be constrained by our observations. In Table~\ref{tab:ionization}, we compute two ionisation fractions: $\chi_\mathrm{ana}$ is purely based on our derived C$^0$ masses and corresponds to the analytically calculated ionisation fraction of a gas parcel that is subject to the full stellar and interstellar radiation (that is, we neglect any optical depth). Other than that, the calculation was performed as described in Sect.\ \ref{sec:results_LTE} (ionisation equilibrium with $n(\mathrm{e}^-)=n(\mathrm{C}^+)$). $\chi_\mathrm{ana}$ should give a reasonable estimate for the disks with low C$^0$ column densities (HD~48370, HD~61005, HD~95086, HD~146897 and HD~181327), but will overestimate the ionisation for other disks. We also calculated $\chi_\mathrm{obs}$ for the targets with \CII{} data. It is simply $\chi_\mathrm{obs}=M(\mathrm{C}^+)/(M(\mathrm{C^0})+M(\mathrm{C}^+))$. Unfortunately, we are unable to provide strong constraints for $\chi_\mathrm{obs}$. The strongest constraint is found for the disk around $\beta$~Pic where $\chi_\mathrm{obs}>0.4$.

\begin{deluxetable}{lLL}
\tablecaption{Estimates of the analytically calculated ionization fraction ignoring optical depth ($\chi_\mathrm{ana}$) and the observed ionization fraction ($\chi_\mathrm{obs}$).\label{tab:ionization}}
\tablewidth{0pt}
\tablehead{ \colhead{star} & \colhead{$\chi_\mathrm{ana}$} & \colhead{$\chi_\mathrm{obs}$}}
\startdata
\textbf{49 Ceti} & 0.2 & >0.1\\
\textbf{$\beta$ Pictoris} & 0.7 & >0.4\\
\textbf{HD 21997} & <0.3 & <1.0\\
\textbf{HD 32297} & 0.2 & >0.1\\
HD 48370 & >0.8 & \textrm{no \CII{}}\\
HD 61005 & >0.7 & \textrm{no \CII{}}\\
\textbf{HD 95086} & >0.9 & \textrm{no \CII{}}\\
\textbf{HD 110058} & >0.0 & \textrm{no \CII{}}\\
\textbf{HD 121191} & 0.1 & \textrm{no \CII{}}\\
\textbf{HD 121617} & 0.3 & \textrm{no \CII{}}\\
\textbf{HD 131835} & 0.3 & <0.9\\
HD 146897 & >0.4 & \textrm{no \CII{}}\\
HD 181327 & >0.5 & \textrm{unconstrained}\\
\textbf{HR 4796} & >0.8 & \textrm{unconstrained}\\
\enddata
\tablecomments{Disks in bold are A-stars part of the \citet{Moor17} sample. If the observations delivered upper limits for both the C$^0$ and C$^+$ mass, or lower limits for both, $\chi_\mathrm{obs}$ cannot be constrained.}
\end{deluxetable}

\section{Discussion}\label{sec:discussion}

In this section, we will concentrate on what we can learn from considering a sample of debris disks with both carbon and CO observations. The discussion of our results with respect to individual targets can be found in Appendix \ref{appendix:individual_targets}. In that appendix, we also compare our results to the literature and discuss any discrepancies.

\subsection{Comparison to population synthesis models}\label{sec:CO_vs_C0_population_synthesis}
In this section we compare our results to the model by \citet{Marino20} that considers the production and evolution of secondary gas in debris disks. The \citet{Marino20} model is one dimensional and assumes that CO is produced in a collisional cascade of planetesimals that releases CO ice into the gas phase. The radial variation of the CO production rate is modelled as a Gaussian centered at $r_\mathrm{belt}$. The CO production rate is proportional to the rate at which mass is input by catastrophic collisions of the largest bodies in the cascade. This rate decreases with time as the disks depletes its mass. The model considers CO photodissociation as well as CO self-shielding and shielding by C$^0$ and follows the viscous radial evolution of CO and C$^0$ assuming a standard $\alpha$-disk model. C$^0$ can only be removed by accretion onto the star, while CO can be removed by both accretion and photodissociation. CO shielding by C$^0$ is assumed to be maximally efficient, that is, C$^0$ is assumed to be in a layer above and below CO, rather than well-mixed \citep[see][for a discussion of how vertial mixing affects the CO lifetime]{Marino22}.

\citet{Marino20} constructed two synthetic populations of A-star debris disks with gas: one assuming a low viscosity ($\alpha=10^{-3}$) and one with a high viscosity ($\alpha=0.1$). To construct each synthetic population, they evolve a sample of model systems with randomly chosen stellar masses, disk masses and disk radii. The systems are evolved to a random age between 3 and 100\,Myr. By comparing their synthetic populations to CO observations, they conclude that a high viscosity ($\alpha=0.1$) is required. Indeed, the low viscosity population produces too many disks with very large ($\gtrsim$1\,M$_\oplus$) CO masses, inconsistent with observations. This is because for low viscosity, removal of gas by accretion is slow, allowing a large column density to accumulate, which leads to efficient shielding.

\citet{Marino20} also compared their model population to observations of C$^0$, but they had only four disks with C$^0$ data available: 49~Ceti, $\beta$~Pic, HD~32297 and HD~131835. Considering the low number statistics, they concluded that the $\alpha=0.1$ model is in reasonable agreement with the C$^0$ data.

\begin{figure*}
\plotone{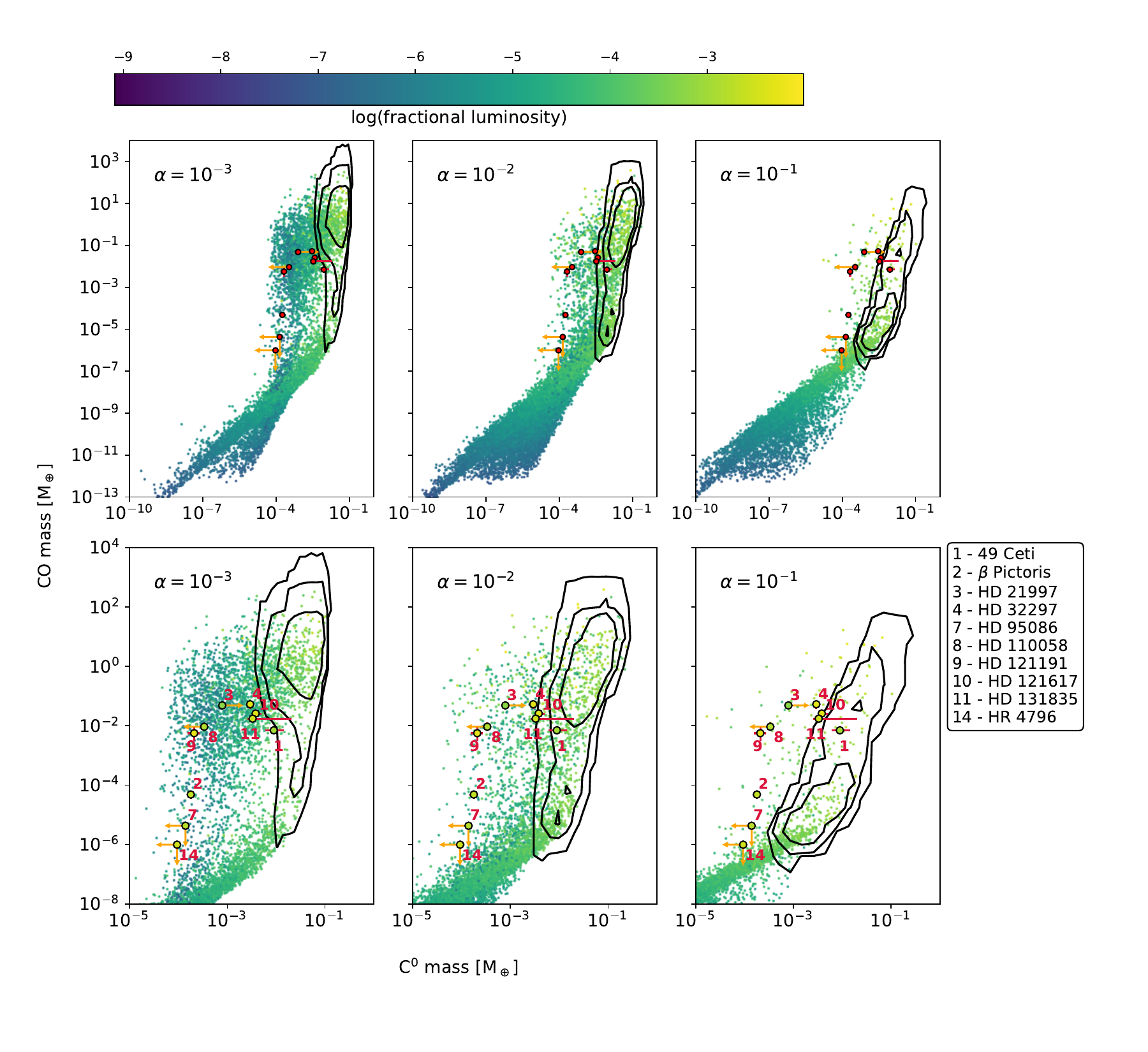}
\caption{Comparison of the A-star population synthesis model by \citet{Marino20} to the observed C$^0$ and CO masses for the A-stars in our sample. Each small dot (colored by the dust fractional luminosity) corresponds to a simulation. The contours enclose 68\%, 95\% and 99.7\% of the sub-population that satisfy the criteria of the A-star debris disks sample by \citet{Moor17}. All ten A-stars in our sample are part of the \citet{Moor17} sample. The top row shows an overview of the location of the observed masses (marked with red dots, with limits indicated by orange arrows) with respect to the synthetic population. The bottom row shows a zoomed view where the observations are colored by fractional luminosity. The columns correspond to different $\alpha$ viscosities assumed for the simulations.\label{fig:obs_mass_vs_Marino20}}
\end{figure*}

We are now in a position to provide an updated comparison of the \citet{Marino20} simulations with observations: we now have 10 disks around A-stars with C$^0$ data. Furthermore, new CO isotopologue data allow a more accurate CO mass measurements compared to the values used by \citet{Marino20}. In particular, our CO masses for 49~Ceti and HD~32297 are larger by factors of ${\sim}50$ and ${\sim}40$, respectively, compared to the values adopted by \citet{Marino20}. Figure \ref{fig:obs_mass_vs_Marino20} shows the comparison of our derived CO and C$^0$ masses to three population synthesis models: the original models presented by \citet{Marino20} with $\alpha=10^{-3}$ and $\alpha=10^{-1}$, and a new model with $\alpha=10^{-2}$. One complication is that the total C$^0$ mass of the simulations is often dominated by low density gas at large radii which is not easily observable, or that might be ionised in reality (the model ignores ionisation). To avoid this issue, the simulated C$^0$ masses shown in Fig.\ \ref{fig:obs_mass_vs_Marino20} correspond to what \citet{Marino20} called the 'observable carbon mass' defined by $\Sigma_\mathrm{C^0}(r_\mathrm{belt})r_\mathrm{belt}^2\pi$ with $\Sigma_\mathrm{C^0}$ the C$^0$ surface density and $r_\mathrm{belt}$ the mid-radius of the gas-producing belt.

In Fig.\ \ref{fig:obs_mass_vs_Marino20}, each small dot (colored by the dust fractional luminosity) represents a single simulation. To take into account observational biases, \citet{Marino20} consider the sub-population of simulated systems that corresponds to the sample proposed by \citet{Moor17}, which is defined by the following criteria: 1) A-type host star, 2) fractional dust luminosity between $5\times10^{-4}$ and $10^{-2}$, 3) a dust temperature below 140 K, 4) a detection with Spitzer or Herschel at $\lambda\geq70$\,$\mu$m and 5) an age between 10 and 50\,Myr. The black contours enclose 68\%, 95\% and 99.7\% of that sub-population, but considering model and observational uncertainties, we shall consider the outermost contour as a reference to compare models and observations. We have ten A-stars in our sample, all of which are part of the \citet{Moor17} sample. They are also shown in Fig.\ \ref{fig:obs_mass_vs_Marino20} and should fall within the black contours if model and data agree. We confirm the finding by \citet{Marino20} that the low viscosity ($\alpha=10^{-3}$) model overpredicts the observed masses. However, in contrast to \citet{Marino20}, we find that the high viscosity model ($\alpha=10^{-1}$) also struggles to reproduce the observed masses. In particular, the model overpredicts the C$^0$ masses. The case with $\alpha=10^{-2}$ turns out to be qualitatively similar to $\alpha=10^{-3}$.

\begin{figure*}
\plotone{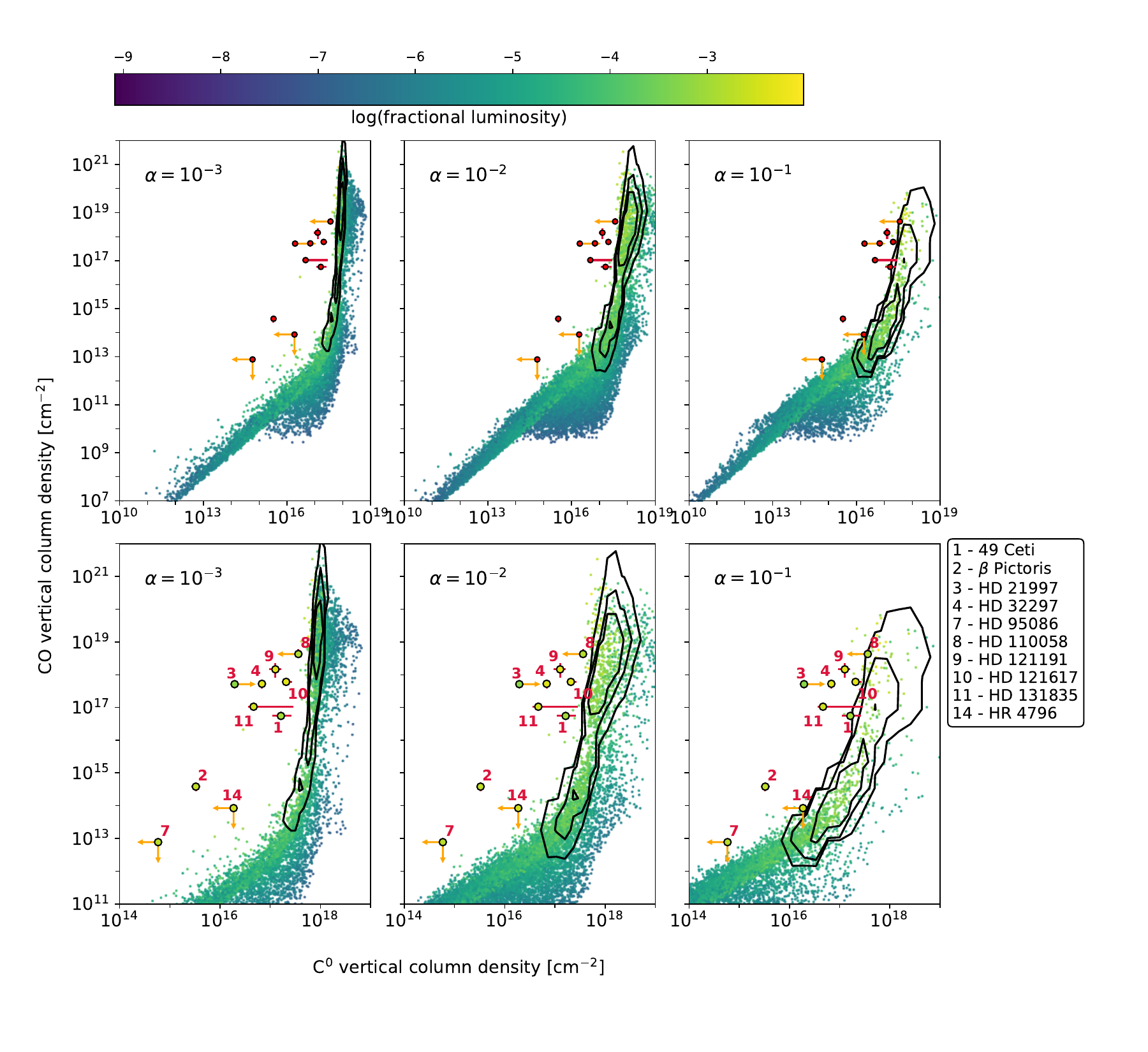}
\caption{Same as Fig.\ \ref{fig:obs_mass_vs_Marino20}, but showing vertical column density instead of mass.\label{fig:obs_column_density_vs_Marino20}}
\end{figure*}

As mentioned before, comparing observed and simulated C$^0$ masses is not trivial. Instead, comparing column densities is more promising because it is the column densities of CO and C$^0$ that determine the CO photodissociation timescale \citep{Marino20}. Thus, in Fig.\ \ref{fig:obs_column_density_vs_Marino20}, we instead compare the synthetic population and the observations in terms of vertical column density. More precisely, we compare the model column density at $r_\mathrm{belt}$ (i.e.\ the radius where CO production peaks) to the observed column density. Again comparing the observations to the black contours, the conclusion remains essentially unchanged: the models tend to overpredict the C$^0$ column density.

One possible explanation for the mismatch between the data and the secondary gas model is that the gas is primordial. We will discuss a primordial scenario in Sect.\ \ref{sec:CO_vs_C0_chemical_model}. Here, we consider possible explanations for the mismatch in the framework of the secondary gas scenario.

\subsubsection{Additional shielding agent}\label{sec:additional_shielding_agent}
One possibility to reconcile models and observations is an additional shielding agent besides C$^0$. Additional shielding would increase the model CO masses while reducing C$^0$ production by photodissociation, thus bringing the model closer to the observations. However, it remains unclear what that shielding agent could be. While H$_2$ can shield CO, it is not expected to be present in sufficient amounts in a secondary gas scenario. Indeed, efficient H$_2$ shielding requires an H$_2$ column density larger than $10^{21}$\,cm$^{-2}$ \citep{Visser09}, and thus a CO/H$_2$ abundance ratio ${\lesssim}10^{-3}$ when referring to the vertical CO column densities of Table \ref{tab:fit_results_LTE}. This CO abundance would be much closer to a primordial ISM-like composition than to a comet-like composition where H$_2$ is negligible.

We also note that the \citet{Marino20} model assumes maximum efficiency of C$^0$ shielding, that is, C$^0$ is implicitly assumed to be in a layer above and below CO. However, depending on the strength of vertical turbulence, CO and C might instead be mixed \citep{Marino22}. In a mixed gas, shielding is reduced and therefore the model CO masses decrease while the carbon masses increase, which further increases the mismatch between model and observations.

\subsubsection{Increased CO content of gas-producing comets}
Another possibility to consider is an increased CO mass fraction of the planetesimals that participate in the collisional cascade. \citet{Marino20} assumed that the planetesimals contain 10\% CO. If the CO content was substantially higher, CO masses would increase. However, it would not decrease C$^0$ masses, and thus not be helpful to resolve the discrepancy between model and data.

\subsubsection{Radiation pressure}
Radiation pressure on the gas was not included in the models by \citet{Marino20}. Could the apparent lack of C$^0$ be explained by blow out due to radiation pressure? Stars with $T_\mathrm{eff}>8000$\,K (spectral type A5V) are able to blow out C$^0$, while for C$^+$ (and CO) $T_\mathrm{eff}>8800$\,K (A2V) is required \citep{Youngblood21}. However, \citet{Kral17} found that a small column density of CO (much smaller than in the disk around $\beta$~Pic) is sufficient to prevent C$^0$ and C$^+$ from being blown out; this shielding was not considered by \citet{Youngblood21}. Thus, it appears unlikely that radiation affects the C content of the disks significantly, except for the disks with low column densities. \citet{Youngblood21} indeed detected a wind driven by radiation pressure from the tenuous gas disk around $\eta$~Tel. Similar processes might be operating in disks such as the one around HR~4796. More detailed modelling of the effects of radiation pressure would be valuable. For example, if C is located at the surface of the disk, separated from CO, it might be more prone to blowout than estimated by \citet{Kral17}.

\subsubsection{High C ionisation fraction}
\citet{Marino20} neglected C ionisation. Could this be the reason that the models overpredict C$^0$ column densities? Most of the systems shown in Fig.\ \ref{fig:obs_column_density_vs_Marino20} have $N(\mathrm{C}^0)\gtrsim10^{16}$\,cm$^{-2}$. Referring to Fig.\ 9 of \citet{Marino20}, those systems would have an ionisation fraction below 0.3, except inside of 10\,au. Therefore, we do not expect that ionisation would strongly impact the comparison of the \citet{Marino20} models to our observations, except for a system such as HD~95086 with a low C$^0$ column density $<10^{15}$\,cm$^{-2}$. However, an observational confirmation of low ionisation fractions is difficult (see Sect.\ \ref{sec:ionisation_fraction}).

\subsubsection{Underestimation of observed \texorpdfstring{C$^0$}{C0} masses}
We can also ask whether the C$^0$ masses (or column densities) are underestimated in our analysis, instead of overpredicted by the models. For example, high optical depth might hide large C masses. To test this idea, we re-run our MCMC fits, forcing the temperature to be between 10 and 20\,K. This indeed results in more lower limits for the C$^0$ column densities (five out of the ten, instead of one for our standard LTE fit). It also results in higher upper limits for the C$^0$ column density of HD~110058 and HR~4796, making them consistent with the \citet{Marino20} model. However, for six out of the ten disks (49~Ceti, $\beta$~Pic, HD~32297, HD~110058, HD~121191 and HD~121617), the low temperature model provides a significantly worse fit by strongly underpredicting some line fluxes.

\subsubsection{Transient events}
Another possibility to explain the mismatch is that the premise of the model (gas production from a steady-state collisional cascade) is not applicable. At least some of the observed systems might have their gas and dust produced in a transient event such as a giant collision or a tidal disruption \citep[e.g.][]{Jackson14,Kral15,Cataldi18,Cataldi20,Schneiderman21}. Unfortunately, there are no model predictions for the gas production from such transient events. Naively, we might expect C-rich, CO-poor disks if most of the CO gas was injected into the system in a short time and then photodissociated. Our sample is not well suited to find such systems because so far, \CI{} observations tended to be conducted for disks known to harbour CO.

\subsubsection{Removal of C by planets or grain surface chemistry}
We propose that C removal from the system by an additional process besides accretion onto the star could explain the mismatch between models and observations. For example, carbon could be removed by accretion onto planets when spreading viscously \citep{Marino20,Kral20_terrestrial_atmospheres}. Here we concentrate on adsorption of C atoms onto dust grains instead \citep{Cataldi20}. We calculated the mean free time of C atoms between collisions with dust grains, assuming a grain size distribution with a slope of -3.5 \citep{Dohnanyi69,Tanaka96}, a minimum grain size equal to the blowout size and a maximum grain size of 1\,mm. The mm dust masses were collected from the literature. We find mean free times of the order of $10^2$--$10^5$\,yr, which is short compared to the system ages. It is also shorter than the viscous timescale $t_\mathrm{vis}$, even for $\alpha=0.1$ for which $t_\mathrm{vis}\gtrsim10^5$\,yr. Therefore, adsorption could indeed be an important C removal process. In fact, the adsorbed C atoms could be recycled into CO, which would both increase the model CO masses and decrease the C$^0$ masses, as required. Indeed, it was recently shown both theoretically and experimentally that adsorbed C atoms can react with amorphous water ice to efficiently form formaldehyde \citep[H$_2$CO,][]{Molpeceres21,Potapov21}. Release of H$_2$CO (or the intermediate product COH$_2$) into the gas phase (for example by photodesorption) and subsequent photodissociation would produce CO. Alternatively, H$_2$CO can be transformed to CO$_2$ by UV photons \citep{Potapov21}, which could then be released to the gas phase and photodissociated to CO. However, the mechanism we propose here depends on the presence of water ice, and \citet{Grigorieva07} showed that UV sputtering might deplete water ice in debris disks. At the moment there is no clear detection of water ice in debris disks, but some indirect observational evidence suggests that grains could be a least partially icy \citep[e.g.][]{Chen08,Lebreton12,Morales13,Morales16}. The James Webb Space Telescope (JWST) might be able to detect water ice features towards debris disks the near future \citep{Kim19}.

However, it remains to be seen whether introducing an additional C removal process to the models can indeed resolve the discrepancy. Removing more C could simply make it more difficult to achieve a shielded state, merely resulting in a smaller proportion of model systems with a large CO mass. But those model systems that manage to become shielded might still have too much C$^0$ compared to observations.

\subsection{Comparison to thermo-chemical model}\label{sec:CO_vs_C0_chemical_model}
One possible way to distinguish primordial, H-rich gas from secondary, H-poor gas is to study the gas chemical composition. However, chemical modelling by \citet{Smirnov-Pinchukov22} showed that molecular emission (besides CO) is expected to be undetectable regardless of the hydrogen content of the gas (with the possible exception of HCO$^+$). Therefore, we instead consider how the CO fraction (that is, the fraction of the total carbon that is in CO) can inform us about the chemistry and hydrogen content of debris disk gas. To this end, we compare our results to the thermo-chemical model by \citet{Iwasaki23}. They consider only A-stars, so we limit the discussion to the A-stars in our sample. \citet{Iwasaki23} compute the chemistry of debris disk gas using a photon-dominated region (PDR) code \citep[the \texttt{Meudon code},][]{LePetit06}. The model solves for the thermal and chemical equilibrium of a stationary, semi-infinite, plane-parallel slab of gas illuminated from one side. It considers various heating and cooling processes, gas phase and grain surface chemistry as well as photo-ionisation, photo-dissociation and self-shielding, taking into account the radiative transfer. In this model, there is no gas production from comets or gas loss from accretion onto the star, and the dust mass (and dust surface) is constant. Thus, the setup is reminiscent of a primordial gas origin. \citet{Iwasaki23} fit an analytical formula to the CO fraction computed with their numerical model. The formula is given by
\begin{equation}\label{eq:CO_fraction_analytical}
    \frac{n(\mathrm{CO})}{n_\mathrm{C}}=\left(1+\left(10^{-14}\eta^{1.8}+6\cdot10^{-11}\eta\right)^{-1}\right)^{-1}
\end{equation}
Here $n(\mathrm{CO})$ is the CO number density, while $n_\mathrm{C}$ is the number density of carbon nuclei (in practice $n_\mathrm{C}\approx n(\mathrm{CO})+n(\mathrm{C}^0)+n(\mathrm{C}^+)$). The parameter $\eta$ is given by
\begin{equation}
    \eta=n_\mathrm{H}Z^{0.4}\chi^{-1.1}
\end{equation}
Here $n_\mathrm{H}$ is the number density of H nuclei and $Z$ is the gas metallicity, where $Z=1$ corresponds to solar metallicity. For a given metallicity $Z$, we can estimate $n_\mathrm{H}$ using our observational results as
\begin{equation}
n_\mathrm{H}\approx\frac{n(\mathrm{CO})+n(\mathrm{C}^0)}{Z\zeta_\mathrm{C}}
\end{equation}
where $\zeta_\mathrm{C}=1.32\times10^{-4}$ is the relative elemental abundance of C adopted by \citet{Iwasaki23}. We neglect $n(\mathrm{C}^+)$ in our estimation because it is generally poorly constrained by our fits. The parameter $\chi$ describes the strength of the UV field. It is given by
\begin{equation}
\chi=\sqrt{\chi_\mathrm{CO}\chi_\mathrm{OH}}
\end{equation}
where $\chi_\mathrm{CO}$ and $\chi_\mathrm{OH}$ are the normalised fluxes of UV photons in the wavelength ranges that are important for the chemistry of CO (91.2--110\,nm) and OH (160--170\,nm). The normalisation constants are $1.2\times10^7$\,cm$^{-2}$\,s$^{-1}$ (the Habing field) for $\chi_\mathrm{CO}$ and $10^{12}$\,cm$^{-2}$\,s$^{-1}$ for $\chi_\mathrm{OH}$. Splitting out the contribution from the ISRF and the star gives
\begin{equation}
    \chi_\mathrm{CO} = \frac{\chi_\mathrm{CO,ISRF}}{2}+\chi_\mathrm{CO,star}(r_\mathrm{mean})
\end{equation}
The factor $1/2$ for $\chi_\mathrm{CO,ISRF}$ accounts for the fact that the ISRF penetrates the modelled gas slab only from one side. \citet{Iwasaki23} find $\chi_\mathrm{CO,ISRF}=1.3$. To compute $\chi_\mathrm{CO,star}$, we use ATLAS9 stellar atmosphere models \citep{Castelli03}, with the exception of $\beta$~Pic that observationally shows additional emission in the UV above the predictions from a standard stellar atmosphere model. Therefore, we use a PHOENIX model as
described in \citet{Fernandez06} complemented with UV data from the Hubble Space Telescope \citep{Roberge00} and the Far Ultraviolet Spectroscopic Explorer \citep{Bouret02,Roberge06}. $\chi_\mathrm{OH,star}$ is computed in the same way as $\chi_\mathrm{CO,star}$, while $\chi_\mathrm{OH,ISRF}$ is negligible. In Table~\ref{tab:chi} we show the computed values of $\chi_\mathrm{CO}$ and $\chi_\mathrm{OH}$.

\begin{deluxetable}{lCCC}
\tablecaption{The normalised flux of UV photons affecting the CO ($\chi_\mathrm{CO}$) and OH ($\chi_\mathrm{OH}$) chemistry of debris disk gas, evaluated at the midplane at $r=r_\mathrm{mean}$.\label{tab:chi}}
\tablewidth{0pt}
\tablehead{ \colhead{star}
           & \colhead{$\chi_\mathrm{CO}$}
           & \colhead{$\chi_\mathrm{CO}^{\mathrm{ext}}$}
           & \colhead{$\chi_\mathrm{OH}$}
           }
\colnumbers
\startdata
49 Ceti & 3.0 & 5.1\times10^{-3} & 1.7\\
$\beta$ Pictoris & 3.3 & 5.6\times10^{-1} & 8.2\times10^{-2}\\
HD 21997 & 8.9\times10^{-1} & 4.9\times10^{-3} & 4.5\times10^{-1}\\
HD 32297 & 7.2\times10^{-1} & 3.1\times10^{-3} & 1.6\times10^{-1}\\
HD 95086 & 6.5\times10^{-1} & 5.7\times10^{-1} & 1.4\times10^{-2}\\
HD 110058 & 2.2 & 6.5\times10^{-5} & 4.6\\
HD 121191 & 1.2 & 1.0\times10^{-3} & 1.4\\
HD 121617 & 1.0\times10^{1} & 8.3\times10^{-4} & 3.4\\
HD 131835 & 8.0\times10^{-1} & 9.8\times10^{-3} & 4.0\times10^{-1}\\
HR 4796 & 2.1\times10^{2} & 8.2\times10^{1} & 9.4\\
\enddata
\tablecomments{(2) Normalised UV flux between 91.2 and 110\,nm without extinction. (3) Same as (2), but with extinction by C$^0$ and CO. (4) Normalised UV flux between 160 and 170\,nm.}
\end{deluxetable}

To account for UV extinction by C$^0$ and CO, we follow \citet{Iwasaki23} and multiply $\chi_\mathrm{CO,ISRF}$ and $\chi_\mathrm{CO,star}$ by a shielding factor. For $\chi_\mathrm{CO,ISRF}$, this shielding factor is given by
\begin{equation}\label{eq:chi_extinction}
    f_\mathrm{shield}^{ISRF} = e^{-\alpha_\mathrm{C}N_\perp(\mathrm{C}^0)/2}\left( 1+\left(\frac{N_\perp(\mathrm{CO})/2}{10^{14}\mathrm{cm}^{-2}}\right)^{0.6} \right)^{-1}
\end{equation}
where $N_\perp$ is the column density perpendicular to the disk midplane and $\alpha_\mathrm{C}$ is the carbon ionisation cross section. For the latter we adopt the same value as \citet{Iwasaki23}: $\alpha_\mathrm{C}=1.777\times10^{-17}$\,cm$^{2}$ \citep{Heays17}. The first term in Equation \ref{eq:chi_extinction} represents extinction by C$^0$, while the second term is a function fitted by \citet{Iwasaki23} to the CO shielding factors tabulated in \citet{Visser09}. For the shielding factor for $\chi_\mathrm{CO,star}$, we replace $N_\perp/2$ by $N_\parallel(r_\mathrm{mean})$ in Equation \ref{eq:chi_extinction}, that is, the column density parallel to the disk midplane from the star to $r_\mathrm{mean}$. We note that $\chi_\mathrm{OH}$ is not affected by extinction. The values for $\chi_\mathrm{CO}$ with extinction taken into account (denoted $\chi_\mathrm{CO}^\mathrm{ext}$) are shown in Table~\ref{tab:chi}.

We then compute the predicted CO fraction for all A-stars in our sample using equation \ref{eq:CO_fraction_analytical} for metallicities ranging from 1 (primordial gas) to 10$^3$ (high metallicity as expected for secondary gas). We emphasize that the high metallicity case does not directly correspond to the secondary gas models that were discussed in the previous section, because there is no CO production from comets included in the \citet{Iwasaki23} model discussed here. Rather, the high metallicity case would correspond to an equilibrium state that secondary gas would attain in the absence of gas production from comets, for example if secondary gas was produced in a single burst rather than continuously.

\begin{figure*}
\plotone{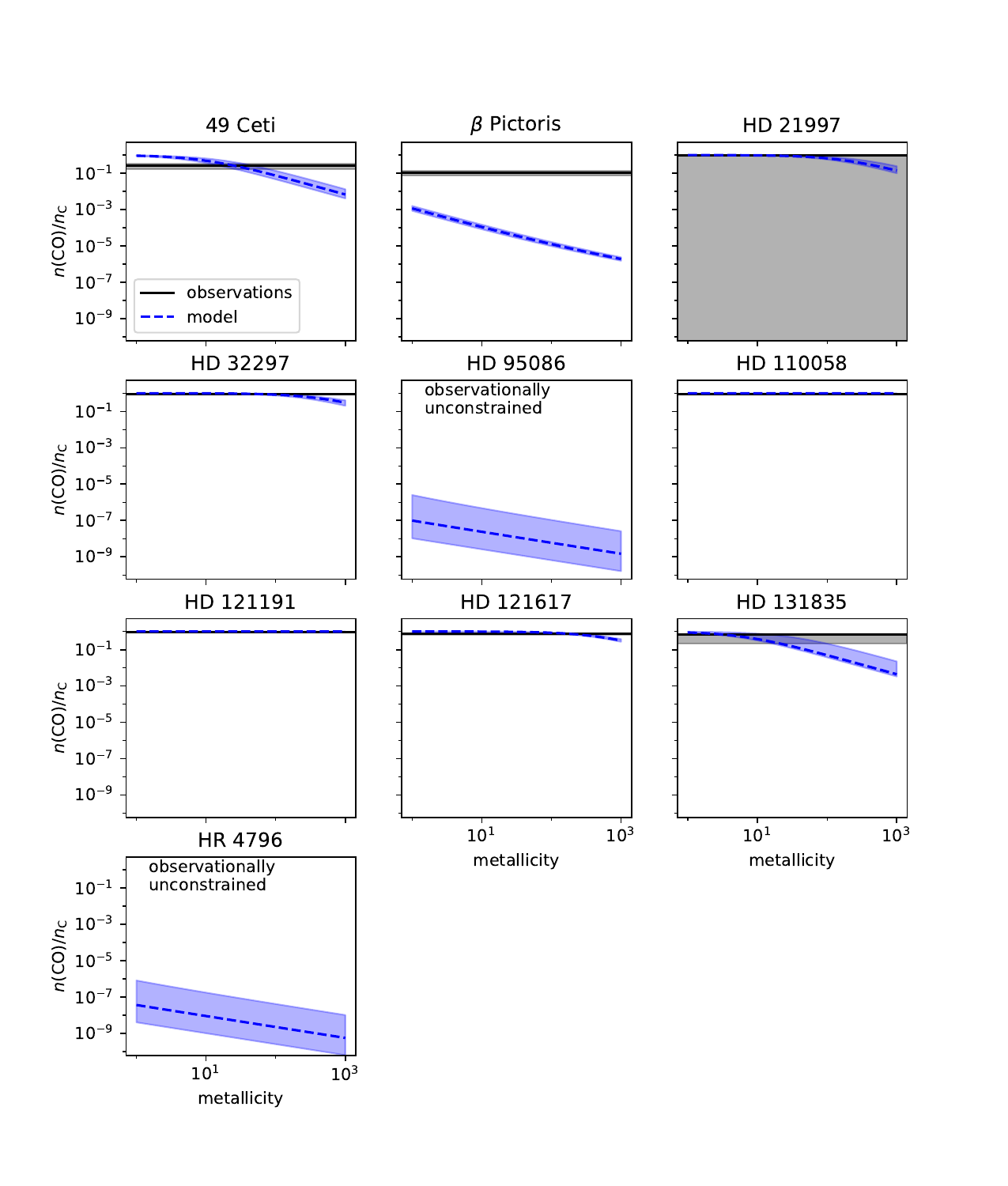}
\caption{Comparison of observed and predicted CO fraction. The blue lines show the CO fraction predicted using the model by \citet{Iwasaki23} as a function of gas metallicity $Z$ (equation \ref{eq:CO_fraction_analytical}). The black lines show the CO fraction estimated from our observations (for HD~21997 the 99\% upper limits is shown, while for HD~110058, the 1\% lower limit is shown). Note that for HD~110058 and HD~121191, the black and blue lines overlap. Gray shading indicates the observationally allowed region (15.9th to 84.1th percentile or the region below/above an upper/lower limit). For HD~95086 and HR~4796 where neither CO nor \CI{} emission is detected, the CO fraction is observationally unconstrained.\label{fig:Iwasaki_model_comparison}}
\end{figure*}

The predicted CO fractions are shown with the blue lines in Fig.\ \ref{fig:Iwasaki_model_comparison}. We also show the CO fraction estimated from our observations with black lines, where we again made the approximation $n_\mathrm{C}\approx n(\mathrm{CO})+n(\mathrm{C}^0)$. Note that for HD~95086 and HR~4796, neither CO nor \CI{} emission is detected and thus the CO fraction remains obsevationally unconstrained.

We see that among the eight targets with an observationally constrained CO fraction, seven are consistent with the predictions from the thermo-chemical models. For most of these objects, both low and high metallicity models are consistent with the data. The only object clearly incompatible with the model predictions is $\beta$~Pic where even for a solar metallicity gas, the gas density is too low to sustain the observed CO fraction purely by chemistry. This is consistent with the view that the gas in the disk around $\beta$~Pic is currently produced from the destruction of cometary material \citep[e.g.][]{Dent14,Matra17_betaPic,Iwasaki23}.

In summary, we find that the CO fraction of the CO-rich debris disks around 49~Ceti, HD~21997, HD~32297, HD~110058, HD~121191, HD~121617 and HD~131835 is consistent with the chemistry expected for a primordial gas. However, the analytical formula we applied assumes a simplified geometry (plane-parallel slab) and a simplified chemical network. Therefore, the uncertainties remain high and we do not interpret our results as a confirmation of the primordial gas origin. Rather, our results encourage further investigation of the primordial scenario. The logical next step would be to apply the full numerical model including a more realistic geometry to each of the disks presented here. \citet{Iwasaki23} already present such detailed models for $\beta$~Pictoris and 49~Ceti, reaching conclusions similar to ours: the disk around $\beta$~Pic cannot be explained by steady-state chemistry, but the model can explain the disk around 49~Ceti if Z=1--10.

\subsection{Dependence of gas masses on the current gas production rate}\label{sec:gas_mass_vs_Mdot_CO}
If the gas is produced from the destruction of solid material in a steady-state collisional cascade, the current gas production rate can be computed using the equations presented by \citet{Matra17_Fomalhaut}. The steady-state condition implies that the rate at which mass is input to the cascade by catastrophic collisions of the largest bodies $\dot{M}_{D_\mathrm{max}}$ equals the sum of the rate at which CO and CO$_2$ are outgassed $\dot{M}_{\mathrm{CO+CO}_2}$ and the rate at which mass is lost via radiation pressure on the smallest grains $\dot{M}_{D_\mathrm{min}}$:
\begin{equation}
    \dot{M}_{D_\mathrm{max}} = \dot{M}_{\mathrm{CO+CO}_2} + \dot{M}_{D_\mathrm{min}}
\end{equation}
Assuming that CO and CO$_2$ are outgassed before reaching the smallest grain size in the system, we have
\begin{equation}
    \dot{M}_{\mathrm{CO+CO}_2} = f_{\mathrm{CO+CO}_2}\dot{M}_{D_\mathrm{max}}
\end{equation}
where $f_{\mathrm{CO+CO}_2}$ is the CO+CO$_2$ fraction (by mass) of the colliding bodies. Combining these two equations yields
\begin{equation}
    \dot{M}_{\mathrm{CO+CO}_2} = \dot{M}_{D_\mathrm{min}} \frac{f_{\mathrm{CO+CO}_2}}{1-f_{\mathrm{CO+CO}_2}}
\end{equation}
However, \citet{Matra17_Fomalhaut} show that $\dot{M}_{D_\mathrm{min}}$ can be estimated from observable quantities as follows:
\begin{equation}
    \dot{M}_{D_\mathrm{min}} = 1.2\times10^3R^{1.5}\Delta R^{-1} f^2L_*M_*^{-0.5}
\end{equation}
where $R$ (in au) is the belt radius, $\Delta R$ (in au) the belt width, $f$ the fractional luminosity, $L_*$ (in L$_\odot$) the stellar luminosity, $M_*$ (in M$_\odot$) the stellar mass and $\dot{M}_{D_\mathrm{min}}$ is in M$_\oplus$\,Myr$^{-1}$.

\begin{figure*}
\plotone{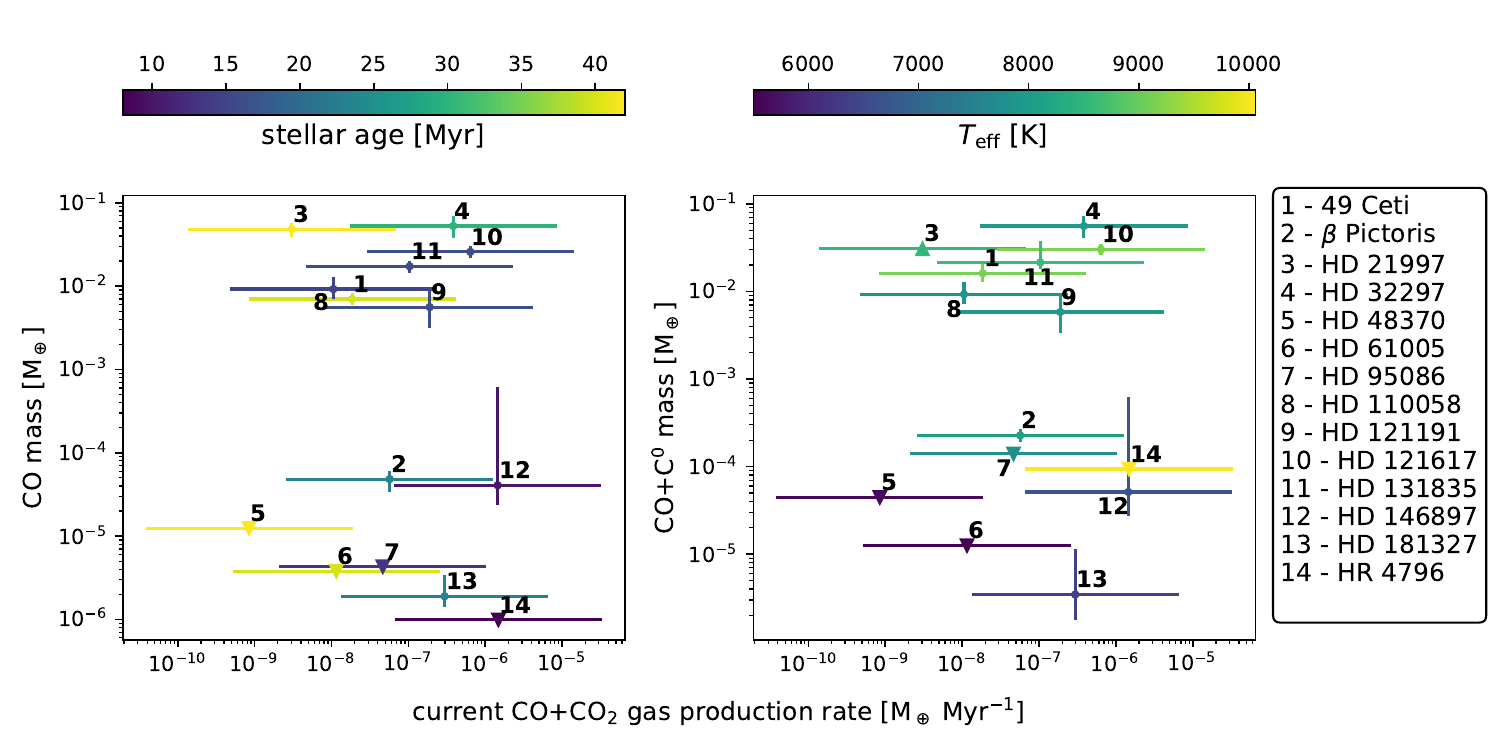}
\caption{The CO (left) and C$^0$+CO (right) masses as a function of the current CO+CO$_2$ gas production rate $\dot{M}_{\mathrm{CO+CO}_2}$. The error bars on $\dot{M}_{\mathrm{CO+CO}_2}$ represent the range of $f_{\mathrm{CO+CO}_2}$ between 0.8\% and 80\%. The data points are colored by the age (left) and effective temperature (right) of the host star. Upper and lower limits are marked by upward and downward triangles.\label{fig:mass_vs_CO_prod_rate}}
\end{figure*}

Naively, we might expect that the current gas production rate $\dot{M}_{\mathrm{CO+CO}_2}$ should correlate with the observed gas mass of a disk. However, this is not the case, as can be seen in Fig.\ \ref{fig:mass_vs_CO_prod_rate} showing the observed CO and CO+C$^0$ masses as a function of $\dot{M}_{\mathrm{CO+CO}_2}$. We follow \citet{Matra19} and consider a range of possible CO+CO$_2$ mass fractions $f_{\mathrm{CO+CO}_2}$ between 0.8\% and 80\%. Fig.\ \ref{fig:mass_vs_CO_prod_rate} demonstrates that $\dot{M}_{\mathrm{CO+CO}_2}$ is not a good predictor for gas-rich debris disks. Fig.\ \ref{fig:mass_vs_CO_prod_rate} also shows that stellar age is not the dividing factor between gas-rich and gas-poor systems. On the other hand, the host stars of gas-rich disks tend to have higher effective temperatures in our sample. This reflects the fact that gas-rich disks have so far only been found around A-type stars, but a more detailed interpretation is complicated because of the biases in our sample.

As pointed out by \citet{Marino20}, it is theoretically expected that the gas viscous evolution is slower than the steady-state collisional evolution of the dust. Thus, the gas can retain a memory of previous states, and therefore the gas content depends on the time-integrated gas production rate of the disk rather than the current rate. Furthermore, a steady-state collisional cascade has the interesting property that the mass at late times (i.e.\ at times exceeding the collisional timescale of the largest planetesimal in the cascade) is independent of the initial disk mass, because massive disks process their mass faster \citep{Wyatt07}. In other words, two disks with the same current mass loss rate (and therefore the same $\dot{M}_{\mathrm{CO+CO}_2}$) can potentially have strong differences in their mass loss rate history, and thus gas content. 

\subsection{Winds instead of Keplerian disks?}\label{sec:wind}
\citet{Kral23} discuss the conditions that determine whether debris disk gas is in Keplerian rotation or in an outflowing wind. They find that for $L_*>20$\,$L_\odot$, a wind driven by the stellar radiation is expected. For lower luminosities, the gas is in Keplerian rotation as long as the gas density is larger than a critical density $n_\mathrm{crit}$. For $n<n_\mathrm{crit}$, a wind driven by the stellar wind is expected. The critical density is given by
\begin{equation}
    n_\mathrm{crit}\approx 7\mathrm{cm}^{-3}\left(\frac{\Delta r}{50\mathrm{au}} \right)^{-1} \left(\frac{\alpha_\mathrm{X}}{\alpha_\mathrm{CO}}\right)^{-2/3}
\end{equation}
where $\Delta r=r_\mathrm{out}-r_\mathrm{in}$ and $\alpha_\mathrm{X}$ is the polarisability of species X.

Among the stars we consider here, only HR~4796 has a luminosity larger than 20\,L$_\odot$. We thus expect a radiative wind in this system. Gas removal with the wind might be the reason why we do neither detect CO nor \CI{} emission from this system. For the other disks, we compared the critical density and the observationally estimated density for CO and C$^0$. We classify a disk as being in the wind regime if both CO and C$^0$ have densities smaller than $n_\mathrm{crit}$. This is the case for the disks around HD~48370, HD~61005 and HD~95086, all of which remain undetected in both CO and \CI{}. Furthermore, the data allow either regime for the disks around HD~146897 and HD~181327. Despite assuming that the gas is in Keplerian rotation (Sect.\ \ref{sec:modelling}), our model still gives the correct total line flux for a given gas mass even in the wind regime, because the emission for low gas density disks is optically thin. However, our flux measurements are affected by our assumption of a Keplerian disk. Indeed, line emission from a wind would come from a different spatial region, although the solid angle of that region should be roughly comparable to the Keplerian case. Furthermore, wind emission should extend over a significantly larger velocity range \citep{Kral23}. We may roughly estimate the factor by which our upper limits on the line emission (and therefore gas mass) should be increased if gas emission arises from a wind. This factor is given by $f_\mathrm{wind}=\sqrt{\Delta v_\mathrm{wind}/\Delta v}$ where $\Delta v$ is the size of the velocity range over which the integrated flux was measured and $\Delta v_\mathrm{wind}$ is the velocity range over which wind emission arises. Using $\Delta v_\mathrm{wind}=20$\,km\,s$^{-1}$ \citep{Kral23}, we find $f_\mathrm{wind}\leq2.3$ for the three disks in the wind regime mentioned above. Such a small increase of the upper limits is inconsequential for our analysis.

\subsection{Outlook: investigating the spatial dimension}
To facilitate a uniform and simple analysis, we have integrated over the spatial and spectral dimension of our data, that is, we compared our model to a single number extracted from the data: the disk-integrated flux. Our approach assumed that all the tracers we consider (CO, C$^0$ and C$^+$) are co-spatial and well-mixed. As far as we can tell from the available images, co-spatiality is a reasonable assumption to derive the total gas masses. Indeed, the CO and C\,I emission morphologies look generally rather similar. For example, the emission of both CO and C\,I appear centrally peaked and with a similar radial extent for HD~21997 (Fig.\ \ref{fig:mom0_HD21997}). Similarly, both CO and C\,I show a ring morphology for HD~121617 (see Fig.\ \ref{fig:mom0_HD121617} for C\,I and Fig.\ 1 in \citet{Moor17} for CO).

However, the fact that a significant fraction of our targets are spatially (and spectrally) resolved clearly suggests a path for future work that will model the data in its entirety \citep[such as was performed by e.g.][]{Kral19,Cataldi20} to compare the spatial distribution of the different gas tracers and the dust continuum. For example, to test the secondary gas production scenario, one could investigate whether the gas has viscously spread with respect to its production location traced by the dust continuum. Similarly, C$^0$ could be more spread out than CO, since the latter can only exist in regions with a high C$^0$ column density \citep[e.g.][]{Marino20}. Comparing the morphologies of different disks could also be instructive. For example, the ring morphology of the C\,I emission towards HD~121617 (Fig.\ \ref{fig:mom0_HD121617}) seems inconsistent with a disk that has viscously spread all the way to the star (accretion disk), although detailed modelling is required to exclude the possibility of an accretion disk that is ionised and thus not observable in \CI{} in the inner region. \citet{Cataldi20} also concluded that there is no accretion disk morphology for the edge-on HD~32297 disk. In that latter case, the velocity information was used to constrain the inner disk regions. Several possible mechanisms could prevent the gas from accreting, for example planets \citep{Marino20}. On the other hand, the disk around HD~21997 is centrally peaked in both CO and C\,I, while the dust continuum shows a ring morphology \citep[see Fig.\ \ref{fig:mom0_HD21997} and][]{Kospal13,Moor13_HD21997}. This disk might be consistent with the picture where gas is produced in a planetesimal belt and then accretes onto the star. Finally, we note that disks produced from transient events (e.g.\ giant collisions) are expected to show asymmetries \citep[e.g.][]{Jackson14,Kral15,Jones23}, providing another motivation to study the spatial distribution of the gas and dust.

\section{Summary}\label{sec:summary}
In this work, we studied the C and CO gas content of a sample of 14 debris disks using new and archival ALMA data of \CI{} and CO emission, complemented by \CII{} data from Herschel. This expands the number of disks with ALMA measurements of both \CI{} and CO by ten disks. We present new detections of \CI{} emission towards HD~21997, HD~121191 and HD~121617. We measure disk-integrated line fluxes and employ a simple disk model to derive masses and column densities in a uniform way, using both LTE and non-LTE calculations. We found that our sample can be divided into two groups: CO-rich disks with $M(\mathrm{CO})>10^{-3}$\,M$_\oplus$ and CO-poor disks with $M(\mathrm{CO})\lesssim10^{-4}$\,M$_\oplus$. For all disks where both the CO and C$^0$ are well constrained, $M(\mathrm{CO})\gtrsim M(\mathrm{C}^0)$, with the notable exception of the disk around $\beta$~Pic. We do not find any CO-poor disks with a large ($M(\mathrm{C}^0)\gtrsim10^{-3}$\,M$_\oplus$) C$^0$ mass.

We compare the C$^0$ and CO masses and column densities to the state-of-the-art models of secondary gas production from a steady-state collisional cascade \citep{Marino20}. We find that the models overpredict the C$^0$ content of debris disk gas. We discuss possible explanations for the discrepancy. We suggest additional C removal by grain surface chemistry as a possible scenario, but new model calculations are needed for confirmation. Our results might also indicate that for some disks the gas originated in a transient event (for example a giant collision or tidal disruption event) rather than in a steady-state collisional cascade. Our work does not exclude the secondary scenario, but suggests that the models need to be refined to explain the gas content of CO-rich debris disks.

We also compared our results to the thermo-chemical model by \citet{Iwasaki23}. This PDR model determines the thermal and chemical equilibrium of a plane-parallel slab of gas subject to the stellar and interstellar UV fields. There is no gas production by comets, nor gas removal by accretion in this model. For low metallicity, this model therefore mimics a primordial gas origin. We use the analytical formula calibrated by numerical simulations presented by \citet{Iwasaki23} to compute the predicted CO fraction of the gas and compare it to the observed CO fraction (approximated as $n(\mathrm{CO})/(n(\mathrm{CO})+n(\mathrm{C^0}))$). Among the eight targets in our sample for which the comparison is possible, seven show a CO fraction consistent with the model predictions. The only target clearly inconsistent with the model is $\beta$~Pic, which confirms that the gas in the $\beta$~Pic disk is secondary. Although promising, this comparison is based on a simplified geometry and chemical network, and therefore more detailed modelling is required before concluding on the possibility of a primordial origin. Future work should apply the full numerical model, including a more realistic geometry, to our targets, as is already presented for two disks in \citet{Iwasaki23}.

Our work shows that observations of C are crucial to complement CO data in order to constrain current models of debris disk gas. Future work should consider a more well-defined sample of disks, such as the volume-limited sample of dust-rich debris disks around young A-type stars by \citet{Moor17}. A well-defined sample would help to better understand the statistical properties of the gaseous debris disk population.

%% not robust enough to handle our new dual anonymous review requirements and thus been replaced with the acknowledgment environment. If you try to 
%% Also note that the acknowledgment environment does not support long amounts of text. If you have a lot of people and institutions to acknowledge, do not use this command. Instead, create a new \section{Acknowledgments}.
\begin{acknowledgments}
We thank the anonymous referee for a careful review that significantly improved the clarity of this manuscript. We would like to acknowledge useful discussions with Germ\'{a}n Molpeceres. We also acknowledge data calibration support by the East Asian ALMA Regional Center. We thank Inga Kamp, Nagayoshi Ohashi, Alycia Weinberger and Yanqin Wu for contributions to the proposal of project 2019.1.01175.S that significantly expanded the number of disks with \CI{} data. G.C.\ was supported by the NAOJ ALMA Scientiﬁc Research grant code 2019-13B. Y.A.\ acknowledges support by Grant-in-Aid for Transformative Research Areas (A) 20H05844 and 20H05847. S.M.\ is supported by the Royal Society as a Royal Society University Research Fellow. A.M.H. is supported by a Cottrell Scholar Award from the Research Corporation for Science Advancement. This paper makes use of the following ALMA data: ADS/JAO.ALMA\#2011.0.00780.S, ADS/JAO.ALMA\#2012.1.00437.S, ADS/JAO.ALMA\#2012.1.00688.S, ADS/JAO.ALMA\#2013.1.00612.S, ADS/JAO.ALMA\#2013.1.00773.S, ADS/JAO.ALMA\#2013.1.01147.S, ADS/JAO.ALMA\#2013.1.01166.S, ADS/JAO.ALMA\#2015.1.00032.S, ADS/JAO.ALMA\#2016.A.00010.S, ADS/JAO.ALMA\#2016.A.00021.T, ADS/JAO.ALMA\#2016.2.00200.S, ADS/JAO.ALMA\#2016.1.01253.S, ADS/JAO.ALMA\#2017.A.00024.S, ADS/JAO.ALMA\#2017.1.01575.S, ADS/JAO.ALMA\#2018.1.00500.S, ADS/JAO.ALMA\#2018.1.00633.S, ADS/JAO.ALMA\#2018.1.01222.S, ADS/JAO.ALMA\#2019.2.00208.S, ADS/JAO.ALMA\#2019.1.01603.S, ADS/JAO.ALMA\#2019.1.01175.S. ALMA is a partnership of ESO (representing its member states), NSF (USA) and NINS (Japan), together with NRC (Canada), MOST and ASIAA (Taiwan), and KASI (Republic of Korea), in cooperation with the Republic of Chile. The Joint ALMA Observatory is operated by ESO, AUI/NRAO and NAOJ. The National Radio Astronomy Observatory is a facility of the National Science Foundation operated under cooperative agreement by Associated Universities, Inc. This work has made use of data from the European Space Agency (ESA) mission {\it Gaia} (\url{https://www.cosmos.esa.int/gaia}), processed by the {\it Gaia} Data Processing and Analysis Consortium (DPAC, \url{https://www.cosmos.esa.int/web/gaia/dpac/consortium}). Funding for the DPAC has been provided by national institutions, in particular the institutions participating in the {\it Gaia} Multilateral Agreement. This research has made use of NASA’s Astrophysics Data System. This research has made use of the SIMBAD database, operated at CDS, Strasbourg, France.
\end{acknowledgments}

%% in the research for the paper.  Each keyword is check against the master list during copy editing.  Individual instruments can be provided in parentheses, after the keyword, but they are not verified.

\vspace{5mm}
\facilities{ALMA, Herschel (PACS)}


%remember to cite the new CAAS reference paper
\software{ALminer (\url{https://github.com/emerge-erc/ALminer}), astropy \citep{astropy13,astropy18,astropy22}, CASA \citep{CASA22}, CMasher \citep{vanderVelden20}, dill \citep[][\url{http://uqfoundation.github.io/project/pathos}]{McKerns12}, MPoL (\url{https://doi.org/10.5281/zenodo.4939048}) numpy \citep{Harris2020}, pythonradex (\url{https://github.com/gica3618/pythonradex}), scipy \citep{Virtanen20}.
}



\appendix

\section{Continuum fluxes}\label{appendix:continuum_fluxes}
Table~\ref{tab:continuum_fluxes} presents the continuum fluxes we measured.

\startlongtable
\begin{deluxetable}{lccl}
\tablecaption{Measured continuum fluxes.\label{tab:continuum_fluxes}}
\tablewidth{0pt} %0pt means natural width
\tablehead{ \colhead{star} & \colhead{frequency} & \colhead{flux}& \colhead{observation ID}\\
 & \colhead{[GHz]} &  \colhead{[mJy]} &}
\startdata
49 Ceti& 225 & $4.2\pm 0.4$& 2016.2.00200.S, 2018.1.01222.S\\
\hline
HD 21997& 487 & $9.0\pm 1.7$& 2018.1.00633.S, 2019.1.01175.S\\
\hline
HD 48370& 225 & $2.5\pm 0.4$& 2016.2.00200.S\\
& 487 & $9.5\pm 4$ (${<}21$)& 2019.2.00208.S\\
\hline
HD 61005& 222 & $4.3\pm 0.5$& 2012.1.00437.S\\
& 487 & $37\pm 4$& 2019.1.01603.S\\
\hline
HD 95086& 108 & $0.34\pm 0.04$& 2016.A.00021.T\\
& 231 & $2.5\pm 0.3$& 2013.1.00612.S, 2013.1.00773.S\\
& 339 & $4.0\pm 0.8$& 2016.A.00021.T\\
& 487 & $5.8\pm 1.7$& 2019.1.01175.S\\
\hline
HD 110058& 226 & $0.48\pm 0.05$& 2018.1.00500.S\\
& 233 & $0.48\pm 0.05$& 2012.1.00688.S, 2018.1.00500.S\\
& 338 & $1.27\pm 0.13$& 2018.1.00500.S\\
& 487 & $2.4\pm 0.3$& 2019.1.01175.S\\
\hline
HD 121191& 487 & $1.0\pm 0.2$& 2019.1.01175.S\\
\hline
HD 121617& 487 & $9.9\pm 1.1$& 2019.1.01175.S\\
\hline
HD 131835& 324 & $5.1\pm 0.5$& 2013.1.01166.S\\
& 351 & $6.1\pm 0.6$& 2013.1.01166.S\\
\hline
HD 146897& 488 & $5.8\pm 1.0$& 2018.1.00633.S\\
\hline
HD 181327& 341 & $12.6\pm 1.3$& 2015.1.00032.S\\
& 487 & $48\pm 5$& 2016.1.01253.S\\
\hline
HR 4796& 487 & $36\pm 4$& 2017.A.00024.S\\
& 683 & $91\pm 11$& 2016.A.00010.S\\
\hline
\enddata
\end{deluxetable}

\section{Robustness tests}\label{appendix:robustness_tests}
Here we discuss how our LTE results depend on the choice of temperature prior and line width.

\subsection{Separating the temperatures of \texorpdfstring{C$^0$}{C0} and \texorpdfstring{C$^+$}{C+}}\label{appendix:test_TC0_TCplus}
For the targets with \CII{} data, we ran additional LTE fits where the temperatures of C$^0$ and C$^+$ are separated. In other words, instead of two temperatures, we now consider three temperatures for CO, C$^0$ and C$^+$, where we impose $T_\mathrm{CO}<T_\mathrm{C^0}<T_\mathrm{C^+}$. We find that, with the exception of $T_\mathrm{C^0}$ for 49~Ceti, the C$^0$ and C$^+$ temperatures are not well constrained. Compared to the fits with two temperatures, we do not find any significant differences in the derived gas masses.

\subsection{Line width}\label{appendix:line_width_tests}
For our standard LTE fits, we assumed a square line profile with a fixed width of 0.5\,km\,s$^{-1}$. We performed additional MCMC runs where we changed the line width. When decreasing the line width to 0.1\,km\,s$^{-1}$, the derived CO masses change significantly (i.e.\ beyond their error bars) for HD~21997 (decrease by a factor ${\sim}2$), HD~121617 (increase by a factor ${\sim}2$) and HD~131835 (decrease by a factor ${\sim}2$). These models reproduce the data similarly well as the standard fits. For the other disks, the masses remain within the error bars.

When instead increasing the line width to 2\,km\,s$^{-1}$, we observe the following significant changes. For 49~Ceti, the C$^0$ mass decreases by a factor ${\sim}3$, but the model provides a significantly worse fit to the $^{13}$\CI{} emission. For HD~21997, the CO mass decreases by two orders of magnitude, but the model provides a much worse fit to the $^{13}$CO and C$^{18}$O data. For HD~121617, the CO mass decreases by a factor ${\sim}1.5$ while the fit quality remains unchanged. For HD~131835, the CO mass decreases by about two orders of magnitude, but the model provides a much worse fit to the $^{13}$CO and C$^{18}$O data. For the other disks, no significant changes are observed. In conclusion, these tests confirm that a line width of 0.5\,km\,s$^{-1}$ is a reasonable choice.

\subsection{Temperature prior}\label{appendix:T_prior}
For our standard LTE fits, we assumed flat priors for the CO and C temperatures between 10 and 200\,K (Table~\ref{tab:fit_parameters}). Here we perform additional MCMC runs where we changed the limits of the temperature priors. If we increase the lower limit to 20\,K, the derived CO and C$^0$ masses do not change significantly, except for $\beta$~Pic, HD~21997 and HD~131835 for which the standard fit indicates a CO temperature below 20\,K. However for these three disks (as well as 49~Ceti which also has a CO temperature below 20\,K in the standard fit), models with $T_\mathrm{CO}\geq20$\,K provide a significantly worse fit to the observed fluxes. This result strengthens our choice of 10\,K as the lowest temperature.

When decreasing the upper limit of the temperature priors from 200 to 100\,K, the derived CO and C$^0$ masses do not change significantly.

\section{Discussion of individual targets}\label{appendix:individual_targets}
In this section we discuss our results for the individual targets of our sample and how they compare to previous studies.

\subsection{49~Ceti}
The edge-on, gas-rich debris disk around 49~Ceti has been the subject of multiple detailed studies over the past years \citep[some of the more recent work includes][]{Higuchi19_13C,Higuchi19,Moor19,Pawellek19,Higuchi20,Klusmeyer21}. \citet{Hughes17} presented two non-LTE models of the CO disk around 49~Ceti with CO/H$_2$ abundances of $10^{-4}$ (representing a primordial gas) and 1 (representing a secondary gas). They used observations of $^{12}$CO 2--1 and 3--2 from the SMA and ALMA respectively. These models yield CO masses of $10^{-3.8}$ and $10^{-3.5}$\,M$_\oplus$ respectively, more than an order of magnitude smaller than our CO mass ($10^{-2.15}$\,M$_\oplus$). A similar discrepancy applies to the column density derived by \citet{Nhung17} based on $^{12}$CO 3--2. We find that the $^{12}$CO 2--1 and 3--2 emission is strongly optically thick ($\tau\sim120$), while \citet{Hughes17} concluded that it is optically thin. A key difference is that we have CO isotopologue data available, while \citet{Hughes17} were relying on $^{12}$CO data only. To test the influence of the CO isotopologue data, we ran additional non-LTE fits (with H$_2$ collisions) considering only our $^{12}$CO data. In that case, we are only able to derive a lower limit on the CO mass ($>10^{-3.4}$\,M$_\oplus$) because of the high $^{12}$CO optical depth.

On the other hand, when including the isotopologue data, our results are roughly consistent with the mass determined by \citet[][$10^{-2.0}$\,M$_\oplus$]{Moor19}, but an order of magnitude lower than the lowest value given by \citet[][$10^{-1.2}$\,M$_\oplus$]{Higuchi20}. Both of these works included CO isotopologues. This highlights the importance of CO isotopologue observations in determining accurate CO masses: the $^{13}$CO and C$^{18}$O masses can be determined accurately thanks to their optically thin emission. The total CO mass then follows from our assumption that the $^{12}$CO/$^{13}$CO and $^{12}$CO/C$^{18}$O ratios are equal to the C isotope ratios in the ISM. We note, however, that our work and the analysis by \citet{Moor19} and \citet{Higuchi20} are based on disk-integrated fluxes, while \citet{Hughes17} modelled spatially and spectrally resolved data with a 3D disk model. Thus, part of the difference in the mass estimates might also be explained by model differences.

Our CO temperature appears well constrained at 15--17\,K for both the LTE and non-LTE model, indicating a low gas temperature. This is consistent with the results by \citet[][8--11\,K]{Higuchi20} and \citet[][14\,K at 100\,au for their model that assumes CO/H$_2=1$]{Hughes17}. As discussed by \citet{Higuchi20}, such a low gas temperature might not be unreasonable, although more theoretical modelling considering the heating and cooling processes of the gas \citep[e.g.][]{Zagorovsky10,Kral16,Kral19} is required. We note that the dust temperature has been measured to be 59\,K \citep{Holland17}, that is, above the CO freeze-out temperature of 20\,K \citep[e.g.][]{Yamamoto17}. Therefore, we do not expect CO freeze-out to occur even though the gas temperature is low. The model by \citet{Kral19} showed that $T_\mathrm{dust}>T_\mathrm{gas}$ can indeed occur in debris disks. Similarly, the thermo-chemical debris disk model (PDR model) by \citet{Iwasaki23} also gives a gas temperature below the dust temperature.

For C$^0$, the temperature of ${\sim}24$\,K as well as the column density we derive are consistent with the results by \citet{Higuchi19_13C}. The fact that $T_\mathrm{CO}<T_\mathrm{C}$ might suggest that C and CO occupy different layers in the disk. The C$^+$ mass is not well constrained because of high optical depth at low temperature, as was already found by \citet{Roberge13} who derived a lower limit consistent with our result.

\subsection{\texorpdfstring{$\beta$}{beta}~Pictoris}
The edge-on disk around $\beta$~Pictoris is amongst the best-studied debris disks. It hosts two giant planets, $\beta$~Pic~b \citep{Lagrange09} and $\beta$~Pic~c \citep{Lagrange19}. The system shows time-variable absorption features \citep[e.g.][]{Beust98,Kiefer14,Welsh16} and transit events \citep{Zieba19,Lecavelier22,Pavlenko22} attributed to exocomet activity. The debris disk can be traced out to ${\sim}2000$\,au in scattered light \citep{Janson21}, while the mm-sized dust grains are located in a ring centred at ${\sim}100$\,au \citep{Dent14}.

Gas emission is detected from CO \citep[e.g.][]{Dent14,Matra17_betaPic}, \CI{} \citep{Cataldi18}, \CII{} \citep{Cataldi14}, \ion{O}{1} \citep{Brandeker16} as well as various heavier elements such as \ion{Fe}{1}, \ion{Na}{1} and \ion{Ca}{2} \citep{Brandeker04,Nilsson12}. Various atomic lines are also seen in absorption \citep[e.g.][]{Lagrange98,Roberge06}, including nitrogen \citep{Wilson19} and hydrogen \citep{Wilson17}, where the latter is hypothesised to come from the dissociation of water originating from evaporating exocomets.

The masses we derive are consistent with the results by \citet[][CO]{Matra17_betaPic}, \citet[][C$^0$]{Cataldi18} and \citet[][C$^+$]{Cataldi14,Cataldi18}. The disk around $\beta$~Pic is the only disk in our sample where we can be sure that the C$^0$ mass ($(1.4^{+0.4}_{-0.3})\times10^{-4}$\,M$_\oplus$) is clearly larger than the CO mass ($(4.8^{+1.2}_{-1.4})\times10^{-5}$\,M$_\oplus$).

The C temperature is not well constrained by our fits. On the other hand, the CO temperature we derive under the LTE assumption ($11.8^{+2}_{-1.1}$\,K) is in general significantly lower than previous estimates from the literature. Indeed, the theoretical model by \citet{Zagorovsky10} predicts a gas temperature of ${\sim}60$\,K at 100\,au. \citet{Kral16} derived a temperature of 50\,K (at 100\,au) from a PDR model fitted to CO, \CI{} and \CII{} data. \citet{Matra17_betaPic} empirically derive ${\sim}160$\,K (again at 100\,au) from the CO scale height. However, our temperature is consistent with the excitation temperature derived by \citet{Matra17_betaPic} from the CO 3--2/2--1 line ratio. It is also roughly consistent with the rotational excitation temperature of $15.8\pm0.6$\,K measured by \citet{Roberge00} from CO absorption. Therefore, our low temperature likely indicates that CO is not in LTE \citep{Matra17_betaPic}, and therefore that this temperature does not correspond to the kinetic temperature.

\subsection{HD~21997}
The debris disk around HD~21997 is among the disks with the highest CO masses in our sample. The disk served as a prototype for the proposed class of ``hybrid disks'', that is, disks where secondary dust and primordial gas co-exist \citep{Kospal13}. ALMA observations of the CO emission were presented by \citet{Kospal13}, who found that the gas and dust are not co-located: there is a dust-free inner gas disk. If the gas was secondary, then this would suggest viscous spreading of the gas with respect to the dust \citep{Kral19,Marino20}.

Compared to the CO analysis by \citet{Kospal13} and \citet{Higuchi20} that included $^{12}$CO~2--1 and 3--2, $^{13}$CO~2--1 and 3--2, and C$^{18}$O~2-1, we added observations of C$^{18}$O~3-2 (as well as additional observations of $^{13}$CO~3--2). The disk around HD~21997 thus has the most CO lines observed in our sample. We also publish the first measurements of \CI{} and \CII{}, and thus the masses of C$^0$ and C$^+$ are constrained for the first time in this paper.

Our derived CO masses of $4.8^{+0.8}_{-0.9}\times10^{-2}$\,M$_\oplus$ (LTE) and
$6.8^{+7}_{-1.9}\times10^{-2}$\,M$_\oplus$ (non-LTE, H$_2$ colliders) are consistent with the ranges reported by \citet[][4--8$\times10^{-2}$\,M$_\oplus$]{Kospal13} and \citet[][5.5--85$\times10^{-2}$\,M$_\oplus$]{Higuchi20}. There is also agreement in the low temperature of the gas: we derive ${\sim}10$\,K (the lowest temperature allowed in our fits), while \citet{Kospal13} and \citet{Higuchi20} derive 6--9\,K and 8--12\,K, respectively.

\subsection{HD~32997}
The debris disk around HD~32297 has been studied extensively in scattered light \citep[][to cite some recent work]{Bhowmik19,Duchene20}, in the far-IR with \textit{Herschel} \citep{Donaldson13} and in the sub-mm/mm regime \citep[e.g.][]{MacGregor18}. We derive a CO mass of $5.2^{+1.7}_{-1.5}\times10^{-2}$\,M$_\oplus$, consistent with the value derived by \citet{Moor19}. This is the largest CO mass in our sample and about two orders of magnitude larger than the CO mass derived by \citet{MacGregor18}. As for 49~Ceti, the reason for this discrepancy is that our work and \citet{Moor19} include CO isotopologue data, in contrast to \citet{MacGregor18}.

This disk is also detected in \CI{} \citep{Cataldi20} and \CII{} \citep{Donaldson13}. Our estimate of the C$^0$ mass is consistent with the value derived by \citet{Cataldi20} and roughly one order of magnitude smaller than the CO mass. Unfortunately, the C$^+$ mass is not well constrained due to high optical depth at low temperature.

\subsection{HD~48370, HD~61005, HD~95086, HR~4796}
The debris disks around HD~48370, HD~61005, HD~95086 and HR~4796 are the disks in our sample that remain undetected in both CO and carbon. Our LTE upper limit on the CO mass in the HD~61005 disk is a factor of ${\sim}6$ higher than the value derived by \citet{Olofsson16} assuming optically thin emission and LTE. Since \citet{Olofsson16} did not publish the line flux upper limit they used for their estimate, it is difficult to find the reason for this difference. For HR~4796, our LTE upper limit on the CO mass is roughly consistent with the upper limit derived by \citet{Kennedy18}. Comparing instead to the LTE and non-LTE results by \citet{Kral20_HD129590}, our LTE upper limit is about an order of magnitude more constraining, while the higher non-LTE upper limit is roughly consistent with their derivation.

The derived upper limits show that these disks all have a CO content smaller than $\beta$~Pic. The same is true for C$^0$, except maybe for HD~95086 where the upper limit on the C$^0$ mass is relatively high. The two G-stars HD~48370 and HD~61005 are the two coolest stars in our sample. \citet{Matra19_luminosity_correlation} suggested that for a fixed fractional luminosity, the production rate of secondary CO gas is proportional to the stellar luminosity, which might be the reason why we do not detect any gas. Moreover, these two stars are also among the oldest (${\sim}40$\,Myr) in our sample, which might be another reason. HD~95086 does not stand out in any particular way in our sample in terms of stellar effective temperature, age, or fractional luminosity. On the other hand, HR~4796 has the highest effective temperature \citep[spectral type A0,][]{Houk82} and is the youngest system (8\,Myr) in our sample. \citet{Kennedy18} suggest that the non-detection of CO emission cannot be solely explained by the intense UV radiation field of an A0 star that limits the lifetime of CO molecule to 40 years or less. In addition, the planetesimals in the HR~4796 system need to be CO-poor, with a CO+CO$_2$ ice mass fraction of ${<}1.8$\%, which is smaller than solar system comets. It is indeed possible that the planetesimals we observe today at a radius of $\sim$80\,au formed interior to the CO snowline, due to the high luminosity of HR~4796 \citep{Marino20}. From Fig.\ 1 in \citet{Matra18_radius_luminosity_relation}, the CO snowline in the HR~4796 protoplanetary disk could have been located as far out as ${\sim}100$\,au. On the other hand, if the CO emission is not in LTE, the CO upper limit becomes less constraining and the planetesimals could have a CO+CO$_2$ content similar to or even larger than solar system comets \citep{Kral20_HD129590}. Another reason for the absence of gas emission from the HR~4796 system could be that gas is removed by radiation pressure, because the luminosity of HR~4796 is larger than 20\,L$_\odot$ and therefore a wind driven by the stellar radiation can be expected \citep{Kral17,Kral23}.

\subsection{HD~110058}
The edge-on debris disk around HD~110058 was discovered by \citet{Kasper15} in scattered light. The disk shows absorption features most probably due to hot circumstellar gas \citep{Hales17,Iglesias18,Rebollido18}. Emission from $^{12}$CO and $^{13}$CO has been detected \citep{Lieman-Sifry16,Hales22}. \citet{Hales22} derive a CO mass of $\log(M_\mathrm{CO}\,[M_\oplus])=-1.16_{-0.77}^{+1.72}$ (95\% credible interval), consistent with our $-2.03_{-0.12}^{+0.14}$. However, \citet{Hales22} derive a significantly lower CO temperature (${\sim}18$\,K at the radius of the debris belt, compared to our ${\sim}100$\,K). They consider a power law for the radial temperature dependence, which is likely more realistic than our assumption of a constant temperature.

We present the first observations of \CI{} towards this target. No \CI{} emission was detected. This is an interesting result: among the disks with a high CO mass (see Fig.\ \ref{fig:masses_and_column_densities}), HD~110058 is the only one without \CI{} detected.  \citet{Hales22} applied the \citet{Marino20} secondary gas model to the debris disk around HD~110058 and successfully reproduced the observed CO mass. However, they predict a C$^0$ mass of ${\sim}10^{-2}$\,M$_\oplus$, which exceeds our upper limit of $M(\mathrm{C}^0)<3.2\times10^{-4}$\,M$_\oplus$. Thus, our data allow us to exclude their secondary gas model.

\subsection{HD~121191}
The debris disk around HD~121191 was first identified by \citet{Melis13}. The disk shows strong CO emission \citep{Moor17}. The CO mass we derive is consistent with the masses published by \citet{Moor17} and the LTE mass by \citet{Kral20_HD129590}, but our lower limit on the non-LTE CO mass is $\sim$30\% larger than the largest non-LTE mass derived by \citet{Kral20_HD129590}. Our work presents the first detection of \CI{} towards this target. We find that the C$^0$ mass is about an order of magnitude smaller than the CO mass.

\subsection{HD~121617}
Dust emission from the debris disk around HD~121617 was first reported by \citet{Mannings98}, while CO emission was first detected by \citet{Moor17}, who estimated a CO mass roughly consistent with our results. Our work presents the first detection of \CI{} emission. Similarly to HD~121191, the C$^0$ mass is smaller than the CO mass by roughly an order of magnitude.

\subsection{HD~131835}
The debris disk around HD~131835, first detected by \citet{Moor06}, is known to show $^{12}$CO \citep{Moor15_APEX,Lieman-Sifry16,Hales19} and \CI{} \citep{Kral19} emission. For our model, we use additional observations of $^{13}$CO and C$^{18}$O that will be presented in more detail in \citet{Moor_inprep}. Compared to the CO mass published by \citet[][${\sim}3.8\times10^{-2}$\,M$_\oplus$ after correcting for the updated distance of the target]{Moor17}, we find a CO mass roughly a factor of 2 smaller (${\sim}1.7\times10^{-2}$\,M$_\oplus$). On the other hand, our CO mass is larger by a factor ${\sim}4$ compared to the CO mass derived by \citet{Hales19} with a 2D thermo-chemical model ($4.6\times10^{-3}$\,M$_\oplus$), potentially because \citet{Hales19} did not consider the CO isotopologue data and/or because they assumed that the dust temperature equals the gas temperature.

Our C$^0$ mass is consistent with the value derived by \citet{Kral19}, although the mass is not well constrained due to optical depth and could be much higher if the gas temperature is low (see the corresponding corner plots in Figures \ref{fig:corner_nonLTE_H2_HD131835} and \ref{fig:corner_nonLTE_e_HD131835}).

\subsection{HD~146897}
The edge-on debris disk around HD~146897 has been resolved at both mm wavelengths with ALMA \citep{Lieman-Sifry16} as well as scattered light \citep[e.g.][]{Goebel18}. The disk shows CO emission \citep{Lieman-Sifry16} as well as absorption by hot circumstellar gas \citep{Rebollido18}. In this work we present ACA observations of \CI{} that did not detect any emission. Thus, HD~146897 is another example of a disk with CO where no \CI{} was detected, although it might be less surprising than in the case of HD~110058 because 1) the CO mass is likely much smaller (although it could be large if the temperature is low as can be seen in the corner plots in Figures \ref{fig:corner_LTE_HD146897}, \ref{fig:corner_nonLTE_H2_HD146897} and \ref{fig:corner_nonLTE_e_HD146897}) and 2) the ACA \CI{} observations are much less sensitive.

\subsection{HD~181327}
HD~181327 is a member of the $\beta$~Pictoris moving group and hosts a thin debris ring that exhibits weak CO emission \citep{Marino16}. We derive a small CO mass of ${\sim}2\times10^{-6}$\,M$_\oplus$, consistent with the analysis by \citet{Marino16}.

Among the disks with a CO detection in our sample, this disk has by far the smallest CO mass (about 1.5 orders of magnitude smaller than the second smallest CO mass found in the $\beta$~Pic debris disk). There are other debris disks with similarly small CO masses such as Fomalhaut \citep{Matra17_Fomalhaut} or TWA~7 \citep{Matra19_luminosity_correlation}, but they were not included in our sample due to a lack of \CI{} data. In this work we present the first observations of \CI{} towards HD~181327. The line remains undetected, as does \CII{} \citep{Riviere-Marichalar14}.

\section{Moment 0 maps}\label{appendix:moment0_maps}

\begin{figure*}[h]
	\plotone{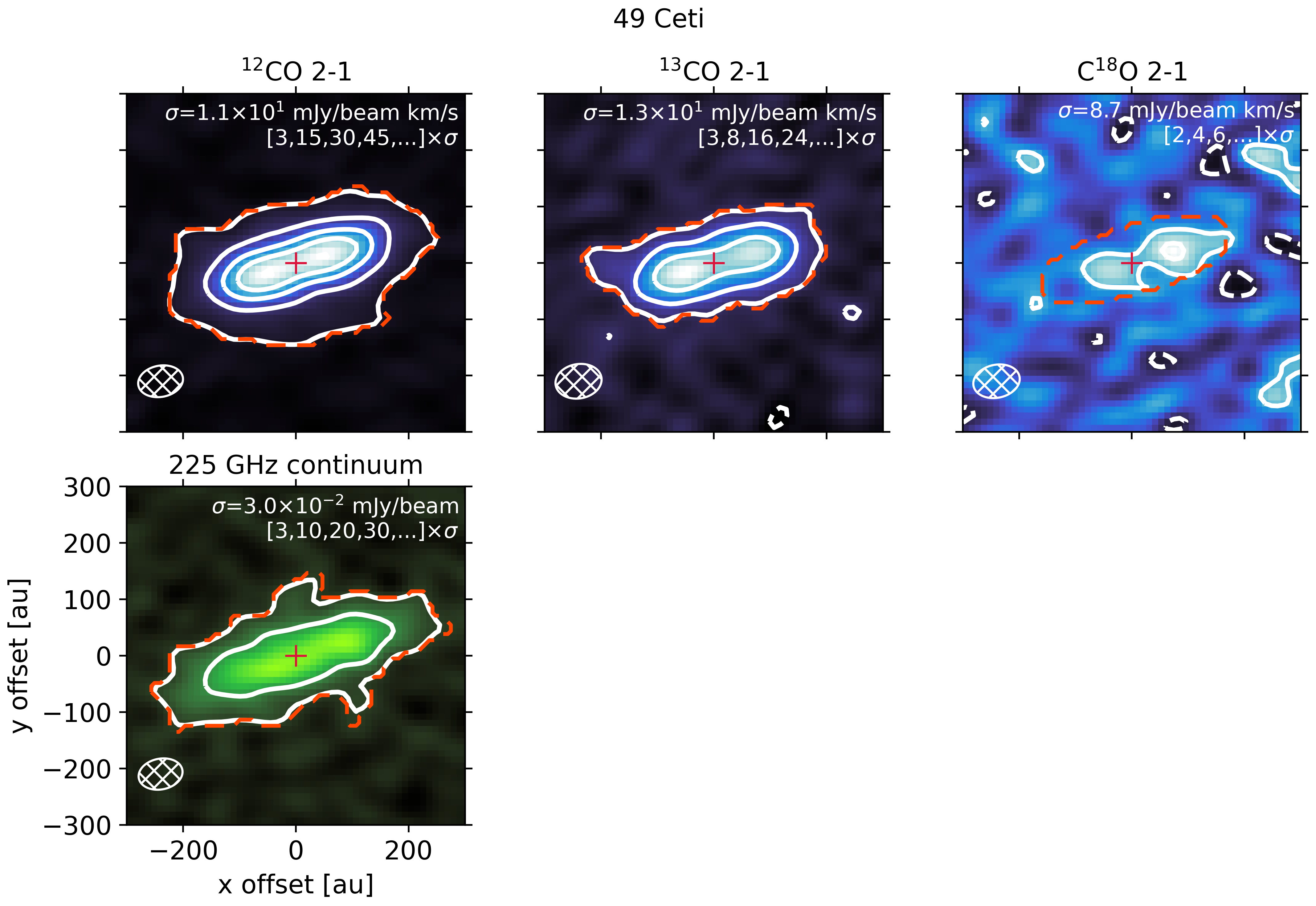}
	\caption{Moment 0 and continuum maps used for flux measurements for the disk around 49~Ceti. The noise and contour levels are indicated in the upper right of each image. The beam size is indicated by the hatched ellipse in the bottom left. The stellar position is marked by the red cross. The orange contours mark the aperture used to measure the flux.\label{fig:mom0_49Ceti}}
\end{figure*}

\begin{figure*}[h]
	\plotone{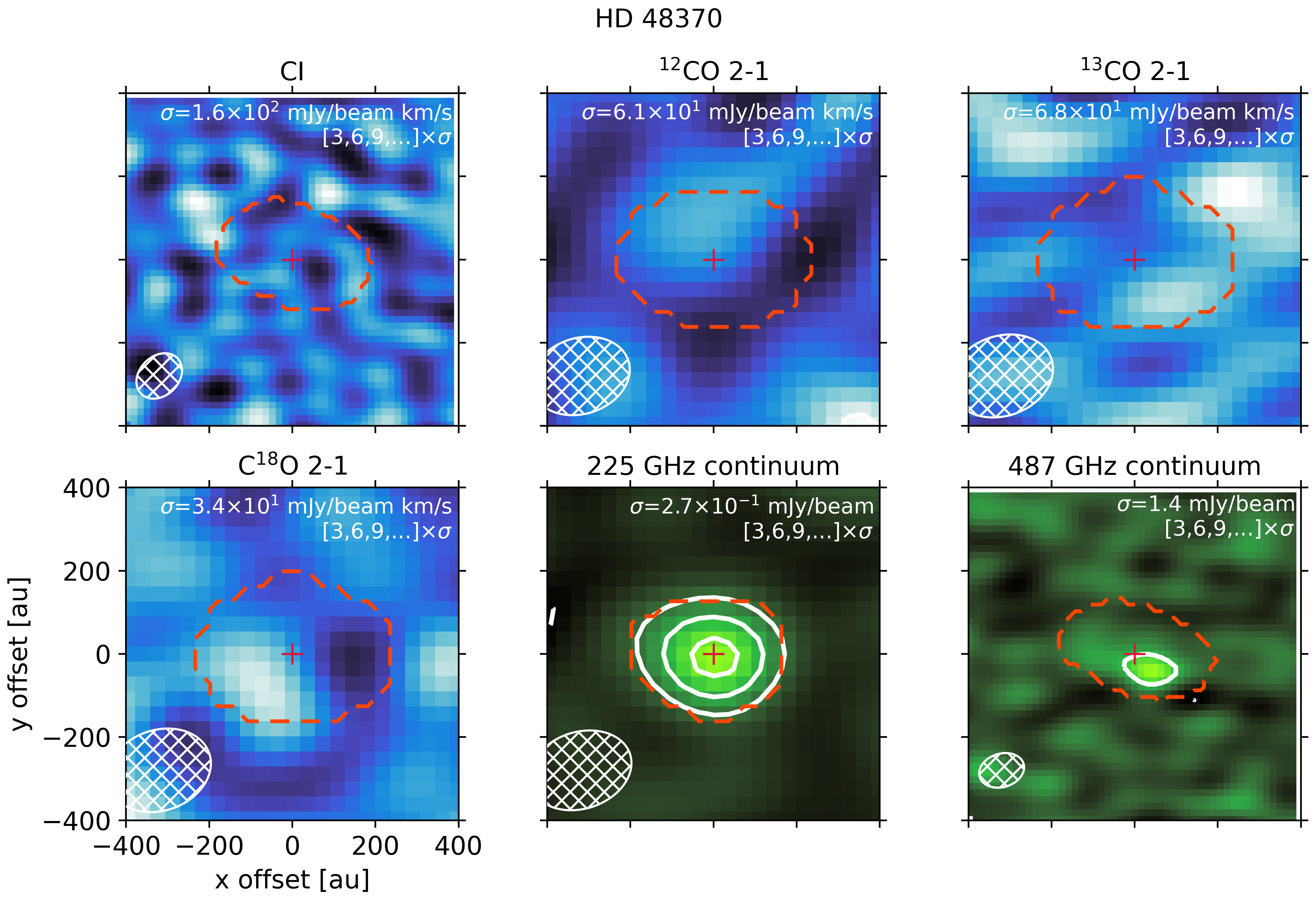}
	\caption{Same as Fig.\ \ref{fig:mom0_49Ceti}, but for HD~48370. \label{fig:mom0_HD48370}}
\end{figure*}

\begin{figure*}[h]
	\plotone{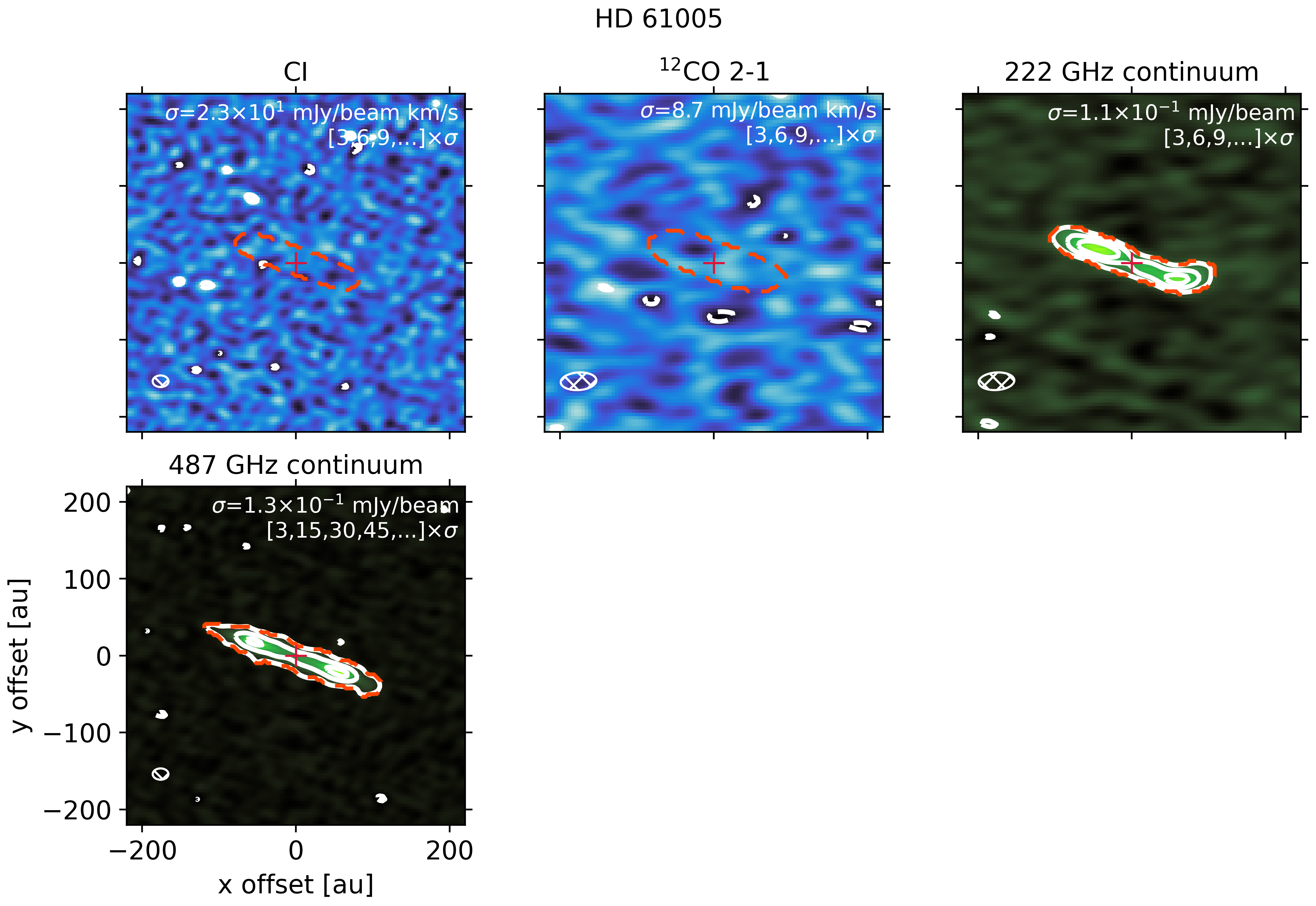}
	\caption{Same as Fig.\ \ref{fig:mom0_49Ceti}, but for HD~61005. \label{fig:mom0_HD61005}}
\end{figure*}

\begin{figure*}[h]
	\plotone{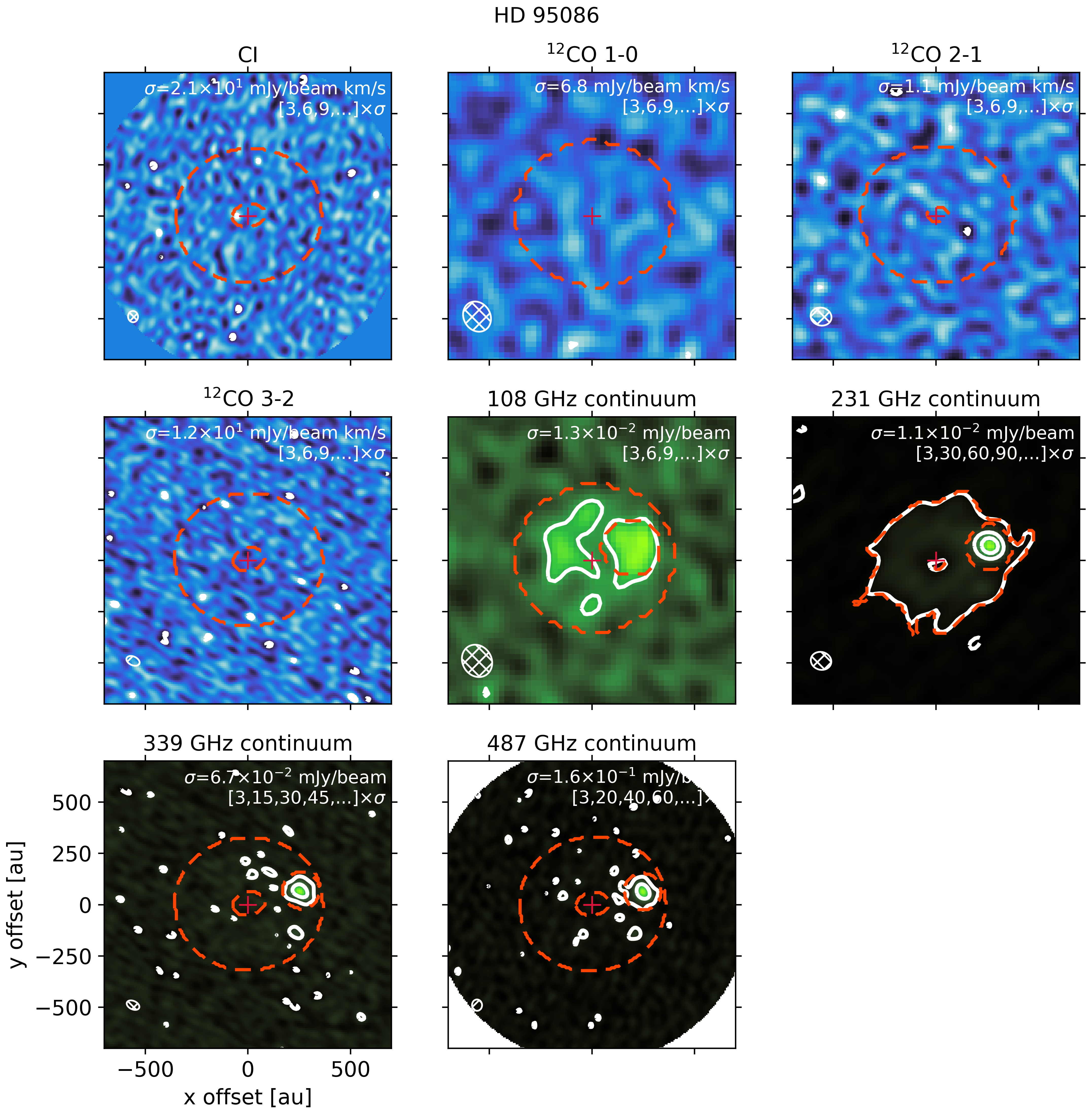}
	\caption{Same as Fig.\ \ref{fig:mom0_49Ceti}, but for HD~95086. Note the continuum source in the north-east part of the disk \citep[probably a submillimetre background galaxy,][]{Zapata18} that was excluded from the flux measurement. \label{fig:mom0_HD95086}}
\end{figure*}

\begin{figure*}[h]
	\plotone{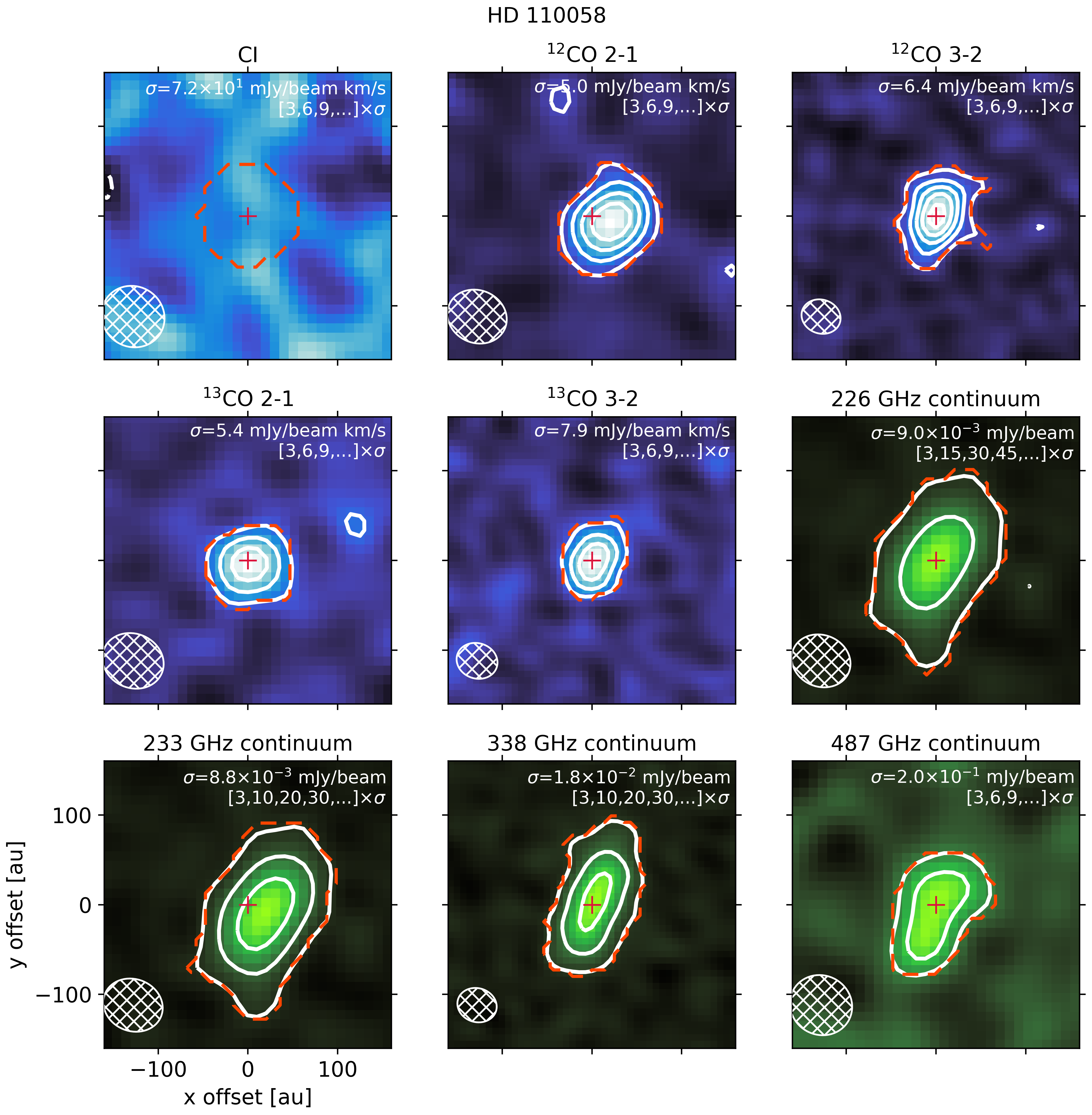}
	\caption{Same as Fig.\ \ref{fig:mom0_49Ceti}, but for HD~110058. There is an offset for $^{12}$CO~2--1 between the disk center and the expected position of the star. \label{fig:mom0_HD110058}}
\end{figure*}

\begin{figure*}[h]
	\plotone{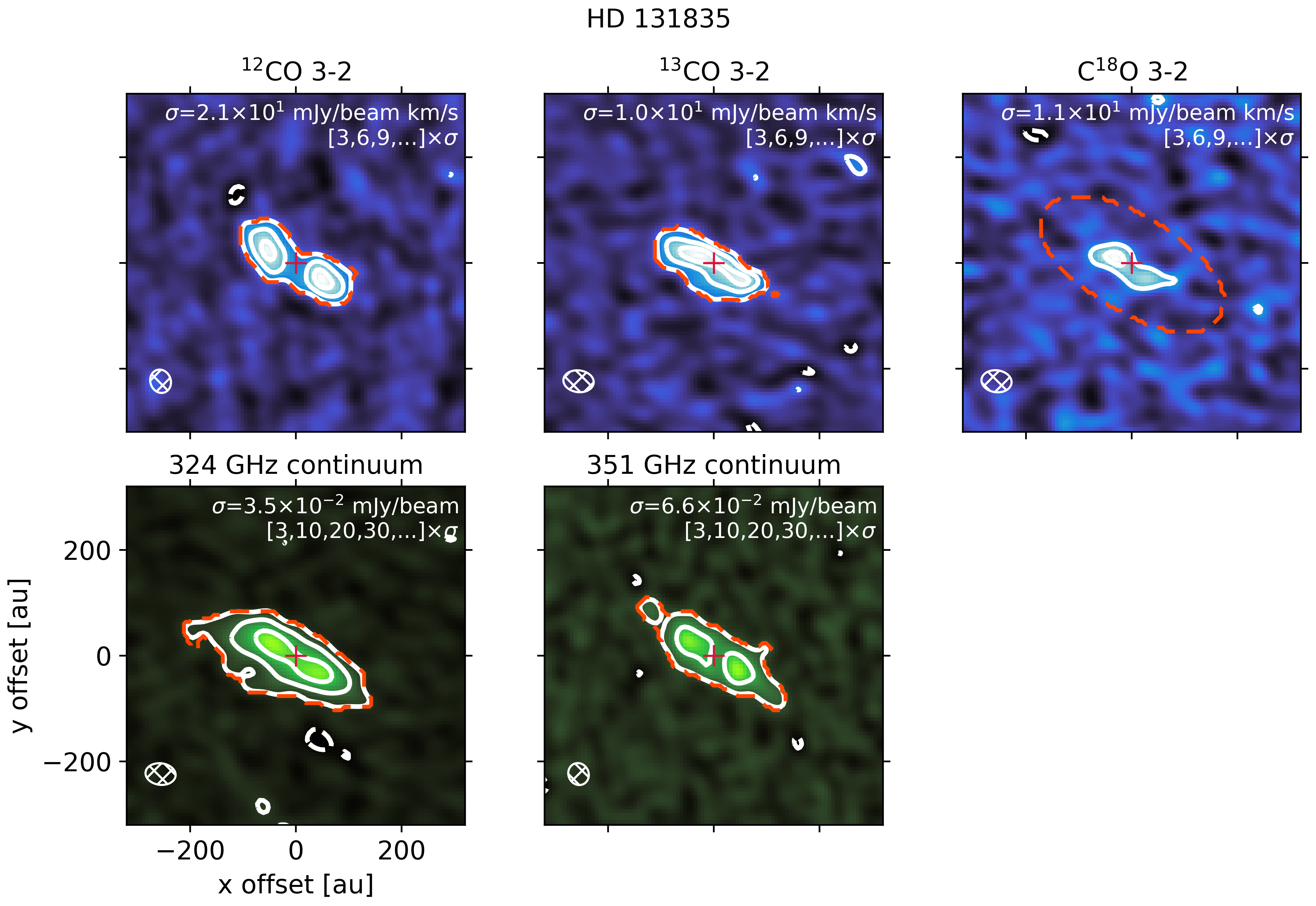}
	\caption{Same as Fig.\ \ref{fig:mom0_49Ceti}, but for HD~131835. There might be slight offset between the disk center and the expected stellar position for $^{13}$CO~3--2 and C$^{18}$O~3--2.\label{fig:mom0_HD131835}}
\end{figure*}

\begin{figure*}[h]
	\plotone{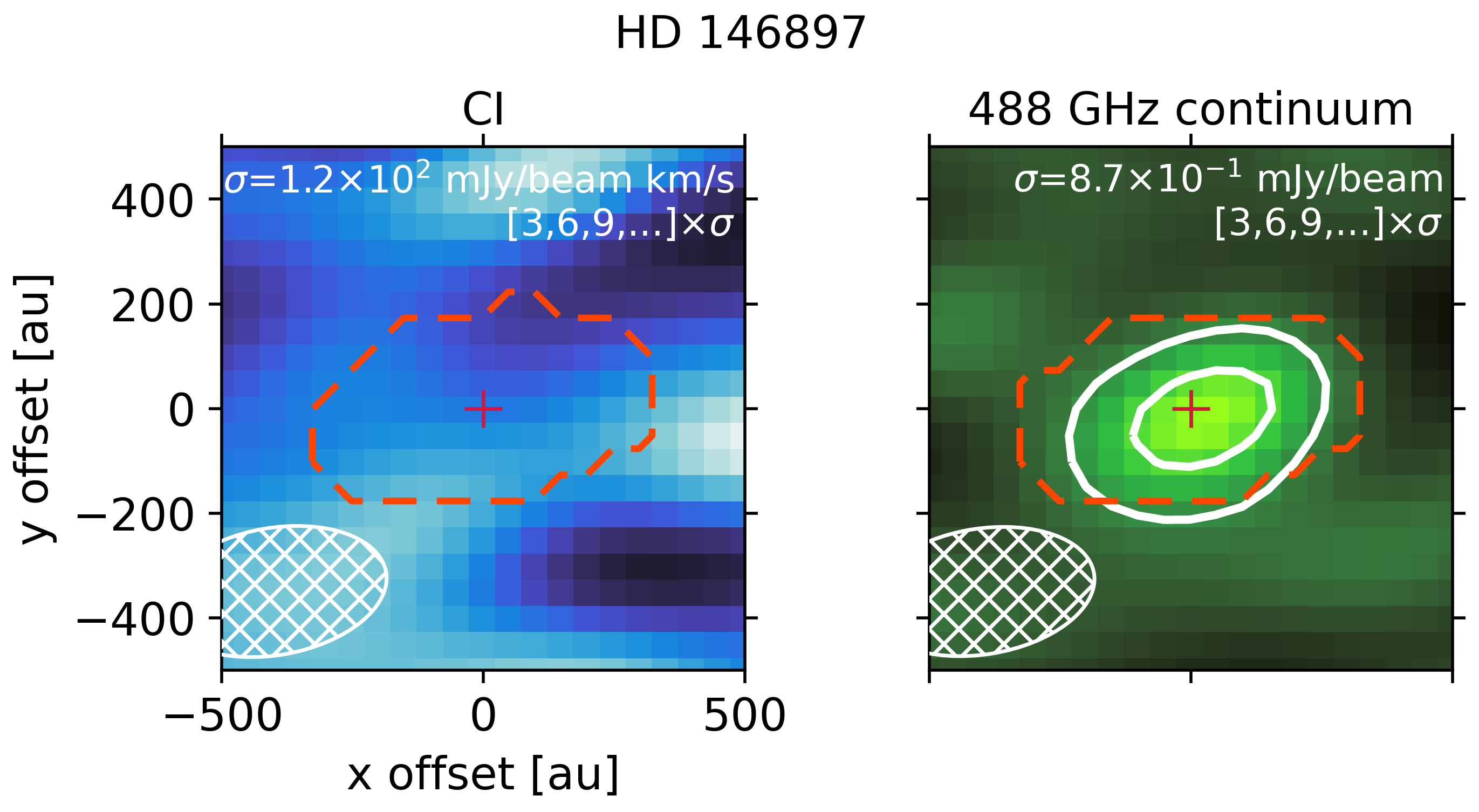}
	\caption{Same as Fig.\ \ref{fig:mom0_49Ceti}, but for HD~146897. \label{fig:mom0_HD146897}}
\end{figure*}

\begin{figure*}[h]
	\plotone{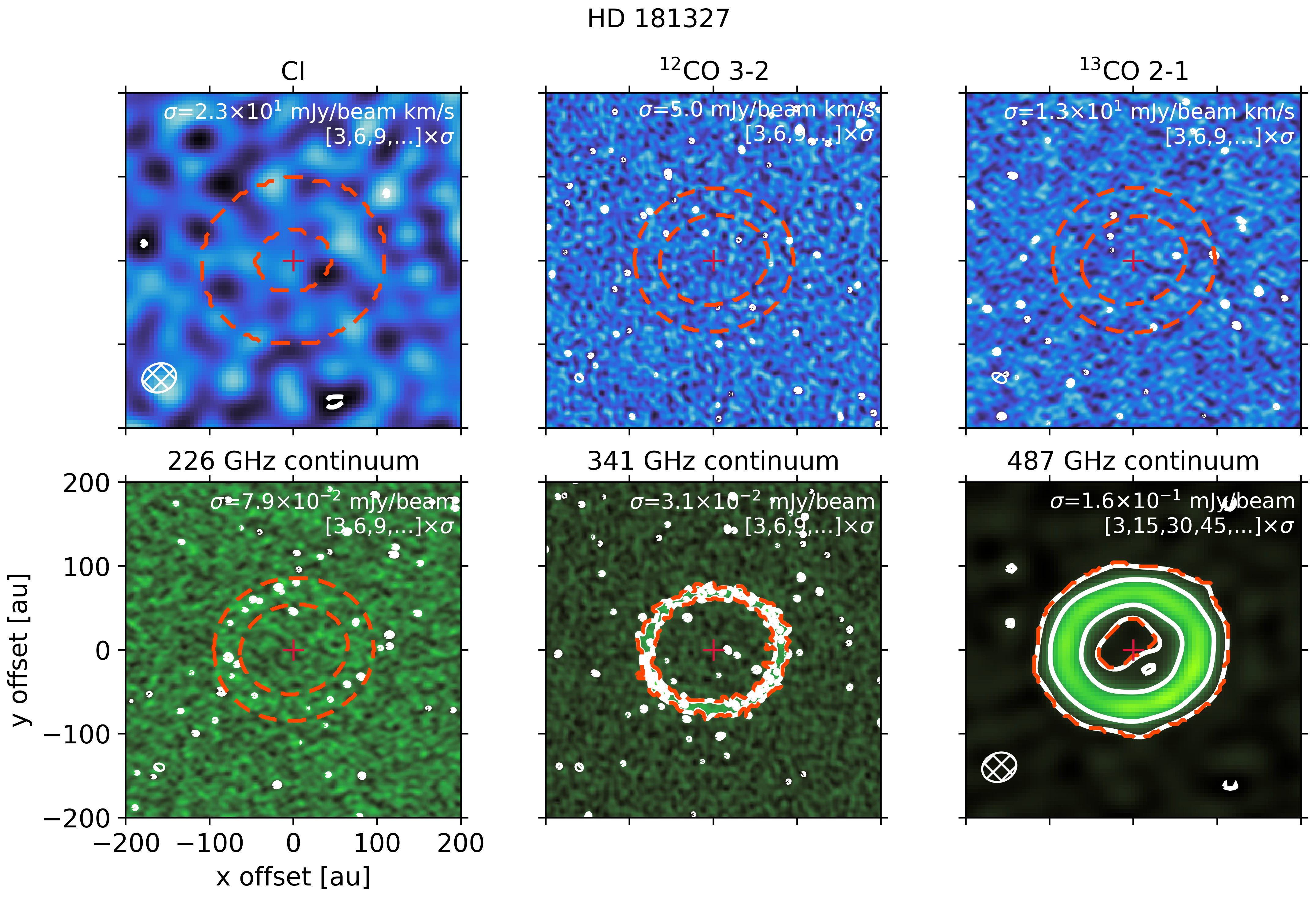}
	\caption{Same as Fig.\ \ref{fig:mom0_49Ceti}, but for HD~181327. \label{fig:mom0_HD181327}}
\end{figure*}

\begin{figure*}[h]
	\plotone{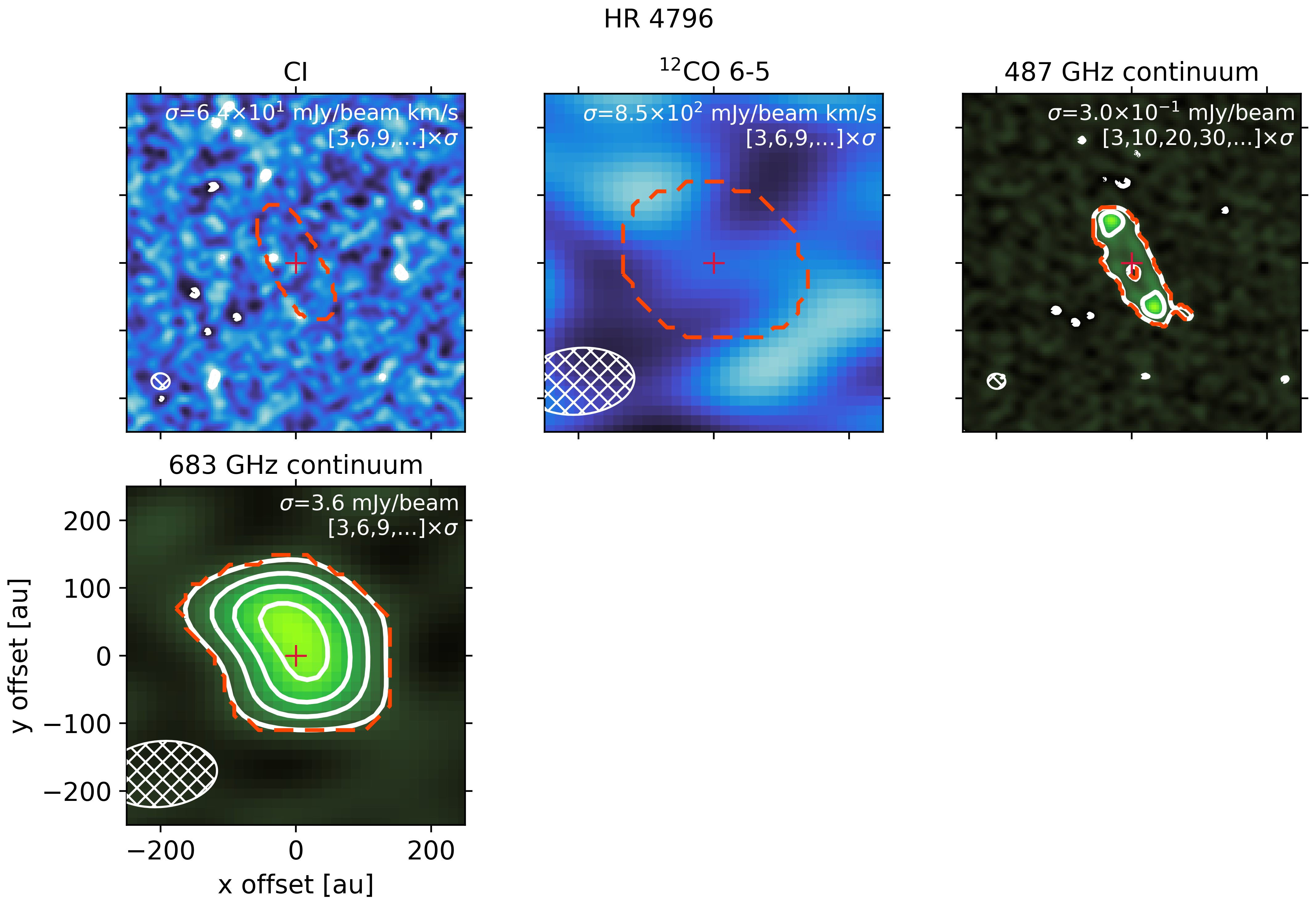}
	\caption{Same as Fig.\ \ref{fig:mom0_49Ceti}, but for HR~4796. \label{fig:mom0_HR4796}}
\end{figure*}

Figures \ref{fig:mom0_49Ceti} to \ref{fig:mom0_HR4796} show the moment 0 and continuum maps that we used to measured disk-integrated fluxes and that were not shown in the main text.

\section{Corner plots}
\subsection{LTE fits}\label{appendix:corner_plots_LTE}
\begin{figure*}[h]
	\plotone{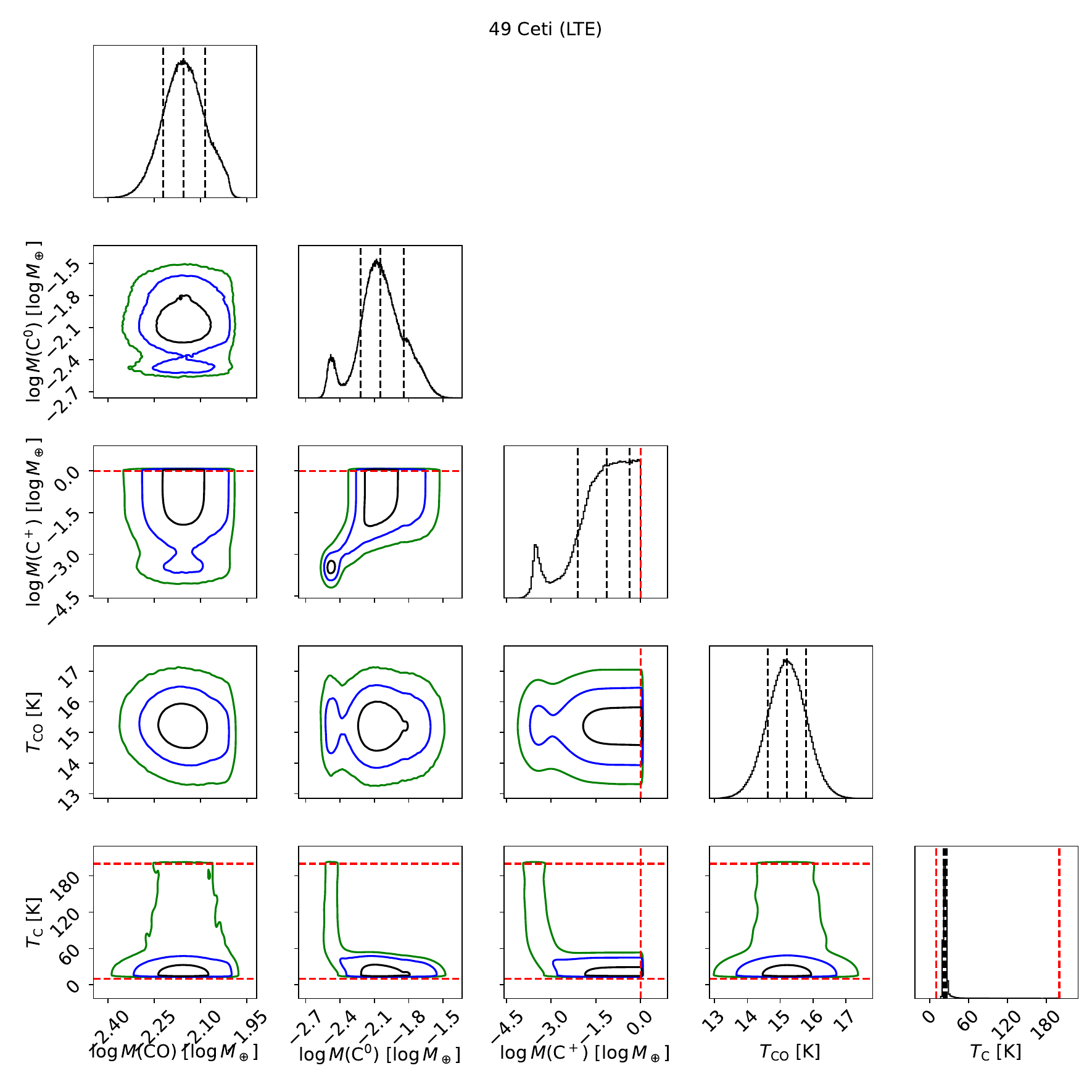}
	\caption{Corner plot showing posterior probability distributions derived assuming LTE, for the disk around 49~Ceti. In the 1D histograms, the black vertical dashed lines indicate the 16th, 50th and 84th percentile. In the 2D histograms, the black, blue and green contours mark the 50th, 90th and 99th percentile. Red dashed lines mark the upper or lower bound of the prior distribution.}\label{fig:corner_LTE_49Ceti}
\end{figure*}

\begin{figure*}[h]
	\plotone{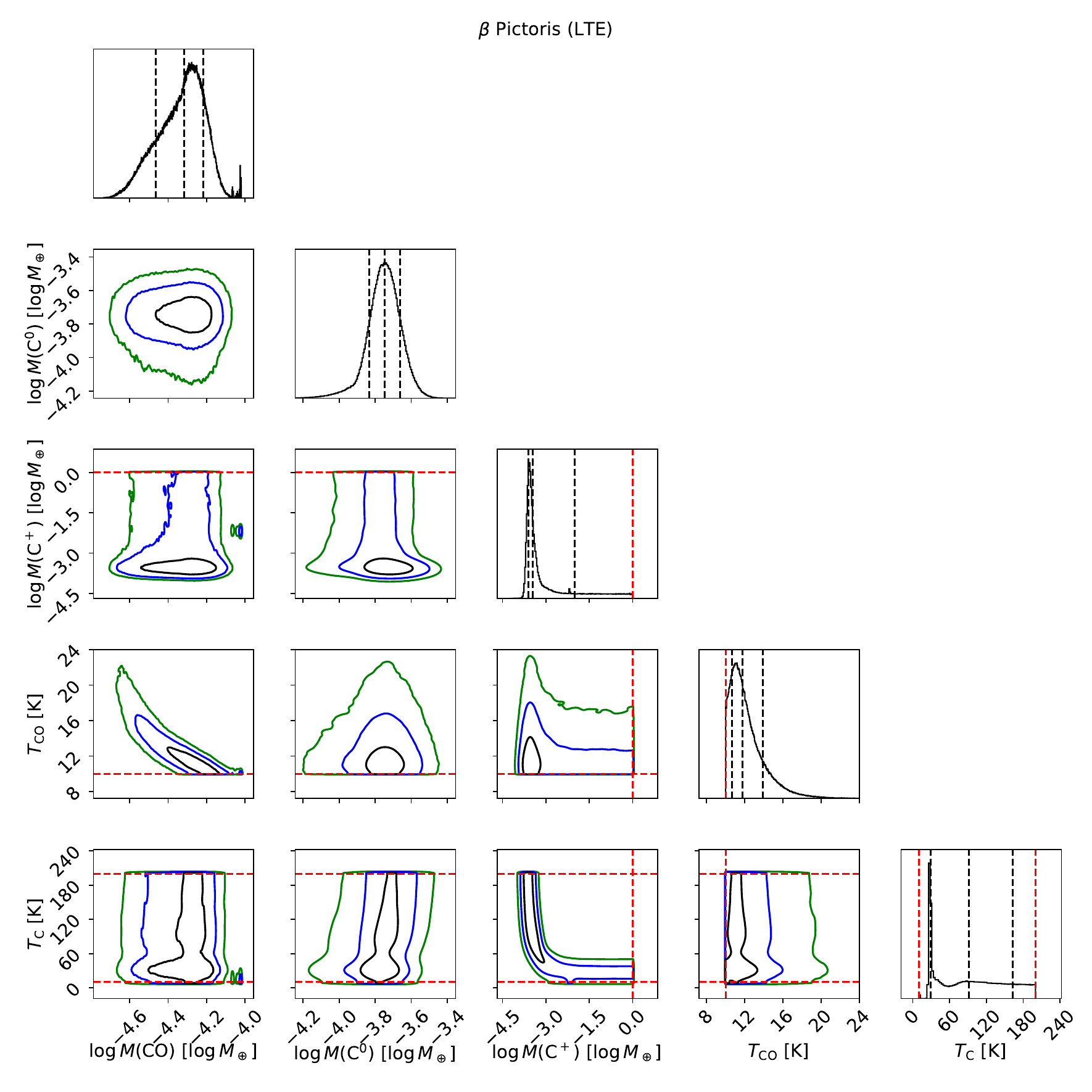}
	\caption{Same as Fig.\ \ref{fig:corner_LTE_49Ceti}, but for the disk around $\beta$~Pic.\label{fig:corner_LTE_betaPic}}
\end{figure*}

\begin{figure*}[h]
	\plotone{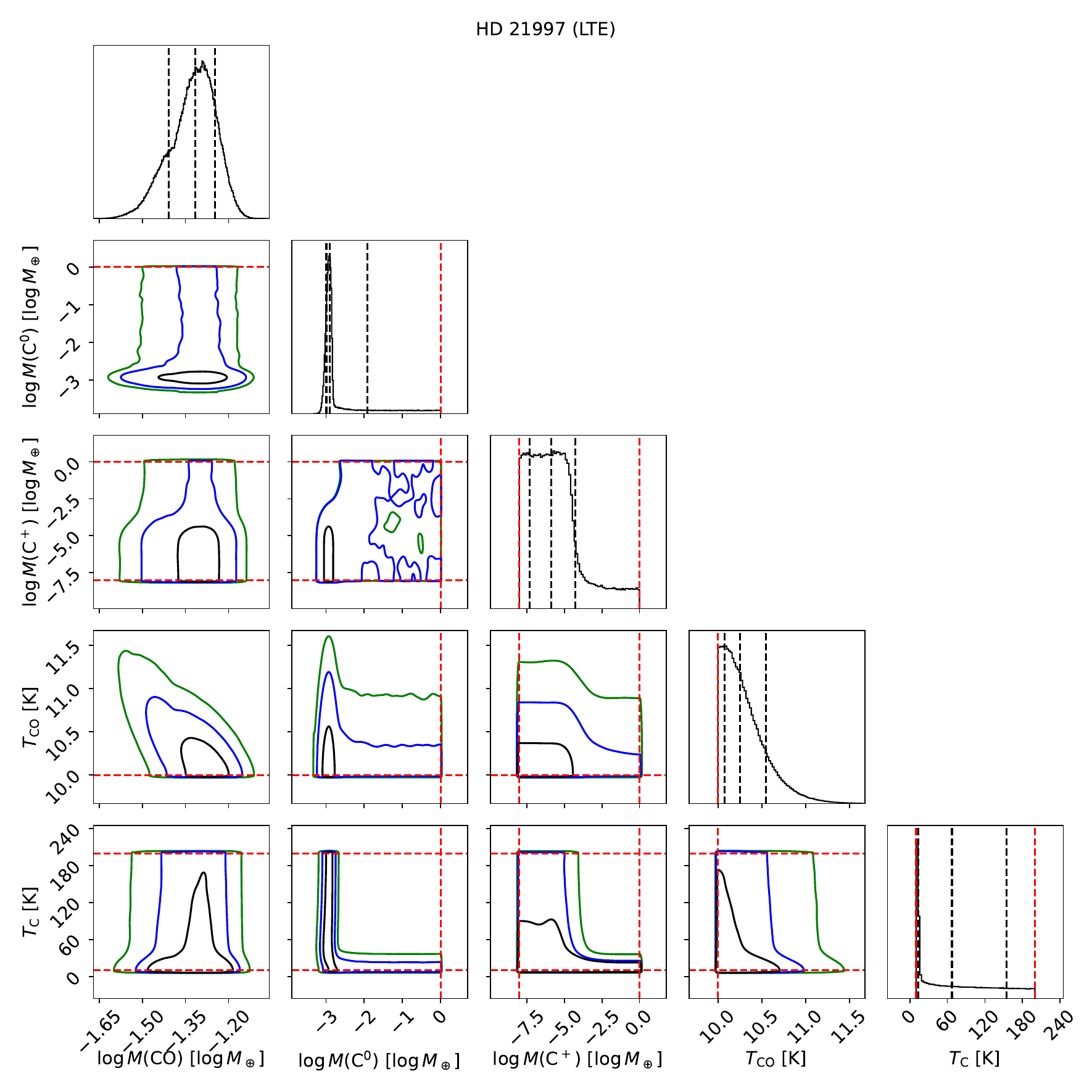}
	\caption{Same as Fig.\ \ref{fig:corner_LTE_49Ceti}, but for the disk around HD~21997.\label{fig:corner_LTE_HD21997}}
\end{figure*}

\begin{figure*}[h]
	\plotone{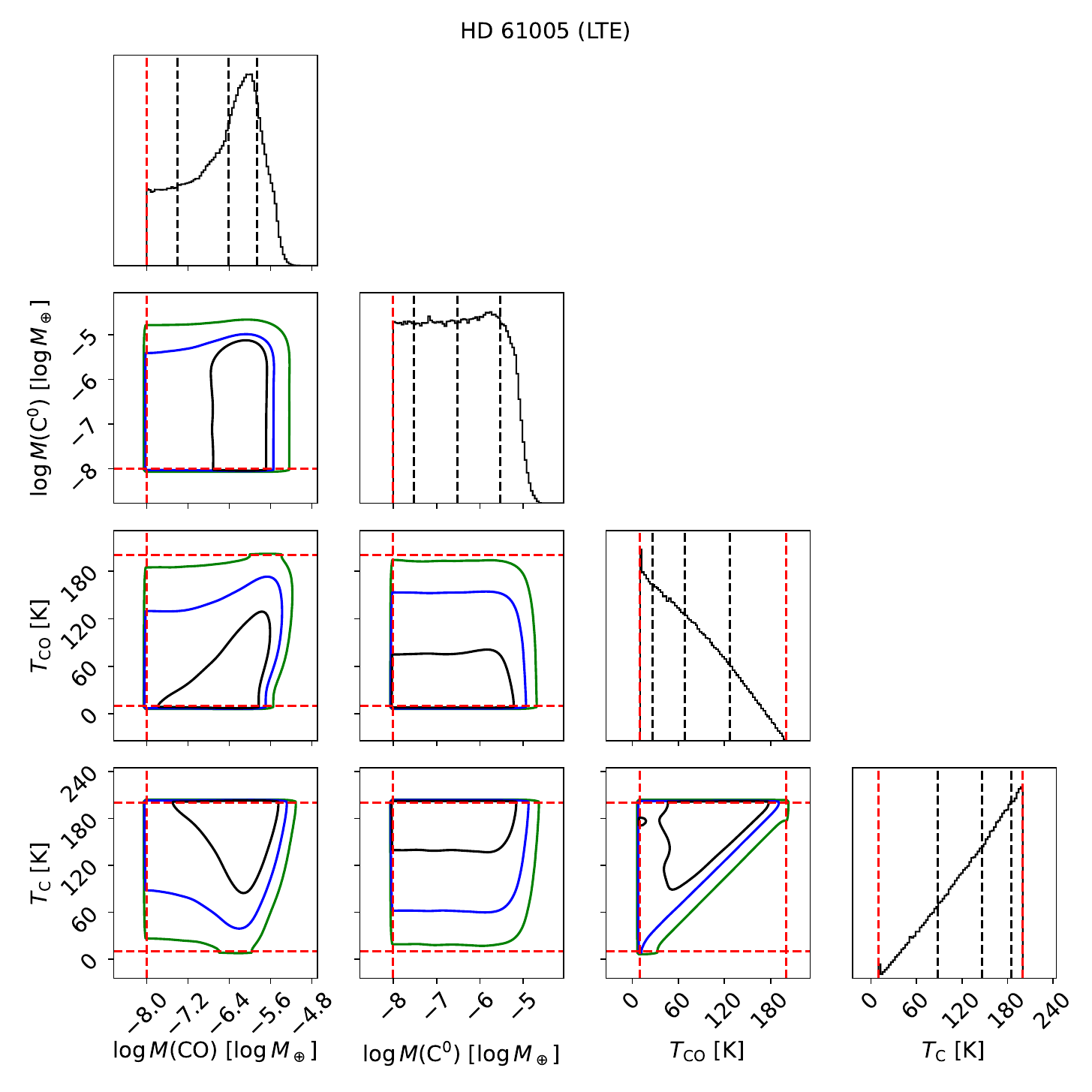}
	\caption{Same as Fig.\ \ref{fig:corner_LTE_49Ceti}, but for the disk around HD~61005.\label{fig:corner_LTE_HD61005}}
\end{figure*}

\begin{figure*}[h]
	\plotone{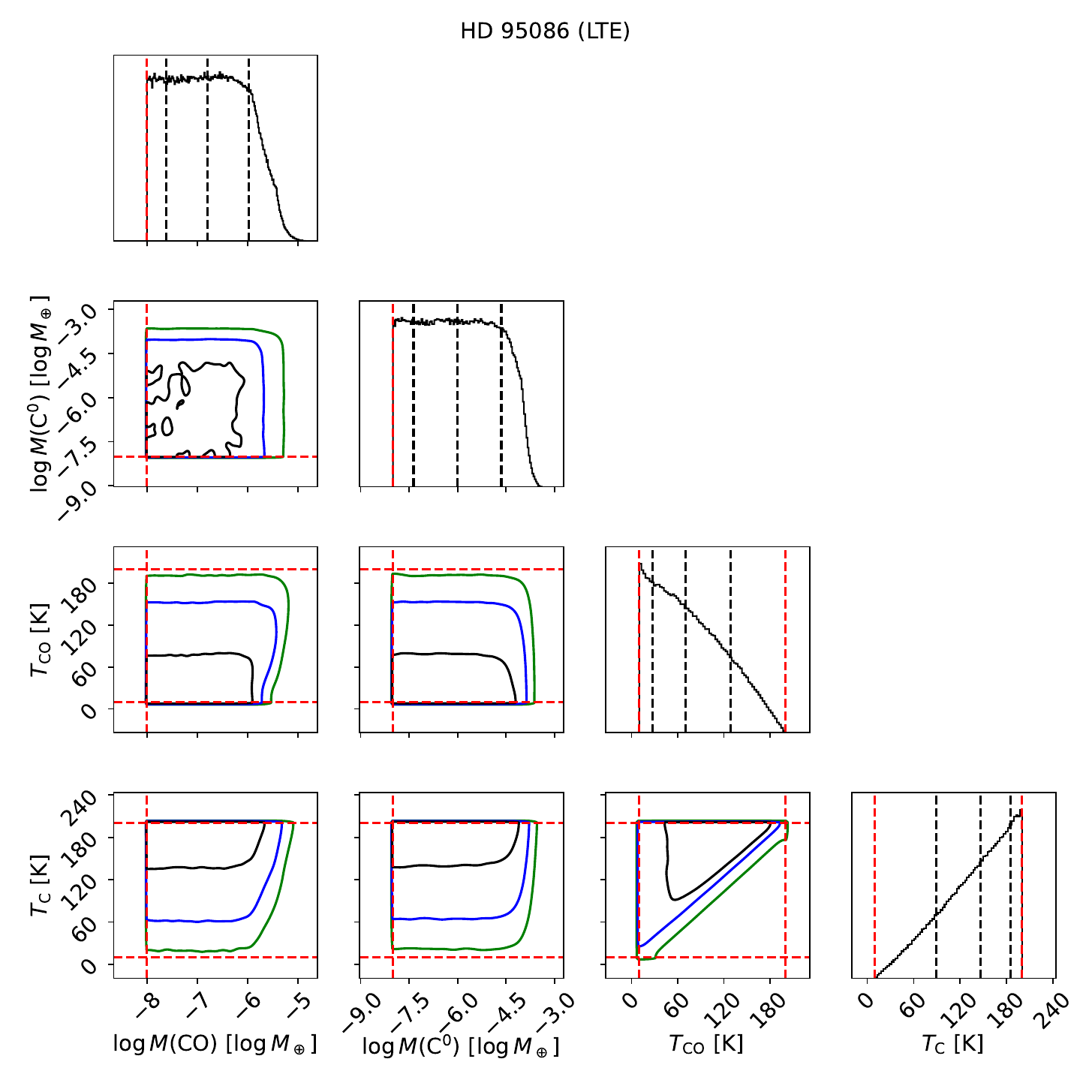}
	\caption{Same as Fig.\ \ref{fig:corner_LTE_49Ceti}, but for the disk around HD~95086.\label{fig:corner_LTE_HD95086}}
\end{figure*}

\begin{figure*}[h]
	\plotone{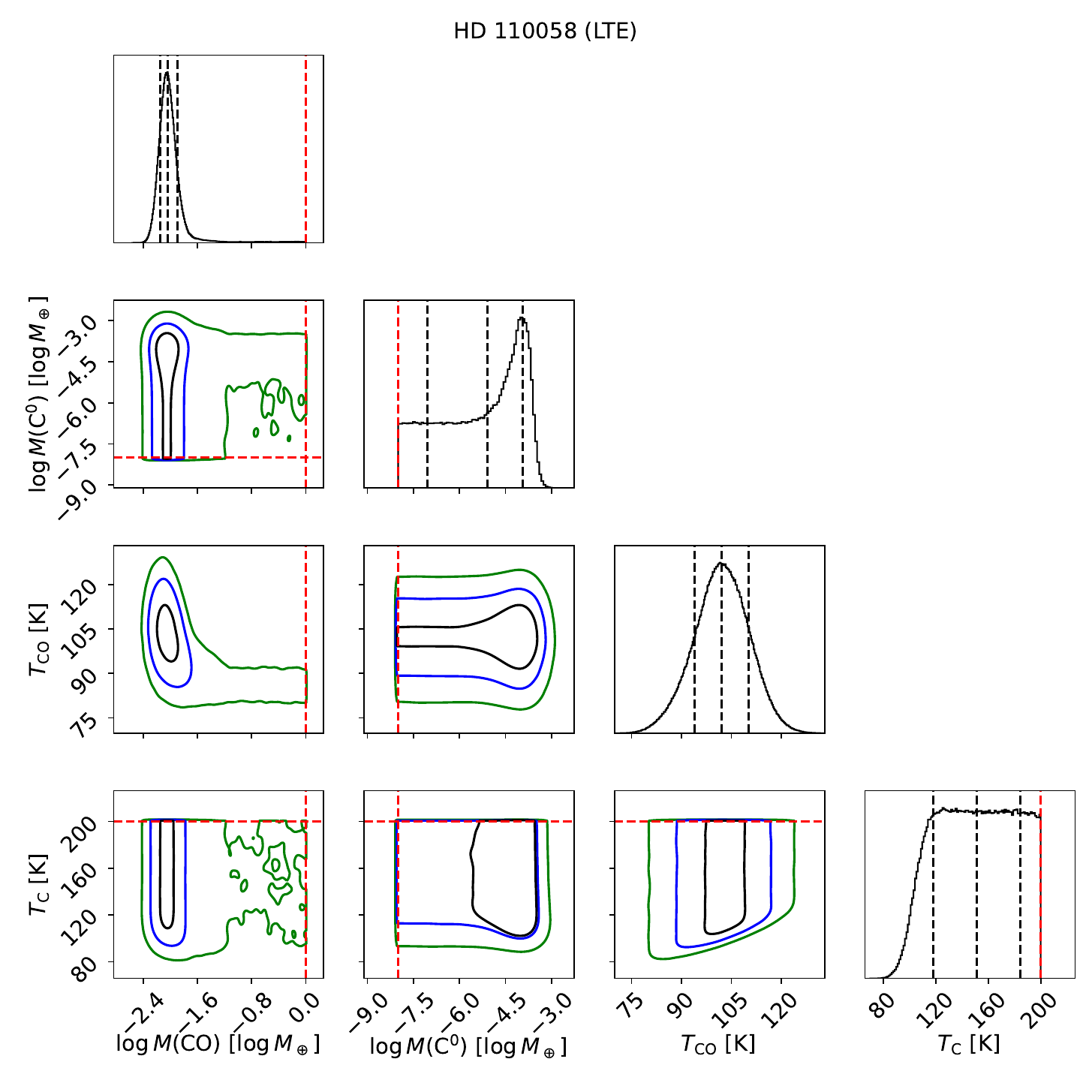}
	\caption{Same as Fig.\ \ref{fig:corner_LTE_49Ceti}, but for the disk around HD~110058.\label{fig:corner_LTE_HD110058}}
\end{figure*}

\begin{figure*}[h]
	\plotone{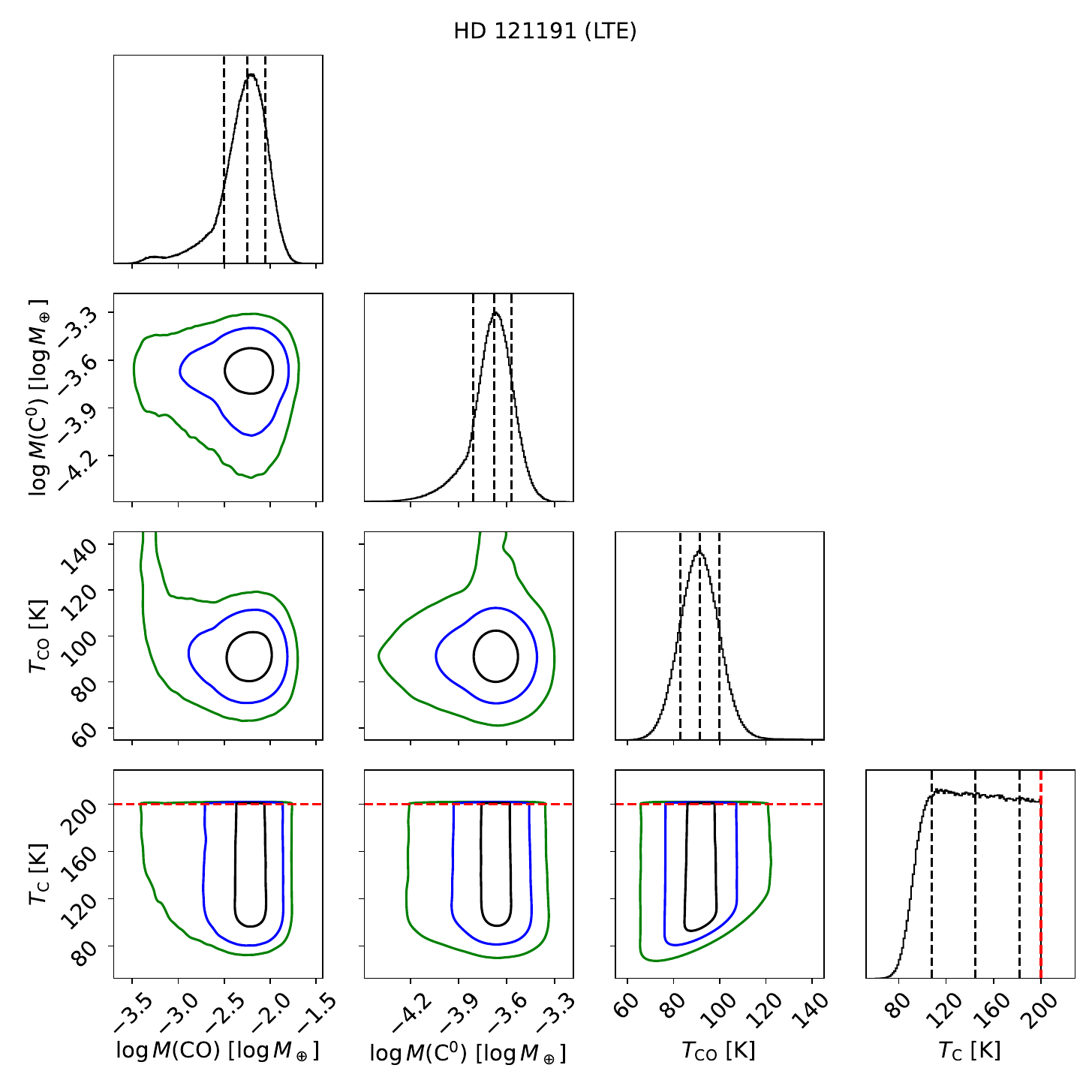}
	\caption{Same as Fig.\ \ref{fig:corner_LTE_49Ceti}, but for the disk around HD~121191.\label{fig:corner_LTE_HD121191}}
\end{figure*}

\begin{figure*}[h]
	\plotone{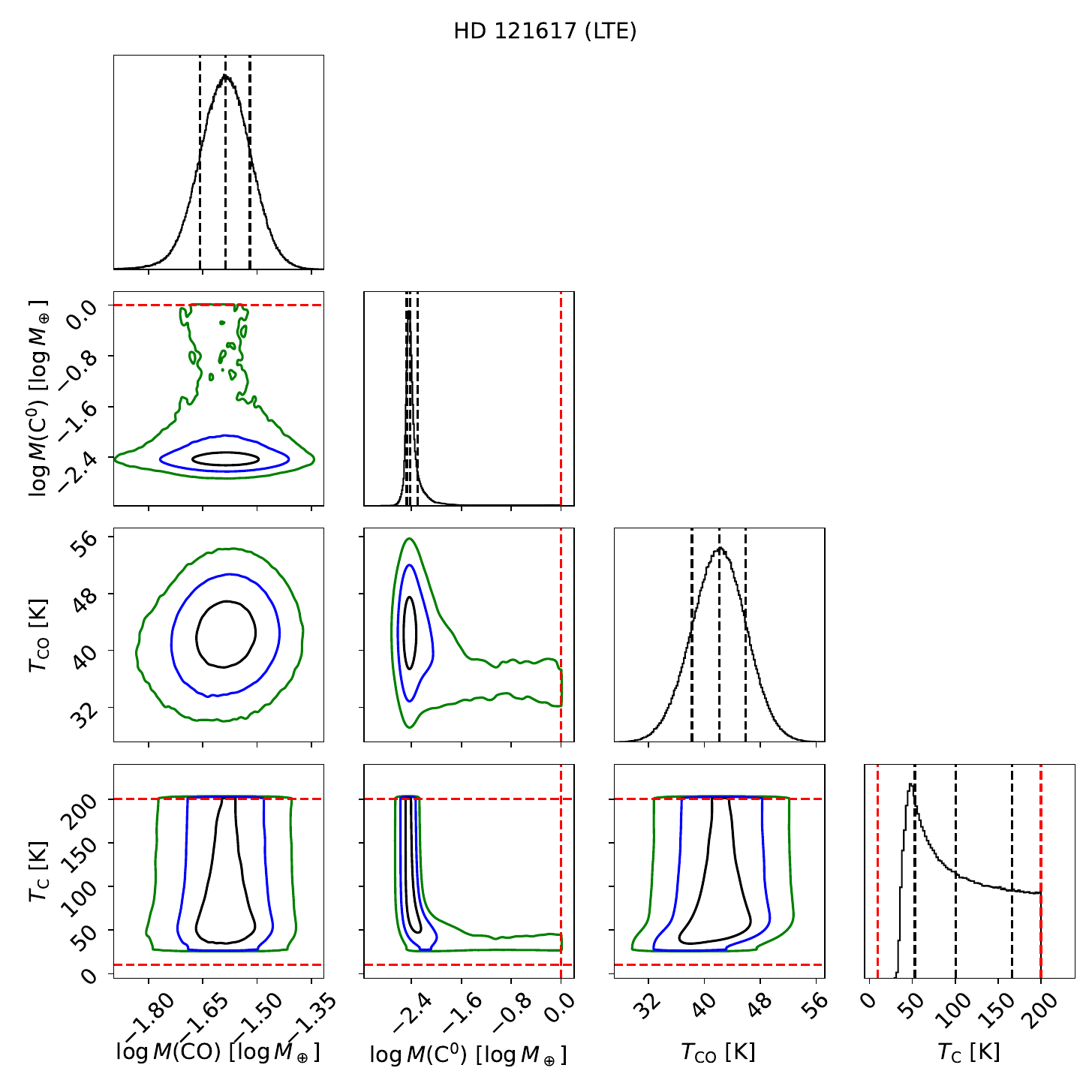}
	\caption{Same as Fig.\ \ref{fig:corner_LTE_49Ceti}, but for the disk around HD~121617.\label{fig:corner_LTE_HD121617}}
\end{figure*}

\begin{figure*}[h]
	\plotone{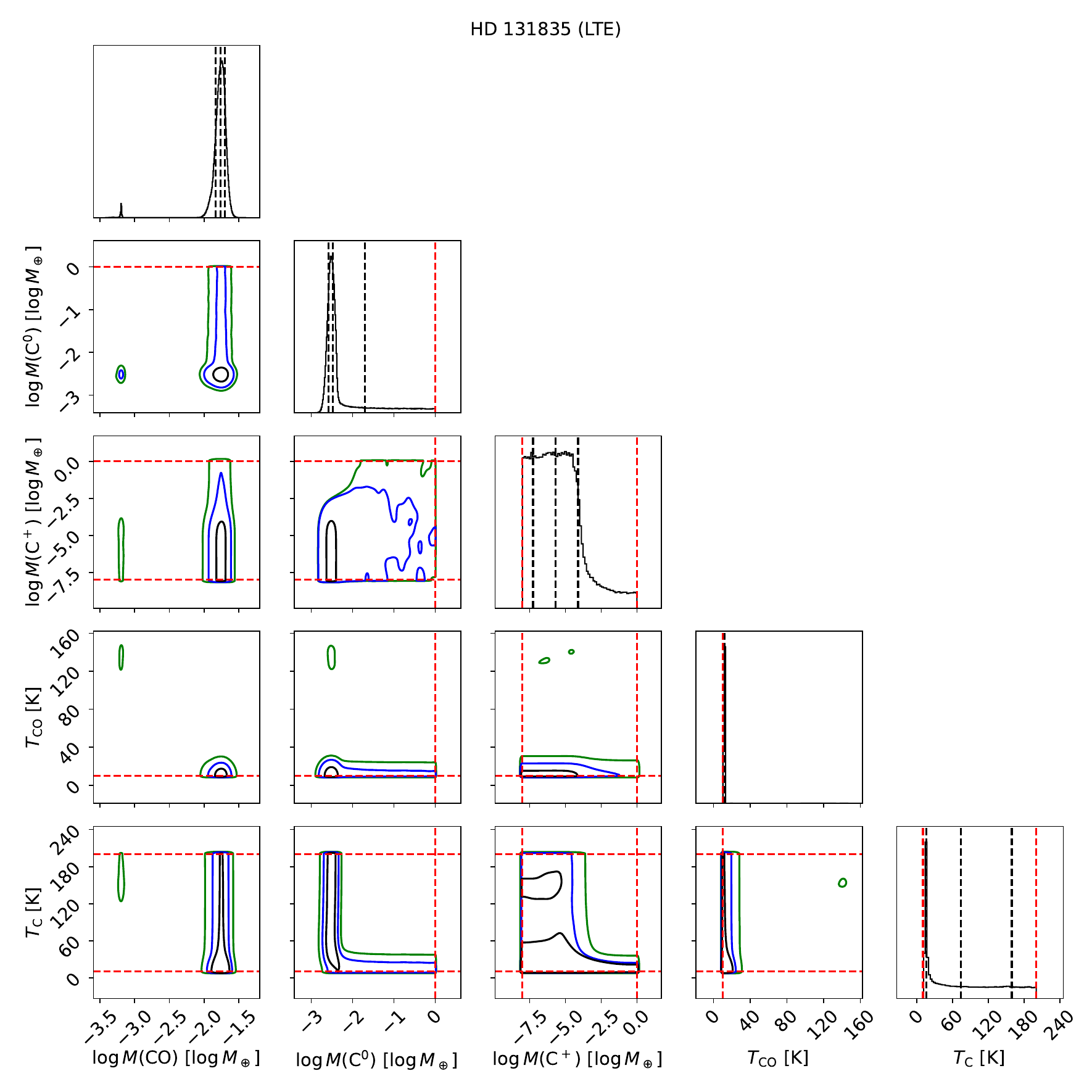}
	\caption{Same as Fig.\ \ref{fig:corner_LTE_49Ceti}, but for the disk around HD~131835.\label{fig:corner_LTE_HD131835}}
\end{figure*}

\begin{figure*}[h]
	\plotone{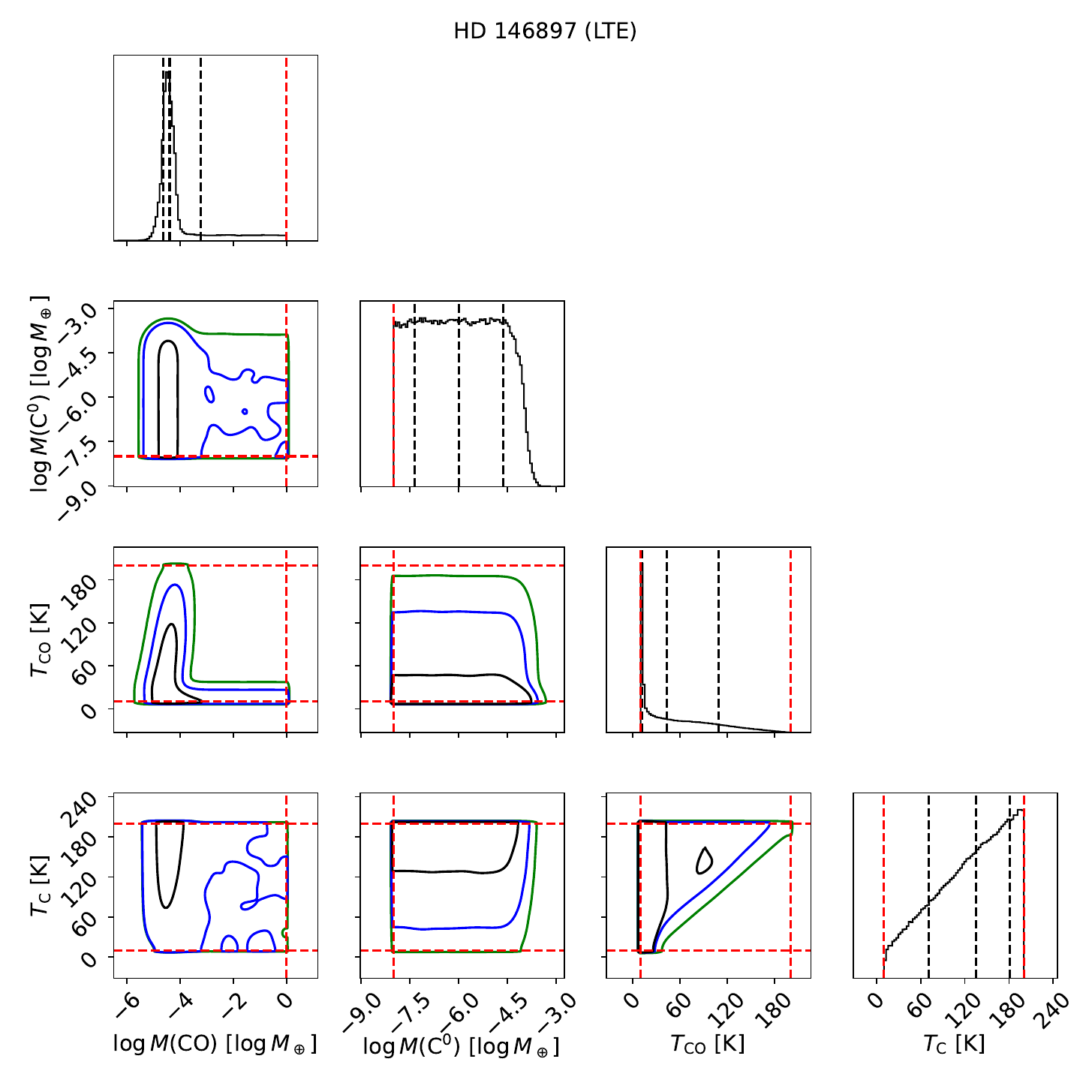}
	\caption{Same as Fig.\ \ref{fig:corner_LTE_49Ceti}, but for the disk around HD~146897.\label{fig:corner_LTE_HD146897}}
\end{figure*}

\begin{figure*}[h]
	\plotone{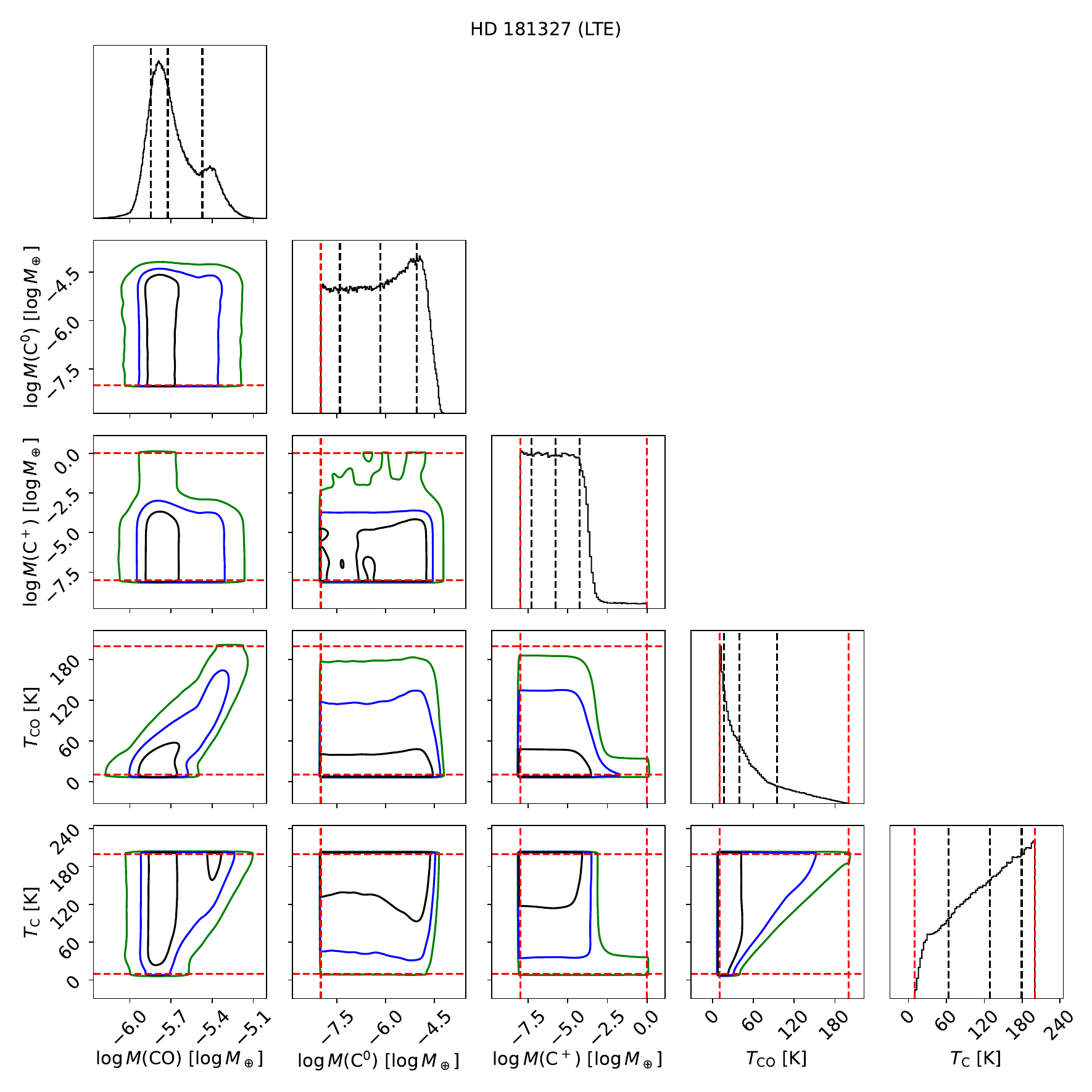}
	\caption{Same as Fig.\ \ref{fig:corner_LTE_49Ceti}, but for the disk around HD~181327.\label{fig:corner_LTE_HD181327}}
\end{figure*}

\begin{figure*}[h]
	\plotone{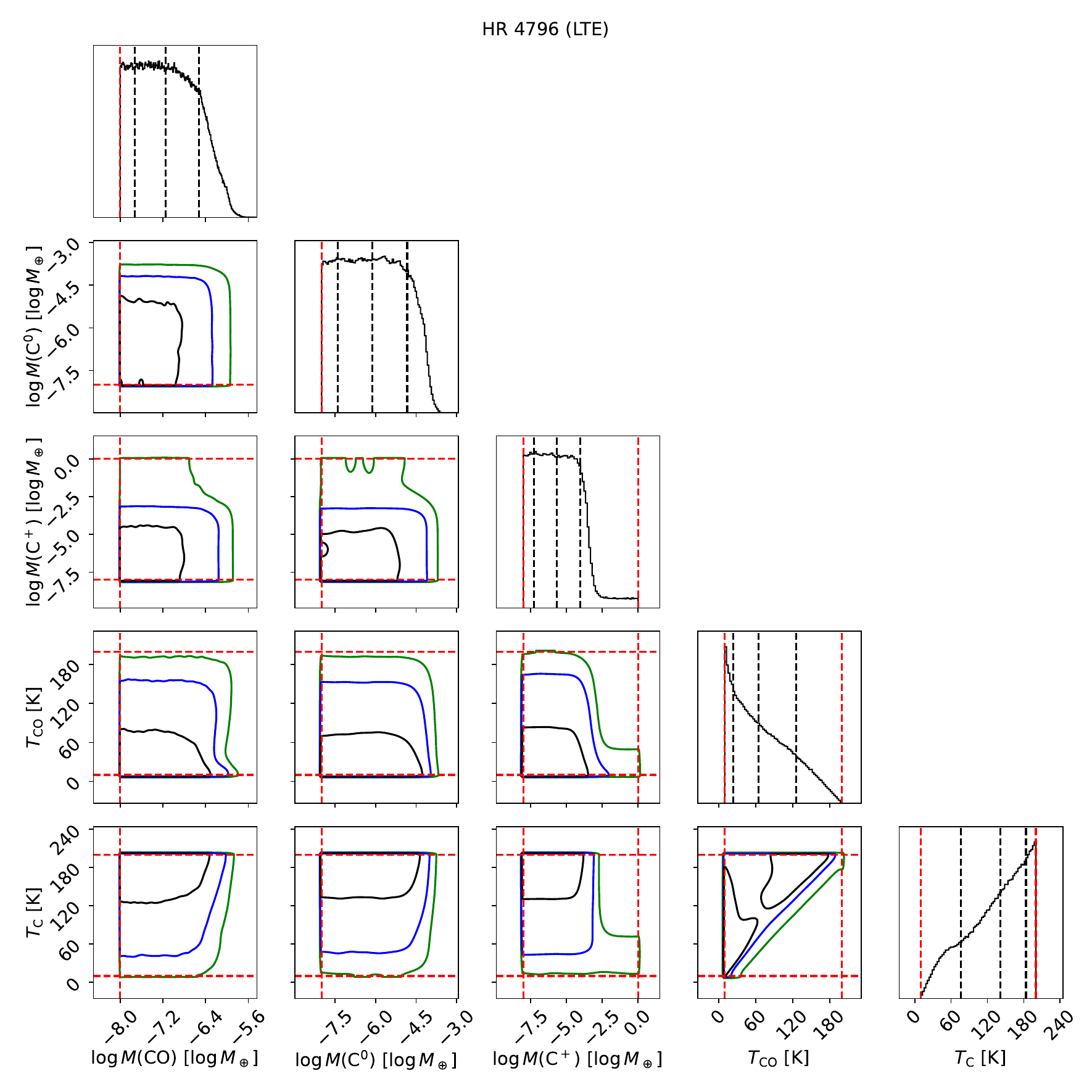}
	\caption{Same as Fig.\ \ref{fig:corner_LTE_49Ceti}, but for the disk around HR~4796.\label{fig:corner_LTE_HR4796}}
\end{figure*}

In Figures \ref{fig:corner_LTE_49Ceti} to \ref{fig:corner_LTE_HR4796} we show the corner plots of the LTE fits that were not shown in the main text.

\subsection{Non-LTE fits with \texorpdfstring{H$_2$}{H2} as collider}\label{appendix:corner_plots_nonLTE_H2}
\begin{figure*}[h]
	\plotone{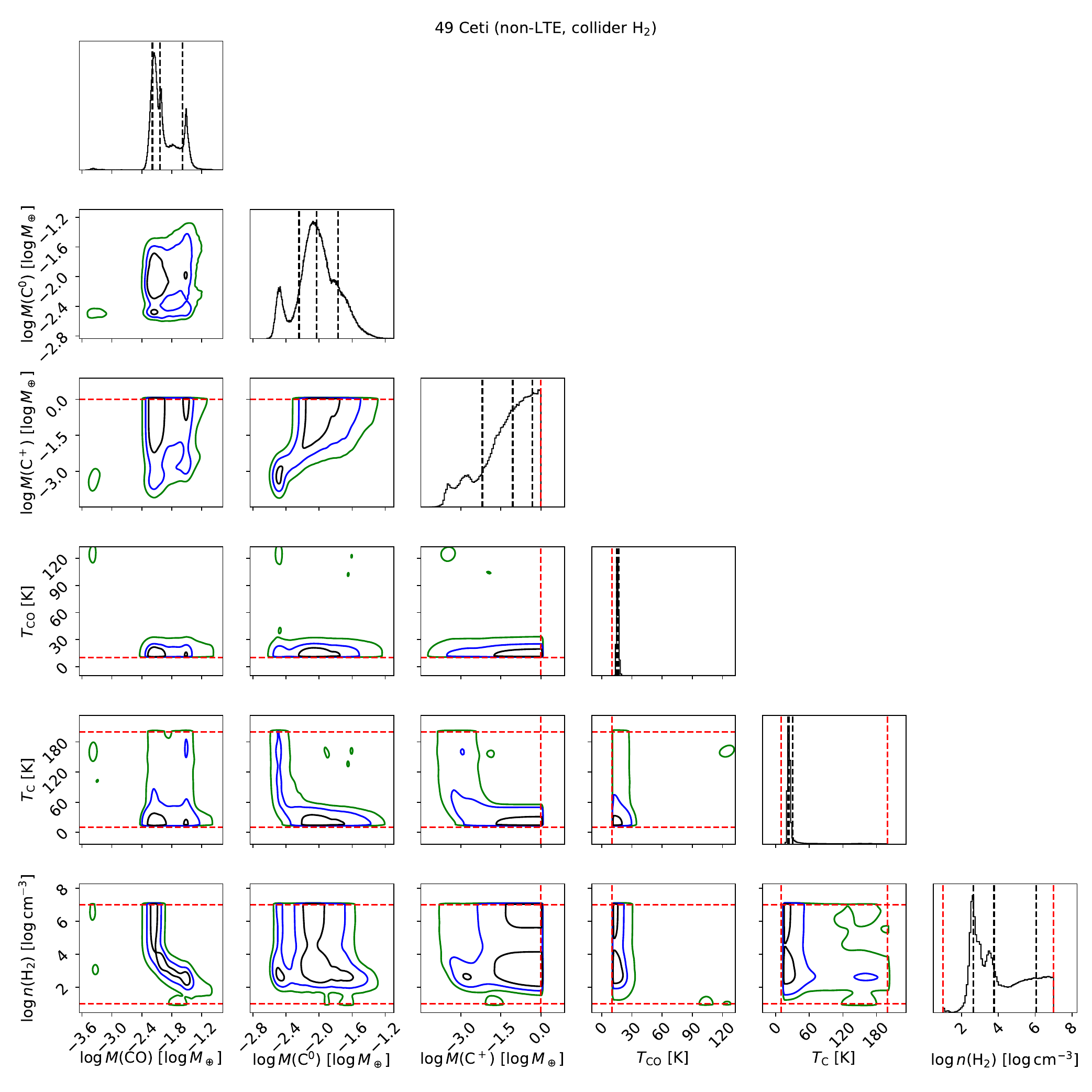}
	\caption{Corner plot showing posterior probability distributions for the non-LTE case with H$_2$ collisions for the disk around 49~Ceti. In the 1D histograms, the black vertical dashed lines indicate the 16th, 50th and 84th percentile. In the 2D histograms, the black, blue and green contours mark the 50th, 90th and 99th percentile. Red dashed lines mark the upper or lower bound of the prior distribution.}\label{fig:corner_nonLTE_H2_49Ceti}
\end{figure*}

\begin{figure*}[h]
	\plotone{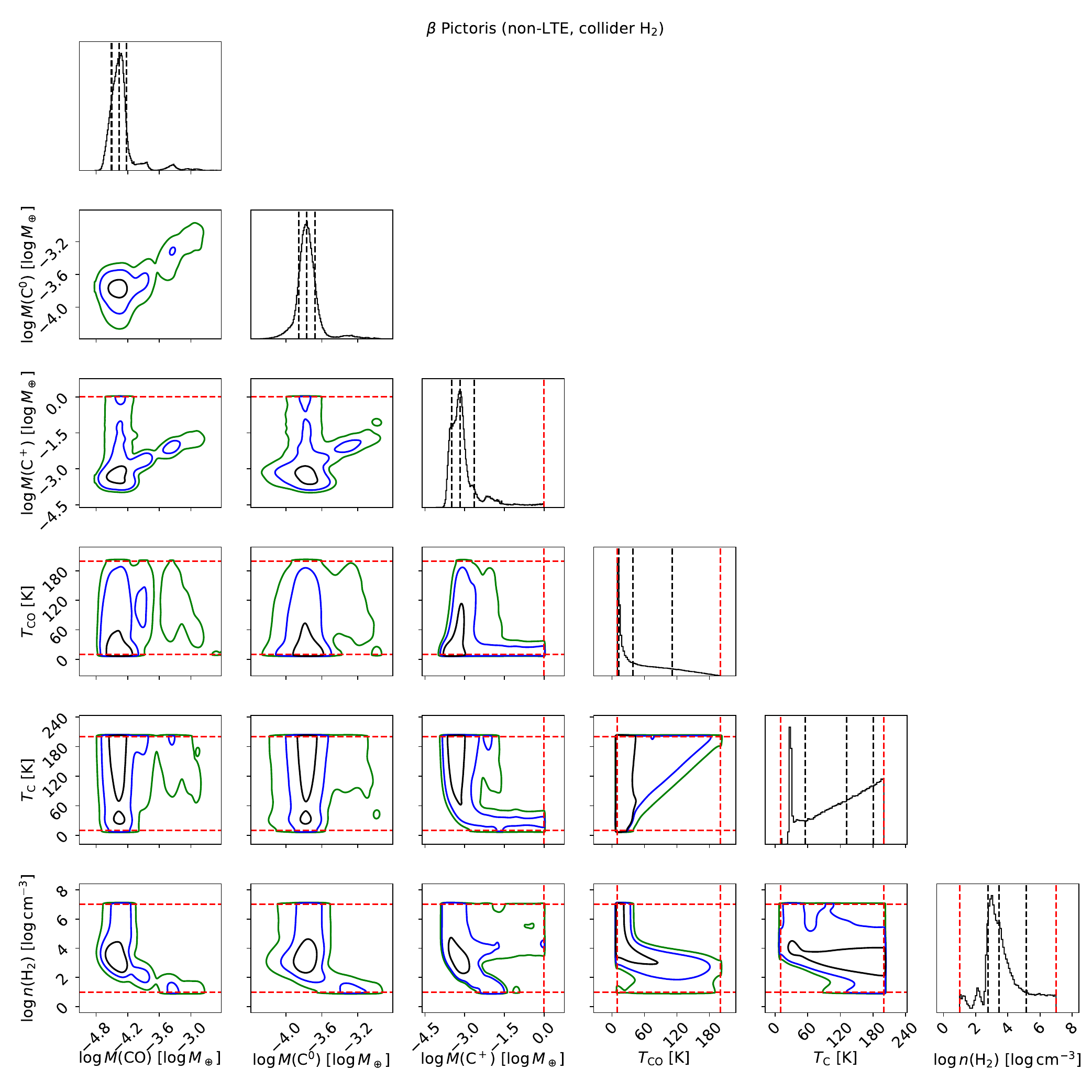}
	\caption{Same as Fig.\ \ref{fig:corner_nonLTE_H2_49Ceti}, but for the disk around $\beta$~Pic.\label{fig:corner_nonLTE_H2_betaPic}}
\end{figure*}

\begin{figure*}[h]
	\plotone{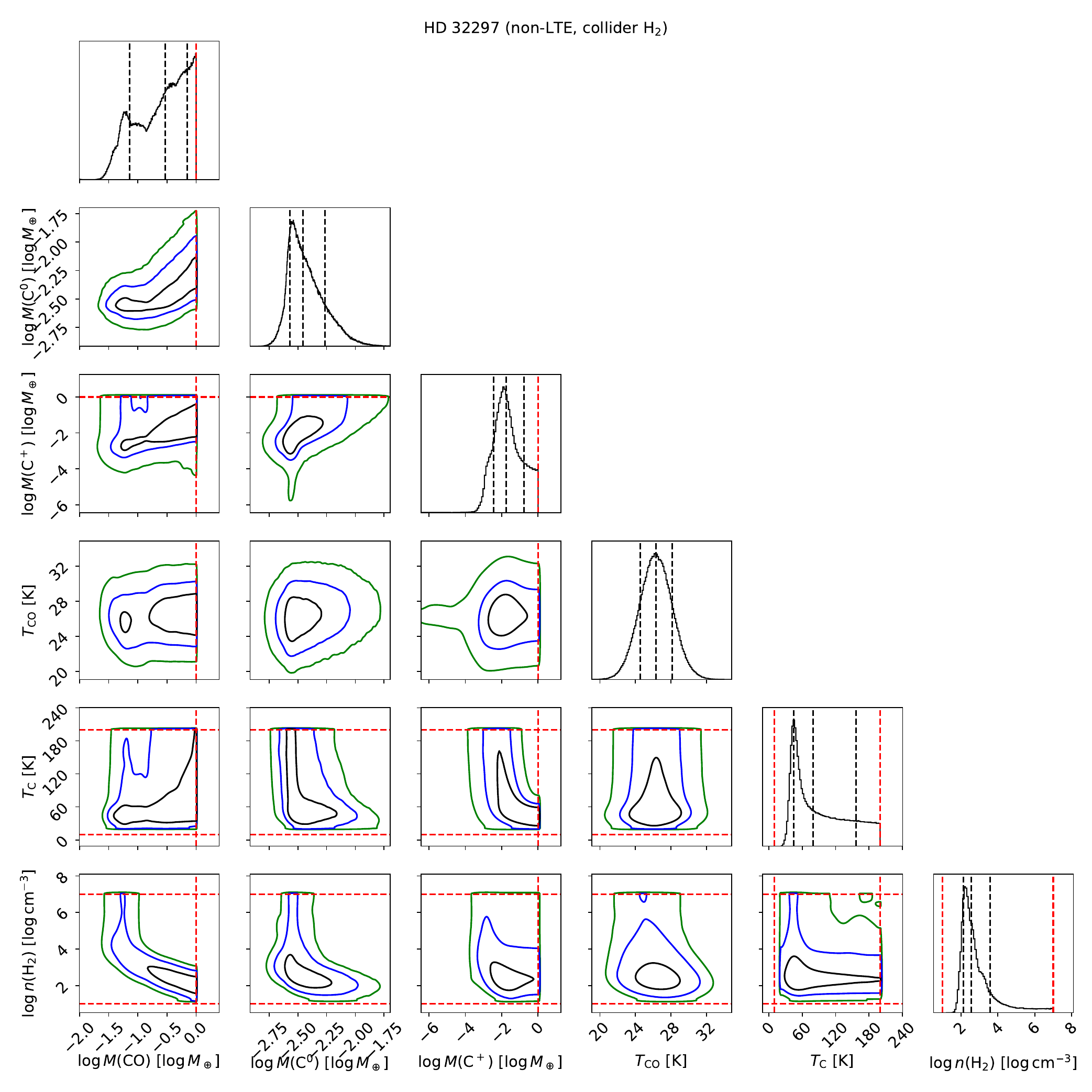}
	\caption{Same as Fig.\ \ref{fig:corner_nonLTE_H2_49Ceti}, but for the disk around HD~32297.\label{fig:corner_nonLTE_H2_HD32297}}
\end{figure*}

\begin{figure*}[h]
	\plotone{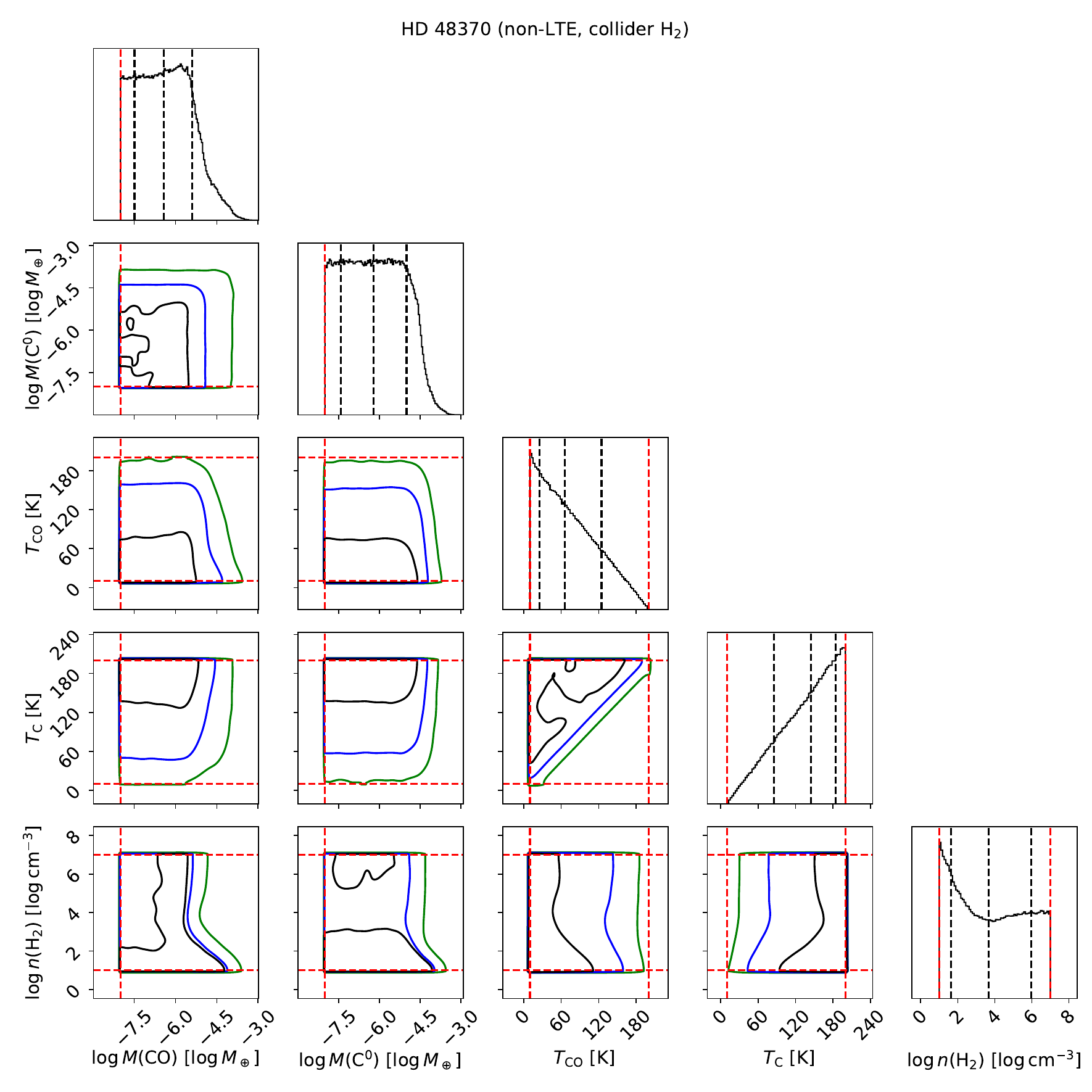}
	\caption{Same as Fig.\ \ref{fig:corner_nonLTE_H2_49Ceti}, but for the disk around HD~48370.\label{fig:corner_nonLTE_H2_HD48370}}
\end{figure*}

\begin{figure*}[h]
	\plotone{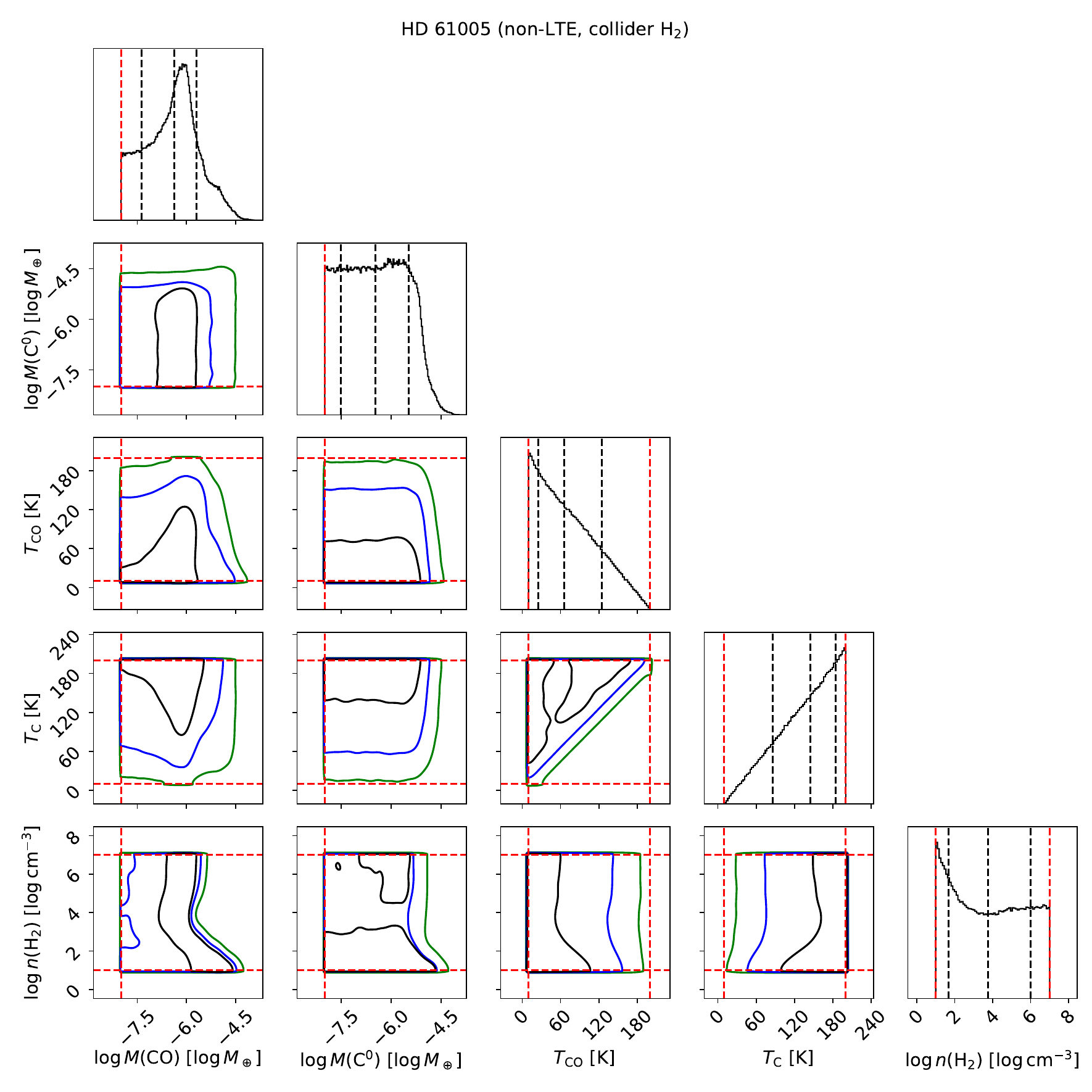}
	\caption{Same as Fig.\ \ref{fig:corner_nonLTE_H2_49Ceti}, but for the disk around HD~61005.\label{fig:corner_nonLTE_H2_HD61005}}
\end{figure*}

\begin{figure*}[h]
	\plotone{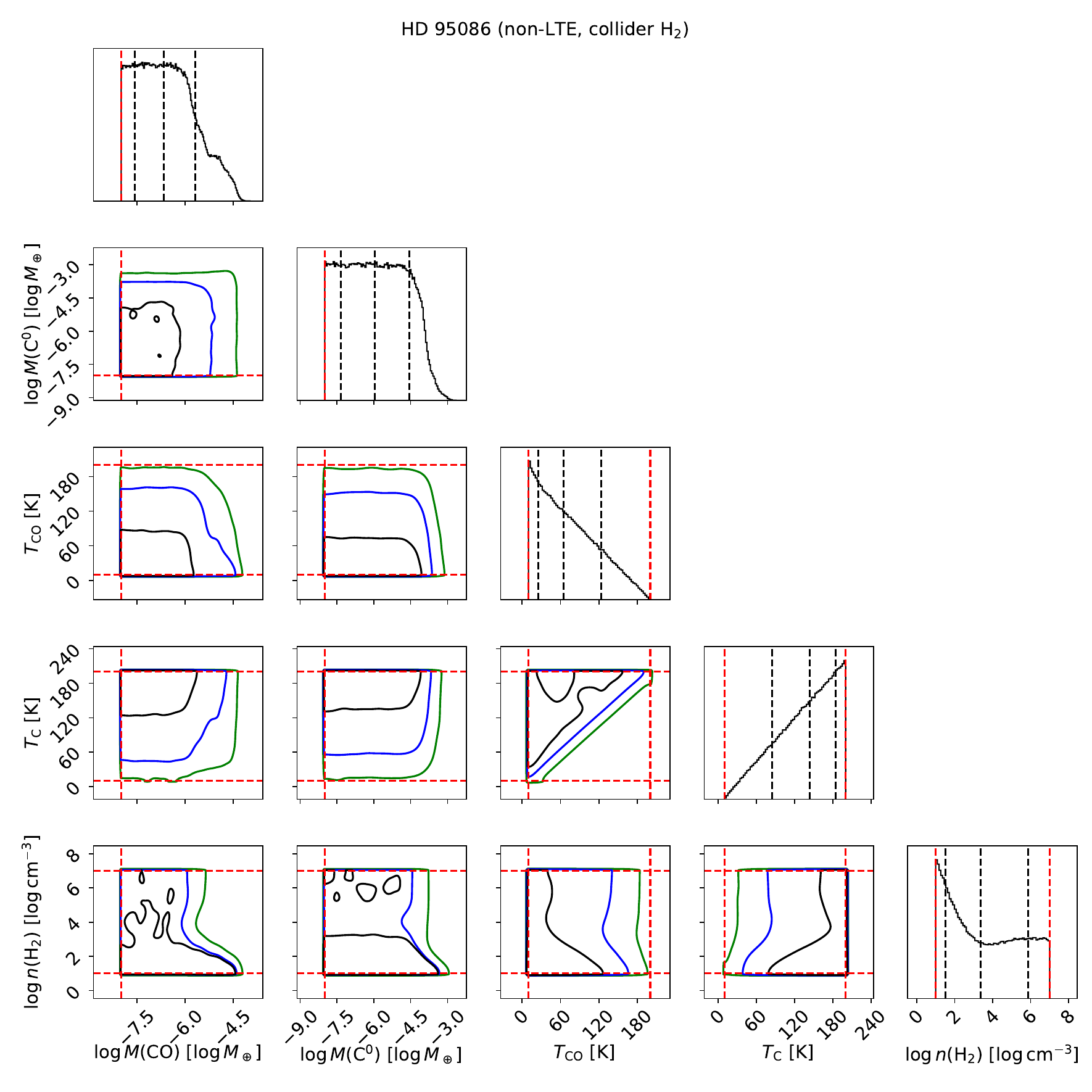}
	\caption{Same as Fig.\ \ref{fig:corner_nonLTE_H2_49Ceti}, but for the disk around HD~95086.\label{fig:corner_nonLTE_H2_HD95086}}
\end{figure*}

\begin{figure*}[h]
	\plotone{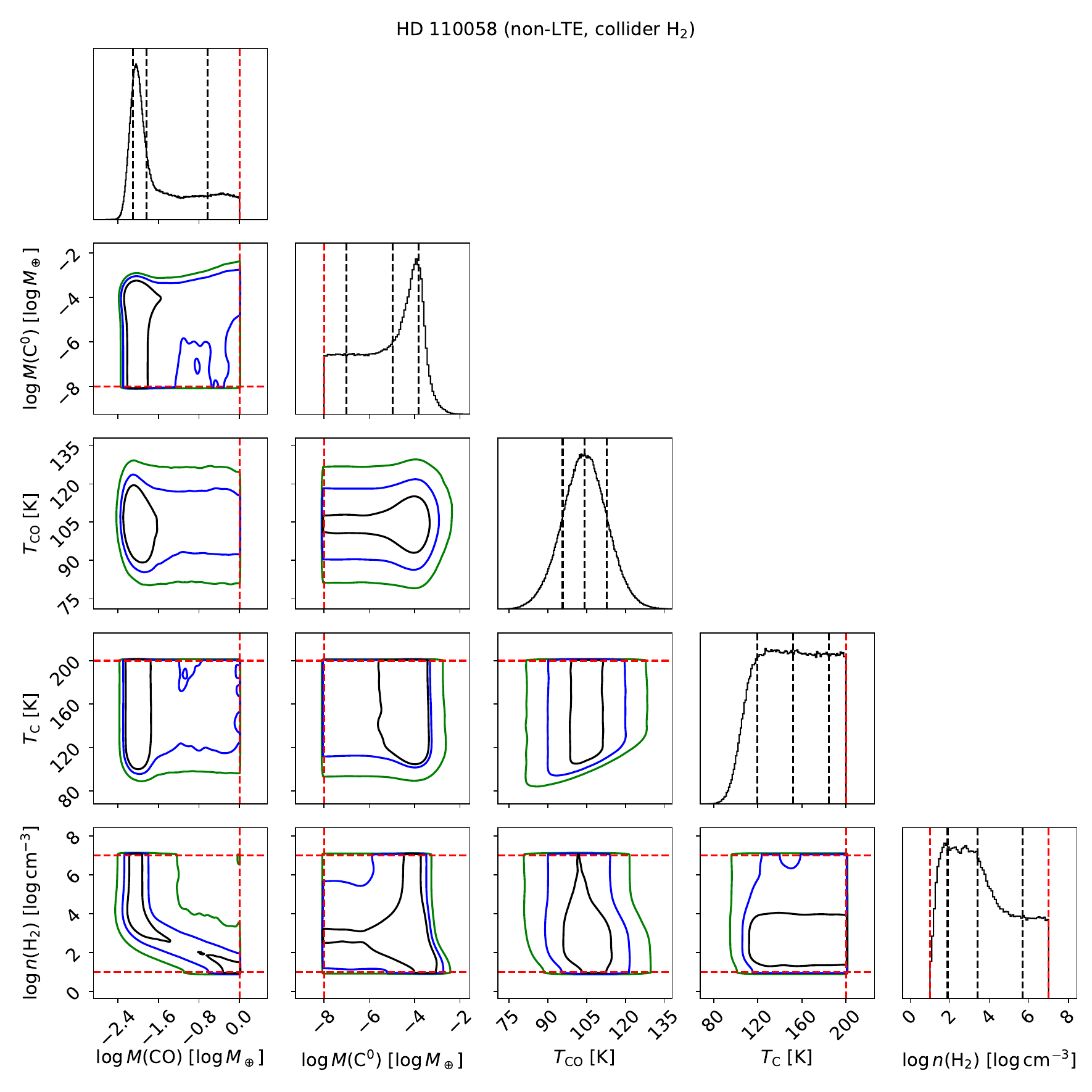}
	\caption{Same as Fig.\ \ref{fig:corner_nonLTE_H2_49Ceti}, but for the disk around HD~110058.\label{fig:corner_nonLTE_H2_HD110058}}
\end{figure*}

\begin{figure*}[h]
	\plotone{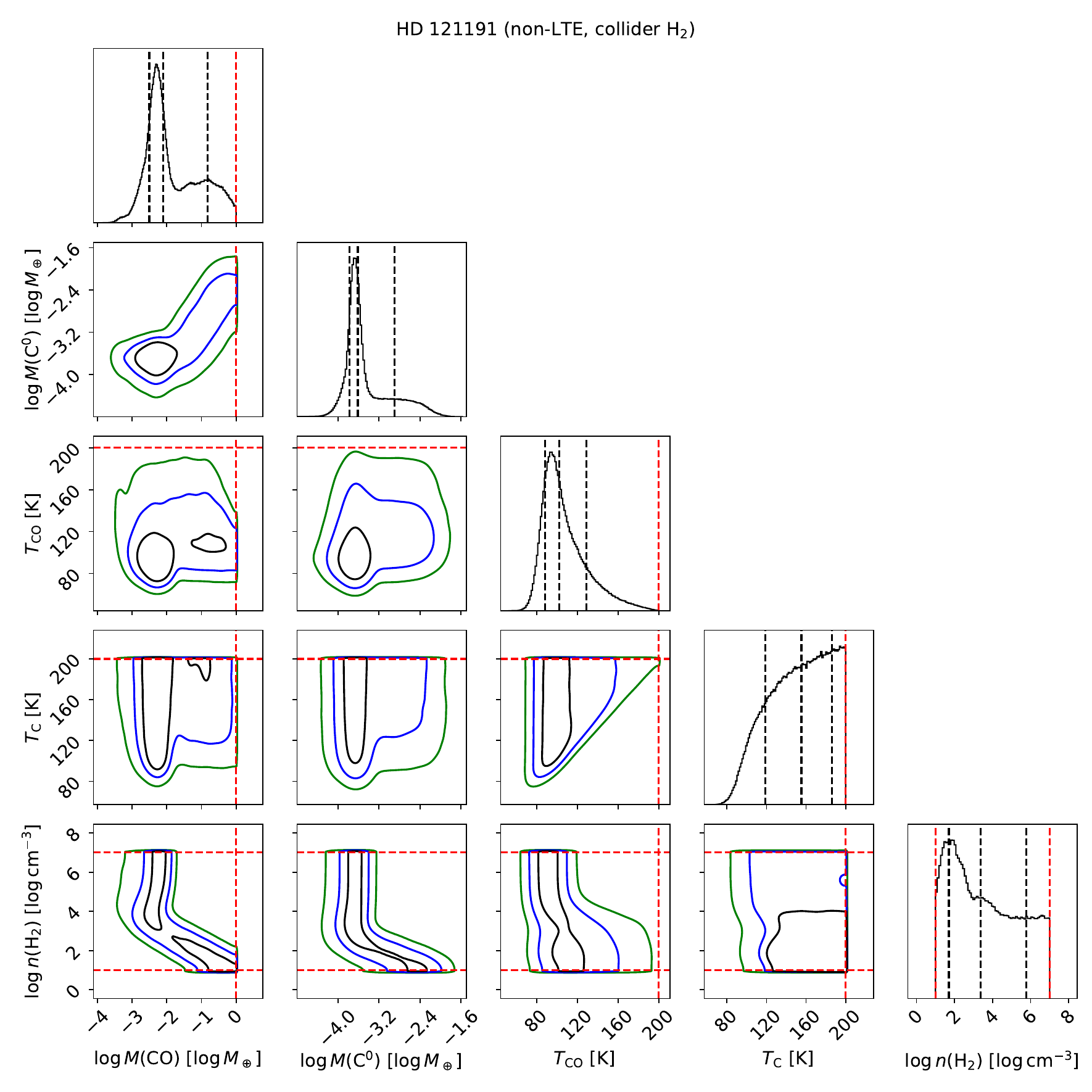}
	\caption{Same as Fig.\ \ref{fig:corner_nonLTE_H2_49Ceti}, but for the disk around HD~121191.\label{fig:corner_nonLTE_H2_HD121191}}
\end{figure*}

\begin{figure*}[h]
	\plotone{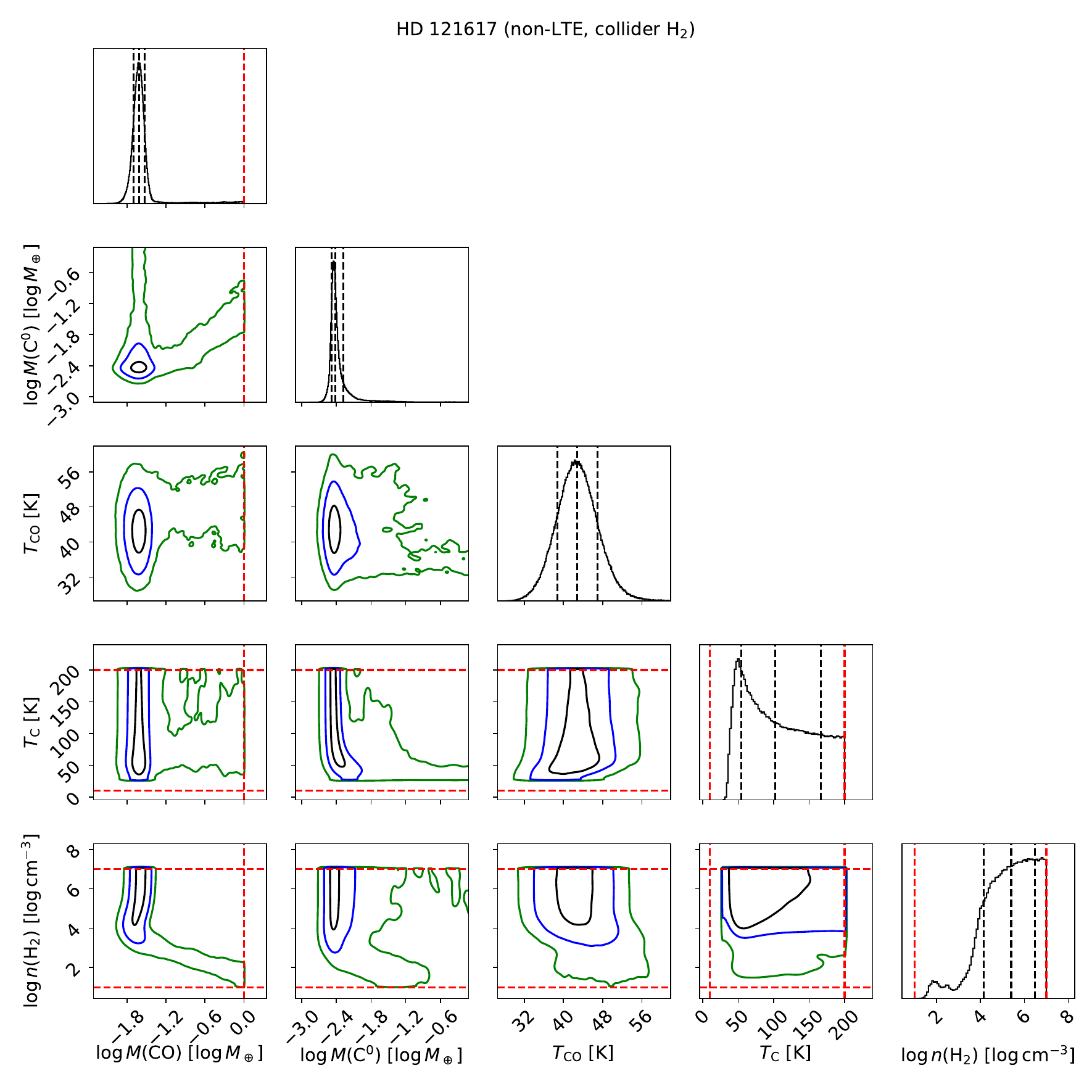}
	\caption{Same as Fig.\ \ref{fig:corner_nonLTE_H2_49Ceti}, but for the disk around HD~121617.\label{fig:corner_nonLTE_H2_HD121617}}
\end{figure*}

\begin{figure*}[h]
	\plotone{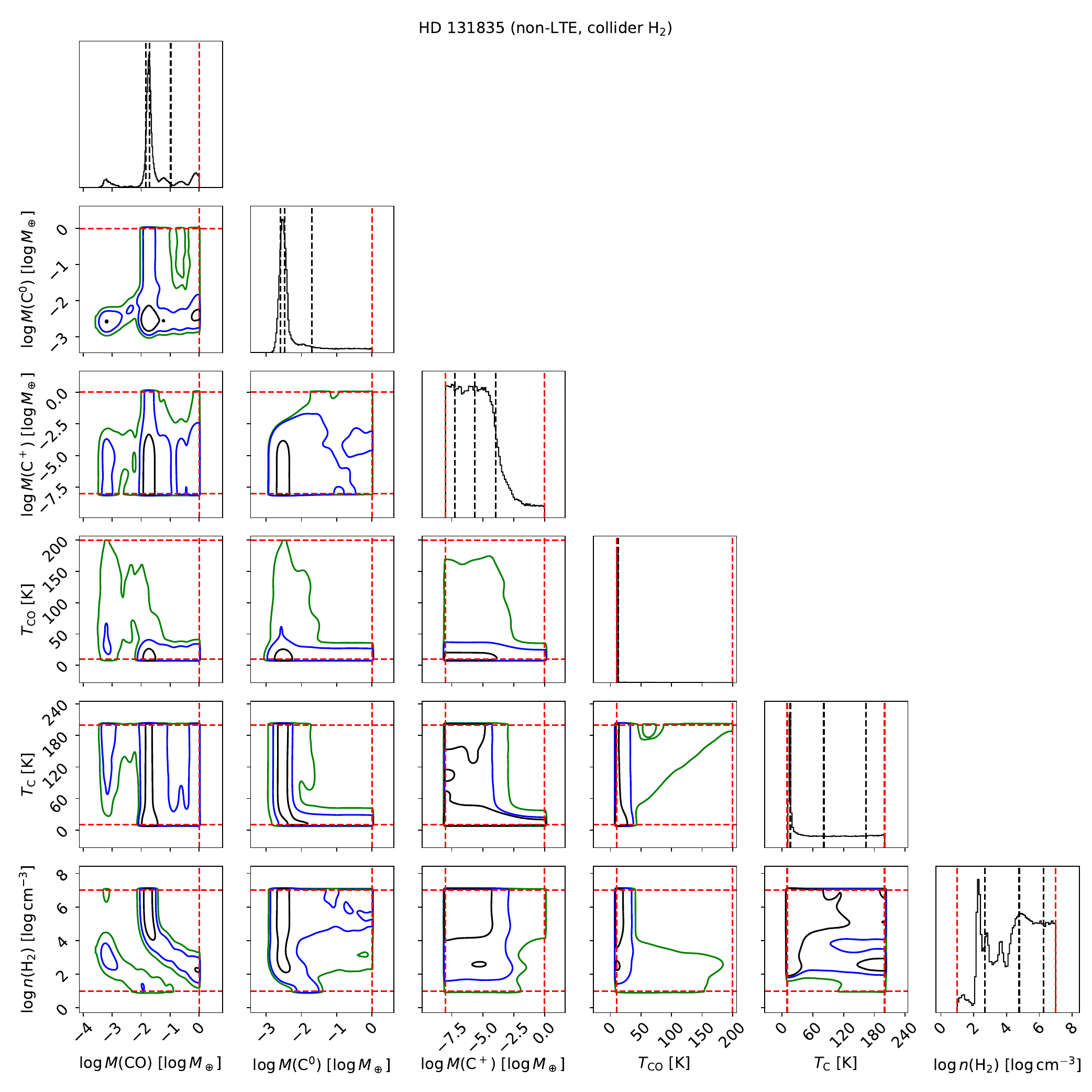}
	\caption{Same as Fig.\ \ref{fig:corner_nonLTE_H2_49Ceti}, but for the disk around HD~131835.\label{fig:corner_nonLTE_H2_HD131835}}
\end{figure*}

\begin{figure*}[h]
	\plotone{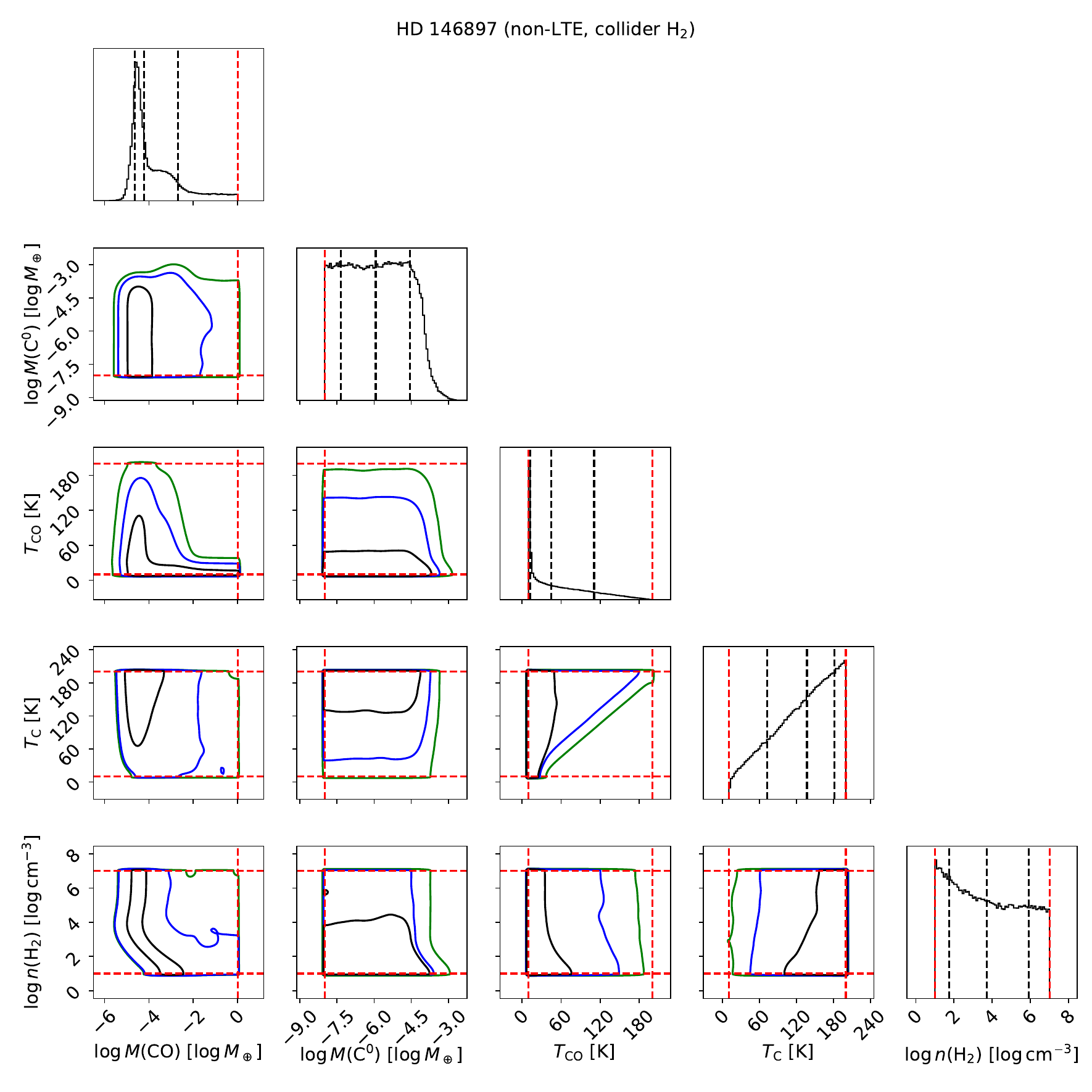}
	\caption{Same as Fig.\ \ref{fig:corner_nonLTE_H2_49Ceti}, but for the disk around HD~146897.\label{fig:corner_nonLTE_H2_HD146897}}
\end{figure*}

\begin{figure*}[h]
	\plotone{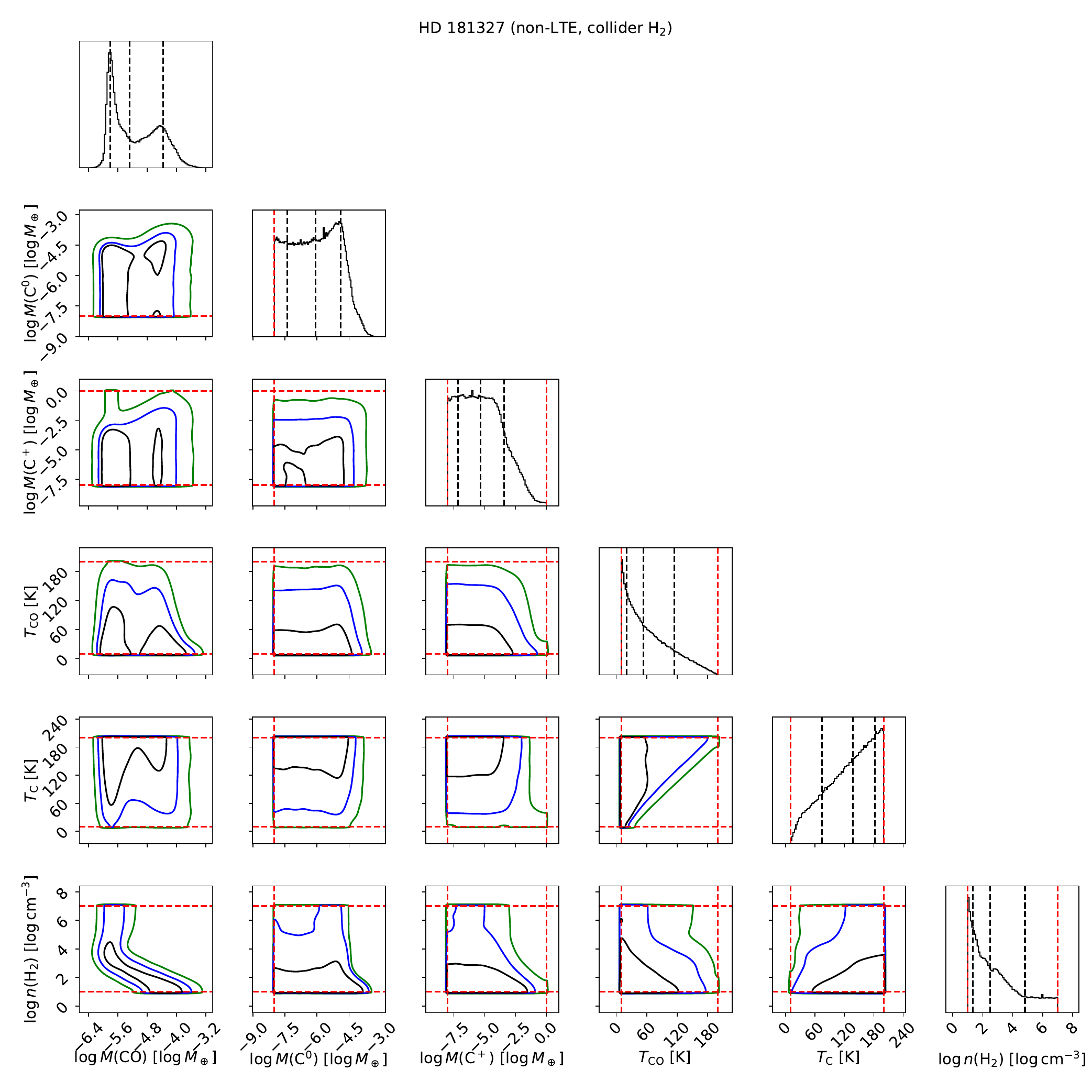}
	\caption{Same as Fig.\ \ref{fig:corner_nonLTE_H2_49Ceti}, but for the disk around HD~181327.\label{fig:corner_nonLTE_H2_HD181327}}
\end{figure*}

\begin{figure*}[h]
	\plotone{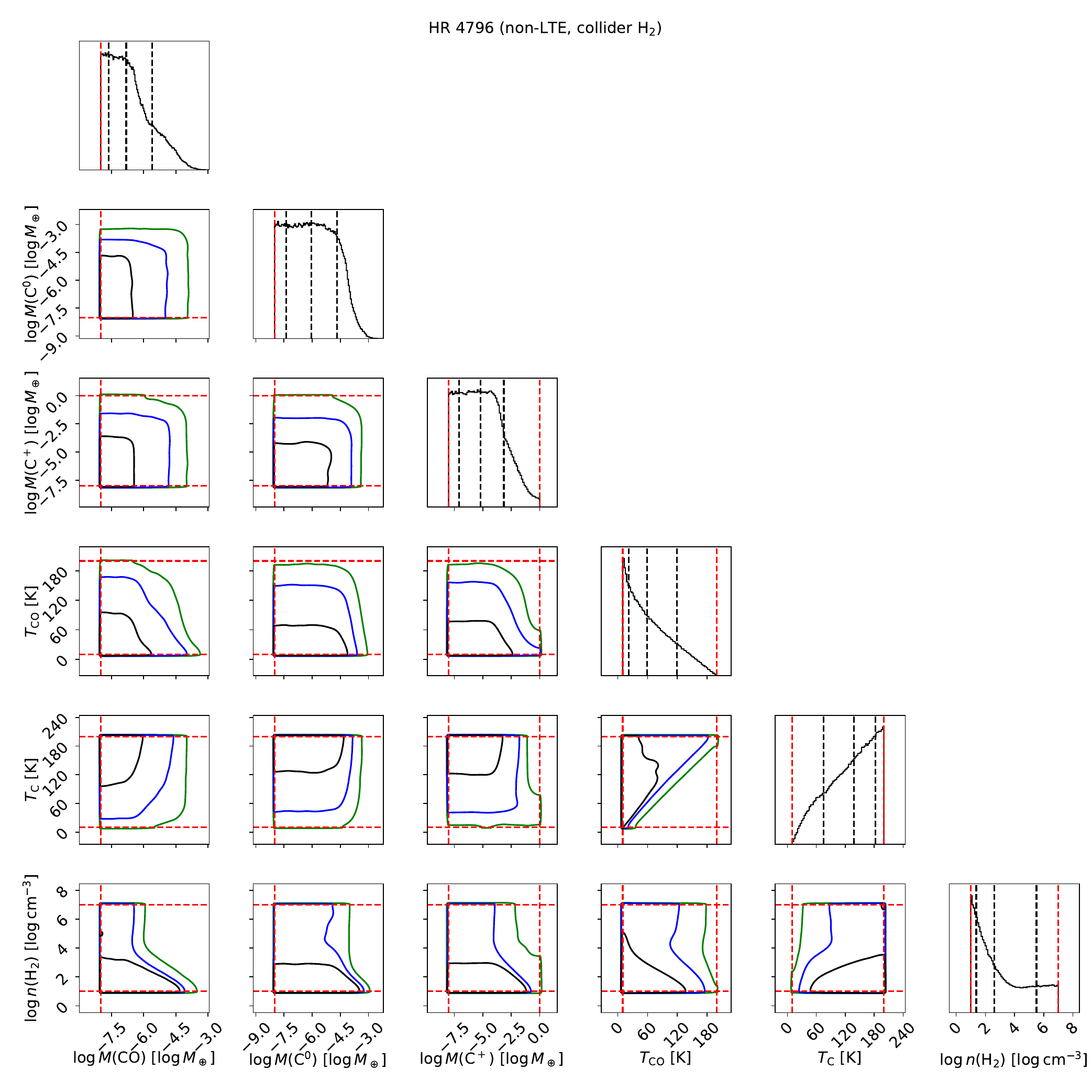}
	\caption{Same as Fig.\ \ref{fig:corner_nonLTE_H2_49Ceti}, but for the disk around HR~4796.\label{fig:corner_nonLTE_H2_HR4796}}
\end{figure*}

In Figures \ref{fig:corner_nonLTE_H2_49Ceti} to \ref{fig:corner_nonLTE_H2_HR4796} we show the corner plots of the non-LTE fits with H$_2$ colliders (for HD~21997, the corner plot was already shown in the main text in Fig.\ \ref{fig:corner_HD21997}).

\subsection{Non-LTE fits with \texorpdfstring{e$^-$}{e-} as collider}\label{appendix:corner_plots_nonLTE_e}
\begin{figure*}[h]
	\plotone{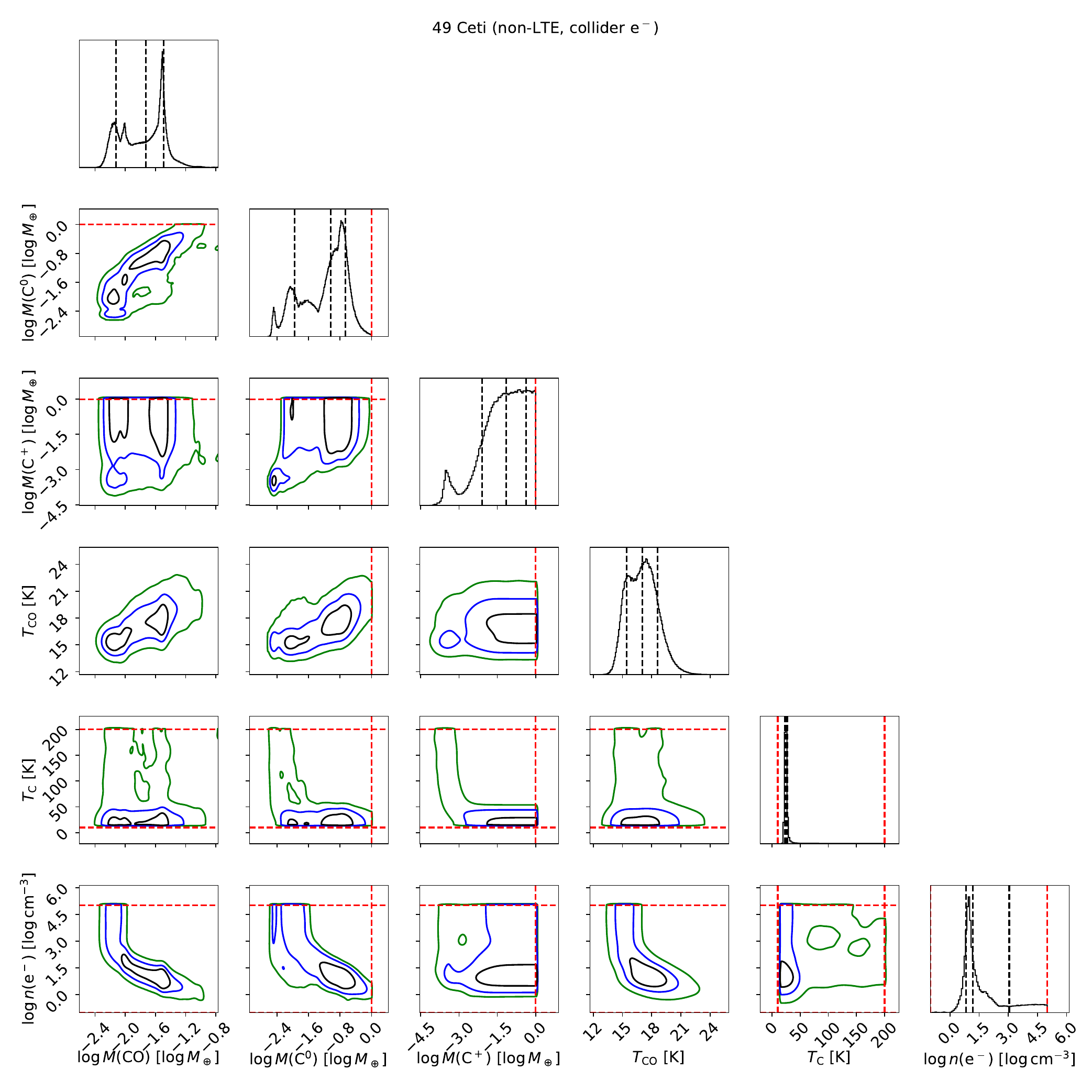}
	\caption{Corner plot showing posterior probability distributions for the non-LTE case with electron collisions for the disk around 49~Ceti. In the 1D histograms, the black vertical dashed lines indicate the 16th, 50th and 84th percentile. In the 2D histograms, the black, blue and green contours mark the 50th, 90th and 99th percentile. Red dashed lines mark the upper or lower bound of the prior distribution.}\label{fig:corner_nonLTE_e_49Ceti}
\end{figure*}

\begin{figure*}[h]
	\plotone{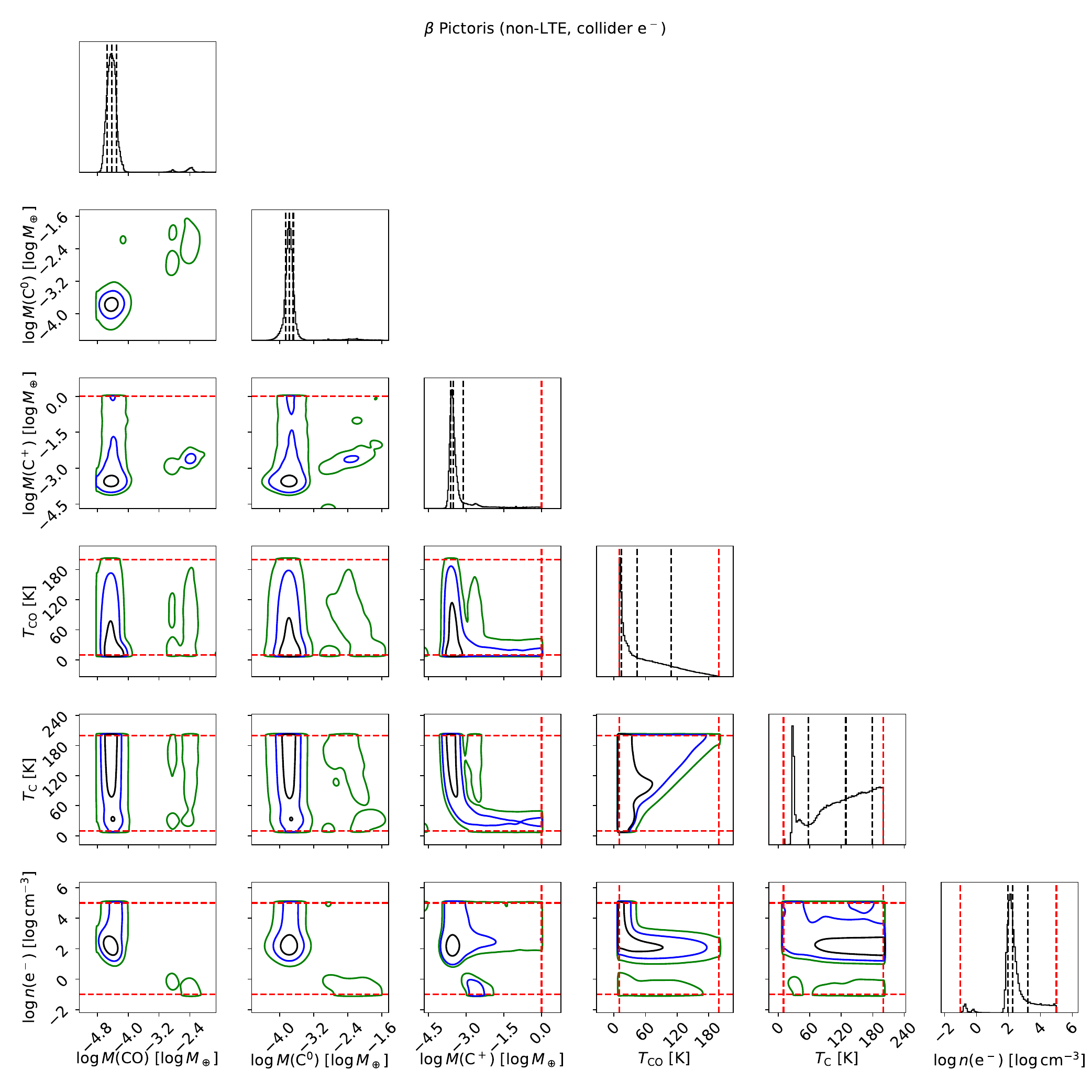}
	\caption{Same as Fig.\ \ref{fig:corner_nonLTE_e_49Ceti}, but for the disk around $\beta$~Pic.\label{fig:corner_nonLTE_e_betaPic}}
\end{figure*}

\begin{figure*}[h]
	\plotone{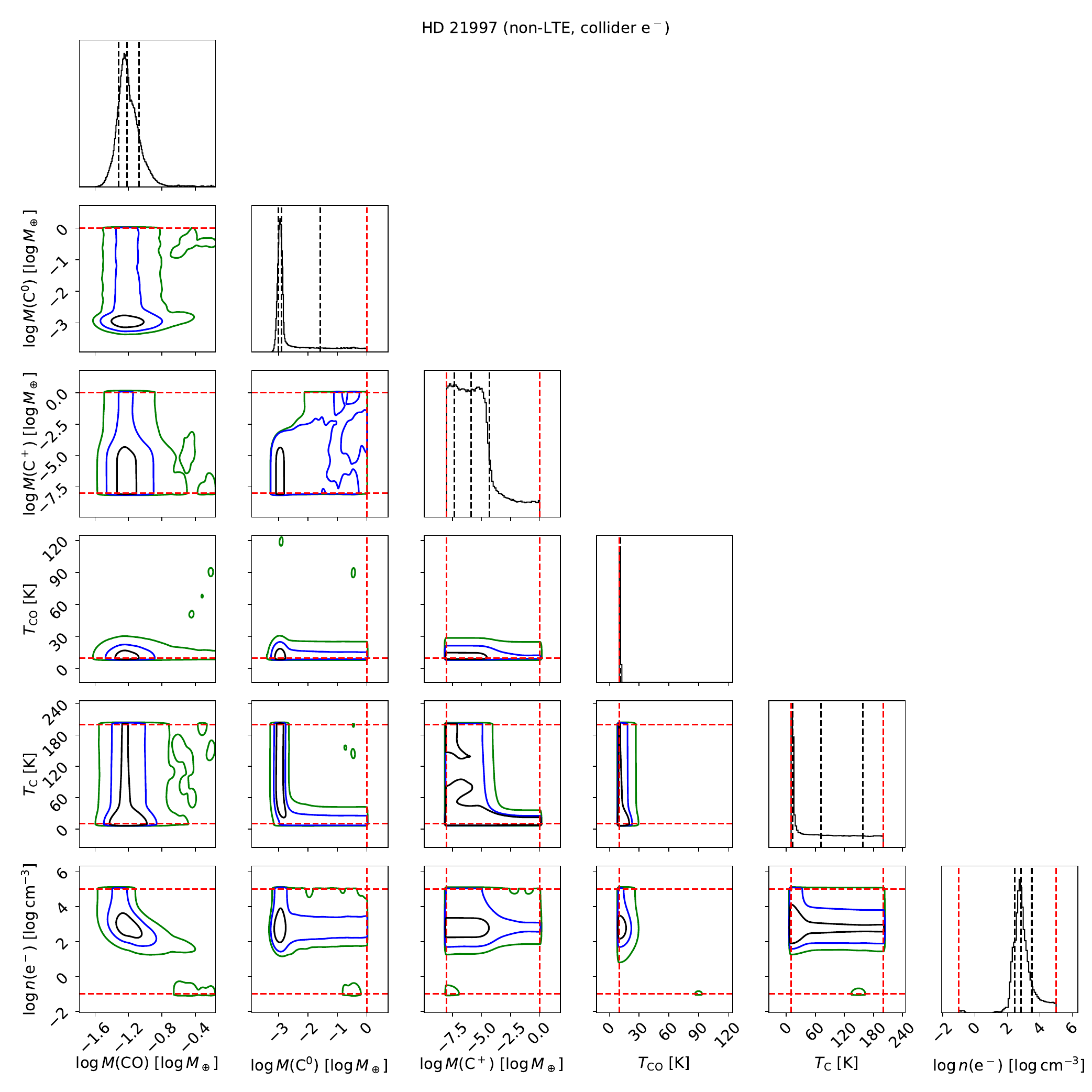}
	\caption{Same as Fig.\ \ref{fig:corner_nonLTE_e_49Ceti}, but for the disk around HD~21997.\label{fig:corner_nonLTE_e_HD21997}}
\end{figure*}

\begin{figure*}[h]
	\plotone{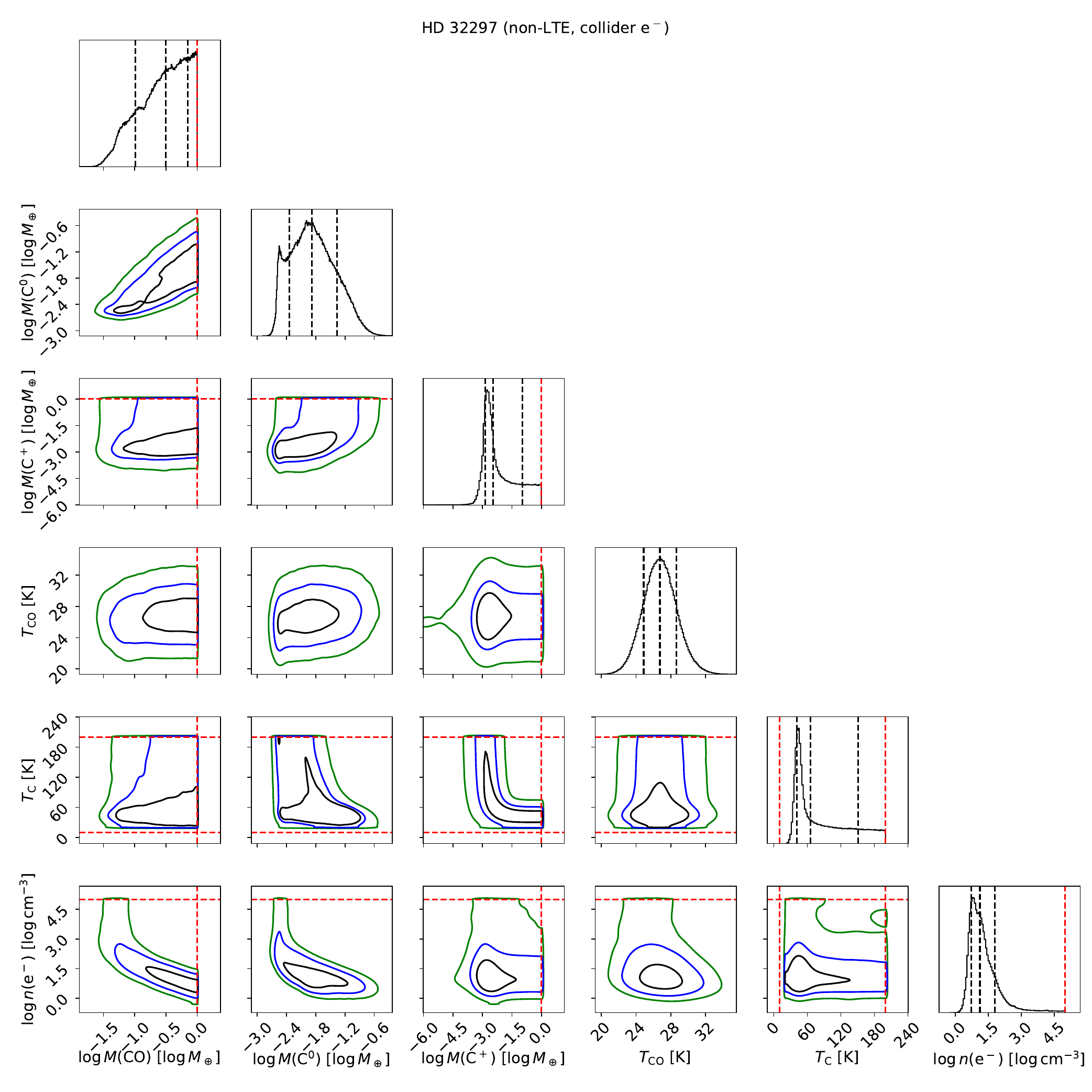}
	\caption{Same as Fig.\ \ref{fig:corner_nonLTE_e_49Ceti}, but for the disk around HD~32297.\label{fig:corner_nonLTE_e_HD32297}}
\end{figure*}

\begin{figure*}[h]
	\plotone{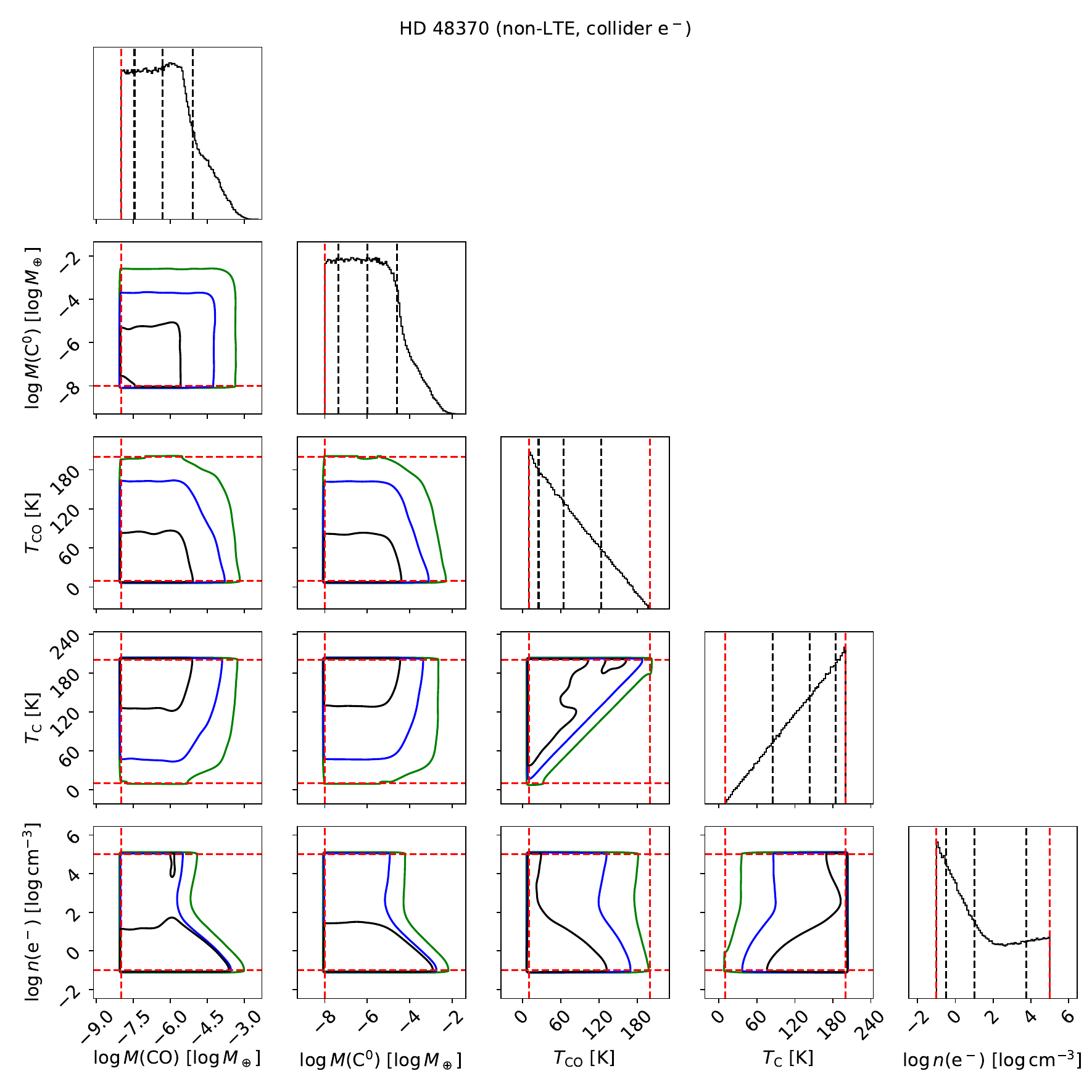}
	\caption{Same as Fig.\ \ref{fig:corner_nonLTE_e_49Ceti}, but for the disk around HD~48370.\label{fig:corner_nonLTE_e_HD48370}}
\end{figure*}

\begin{figure*}[h]
	\plotone{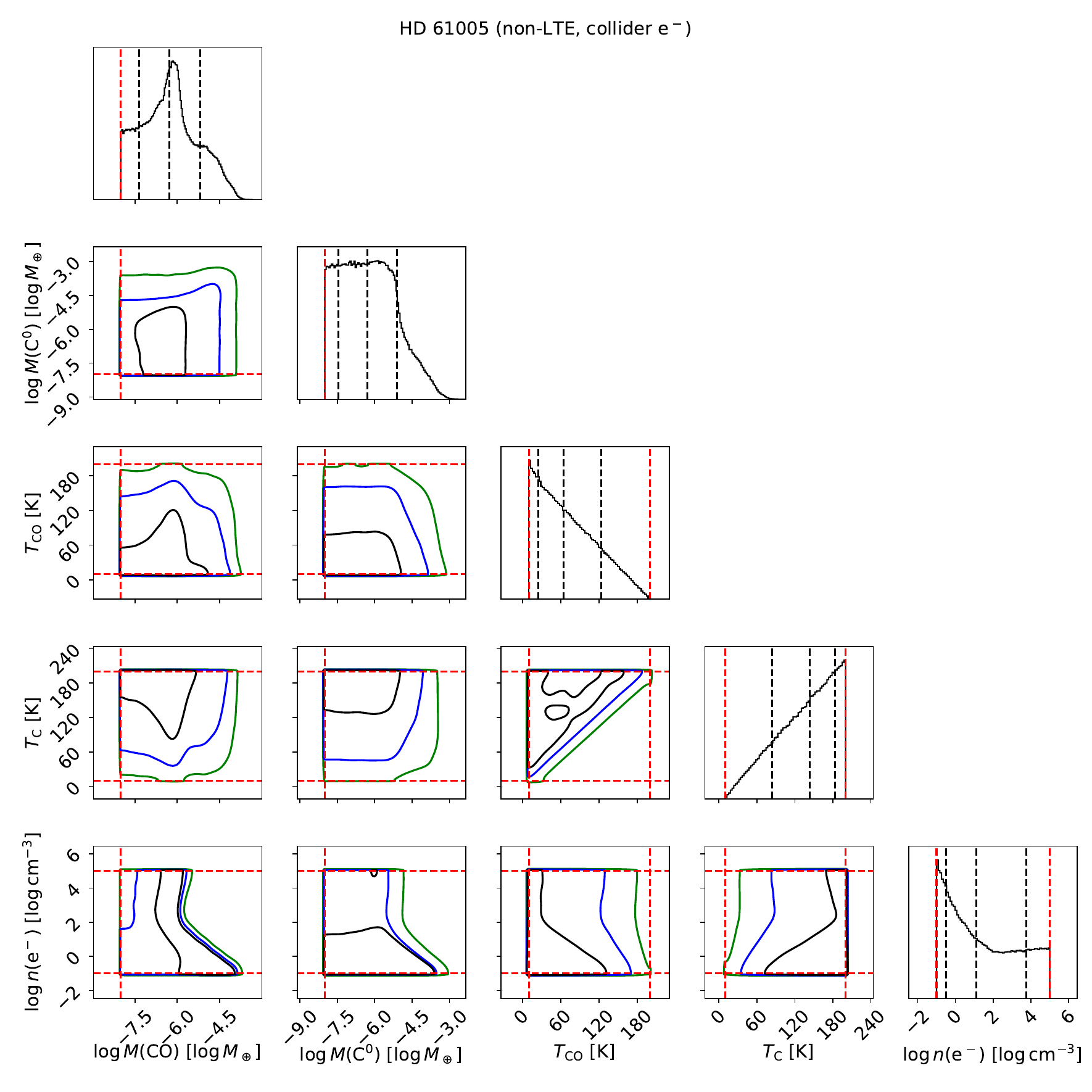}
	\caption{Same as Fig.\ \ref{fig:corner_nonLTE_e_49Ceti}, but for the disk around HD~61005.\label{fig:corner_nonLTE_e_HD61005}}
\end{figure*}

\begin{figure*}[h]
	\plotone{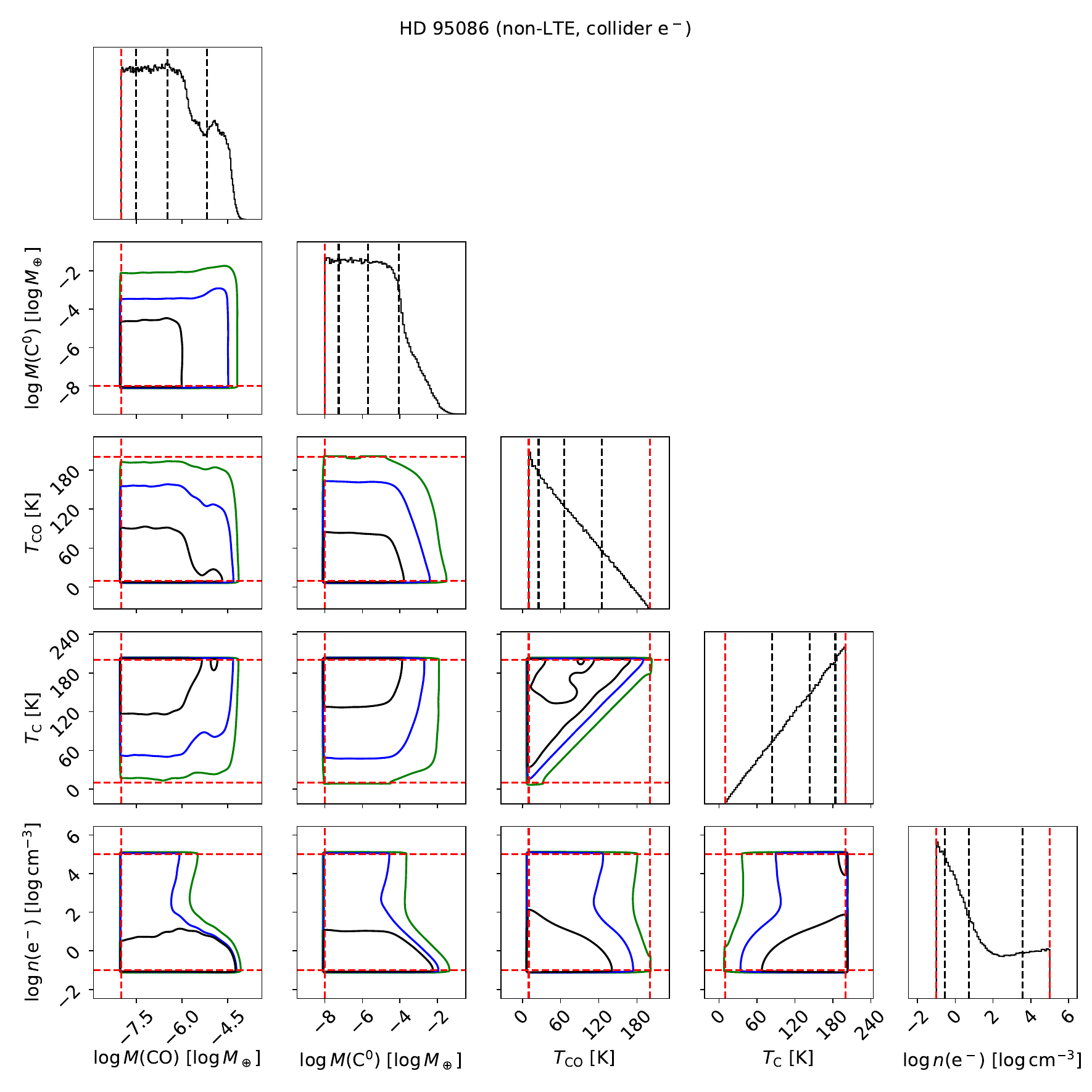}
	\caption{Same as Fig.\ \ref{fig:corner_nonLTE_e_49Ceti}, but for the disk around HD~95086.\label{fig:corner_nonLTE_e_HD95086}}
\end{figure*}

\begin{figure*}[h]
	\plotone{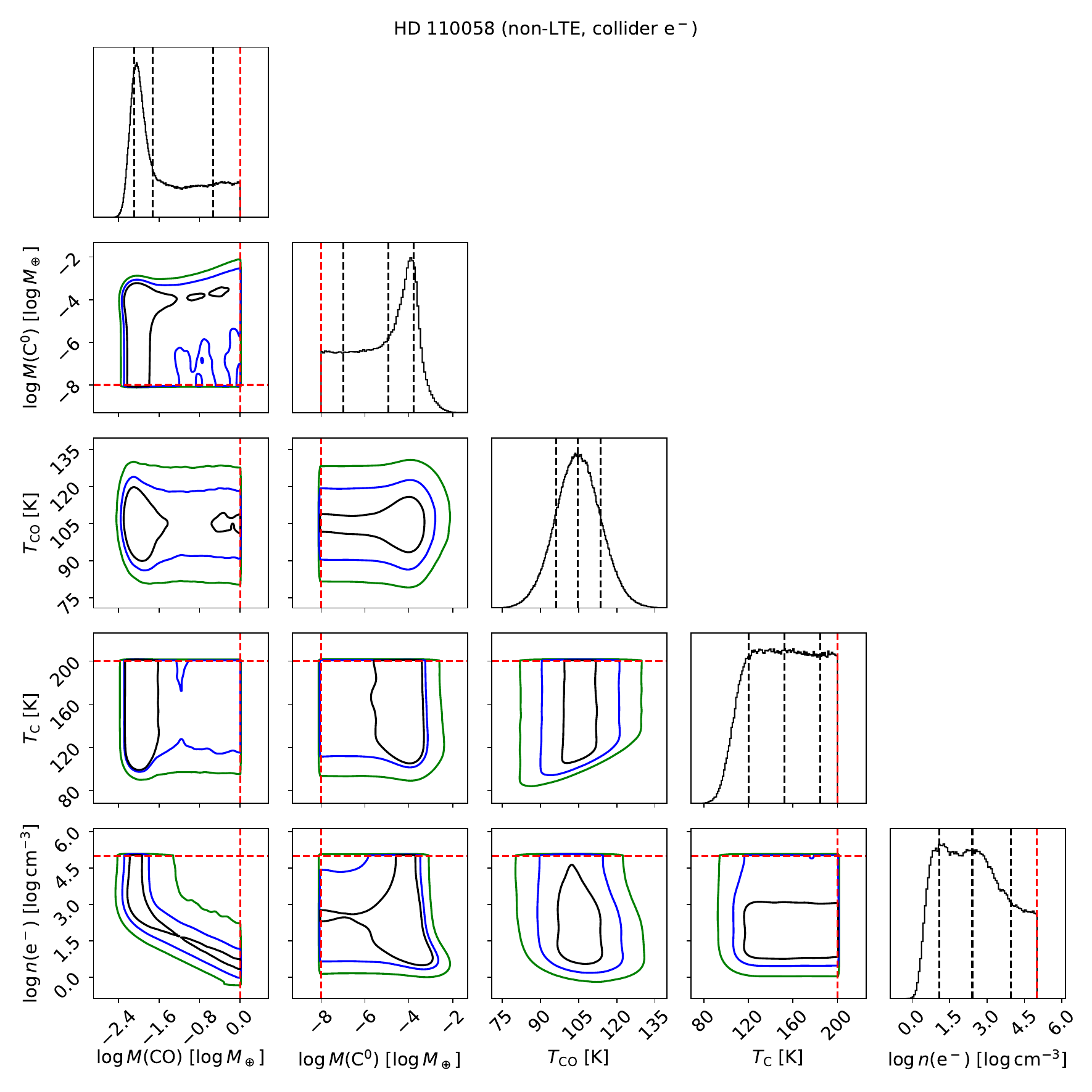}
	\caption{Same as Fig.\ \ref{fig:corner_nonLTE_e_49Ceti}, but for the disk around HD~110058.\label{fig:corner_nonLTE_e_HD110058}}
\end{figure*}

\begin{figure*}[h]
	\plotone{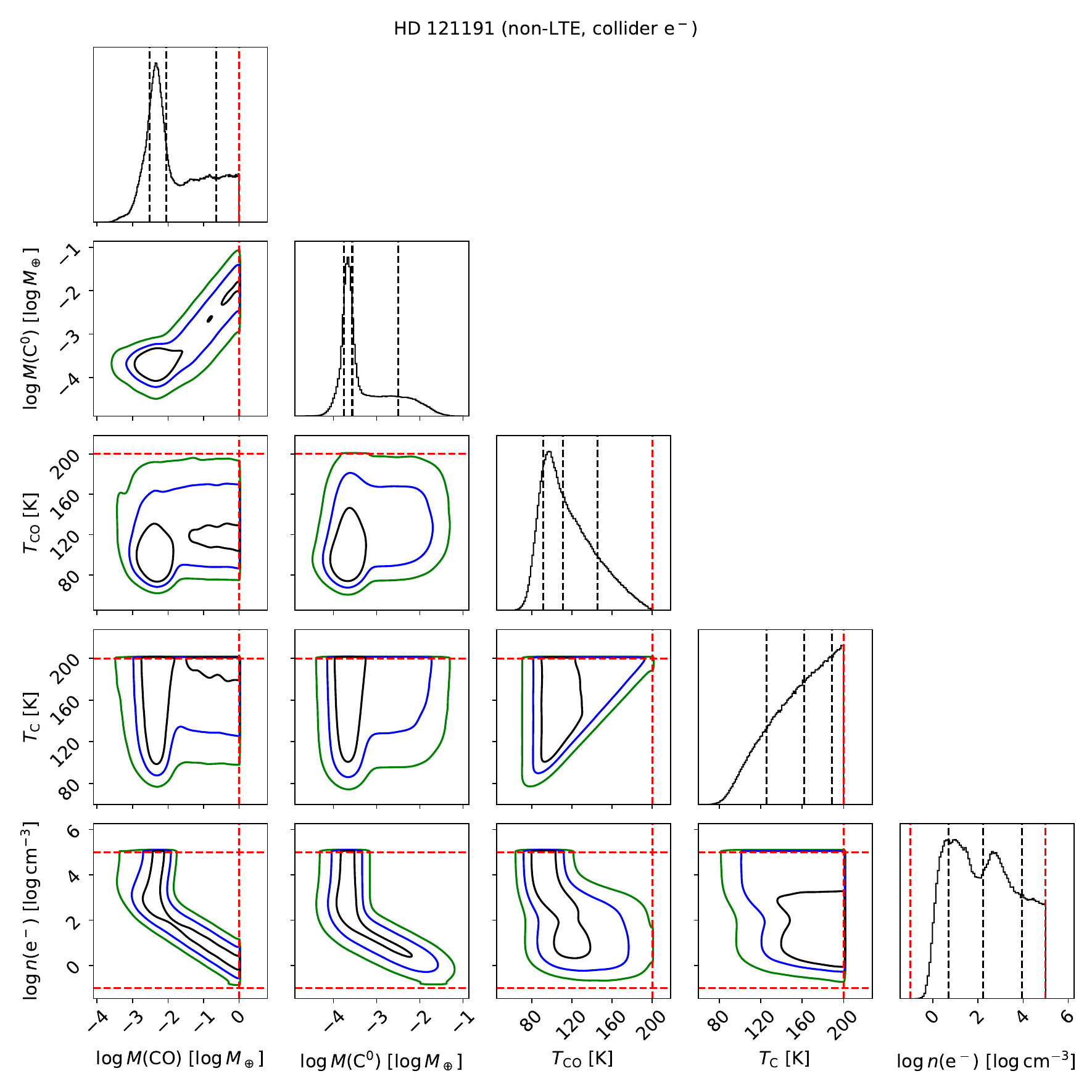}
	\caption{Same as Fig.\ \ref{fig:corner_nonLTE_e_49Ceti}, but for the disk around HD~121191.\label{fig:corner_nonLTE_e_HD121191}}
\end{figure*}

\begin{figure*}[h]
	\plotone{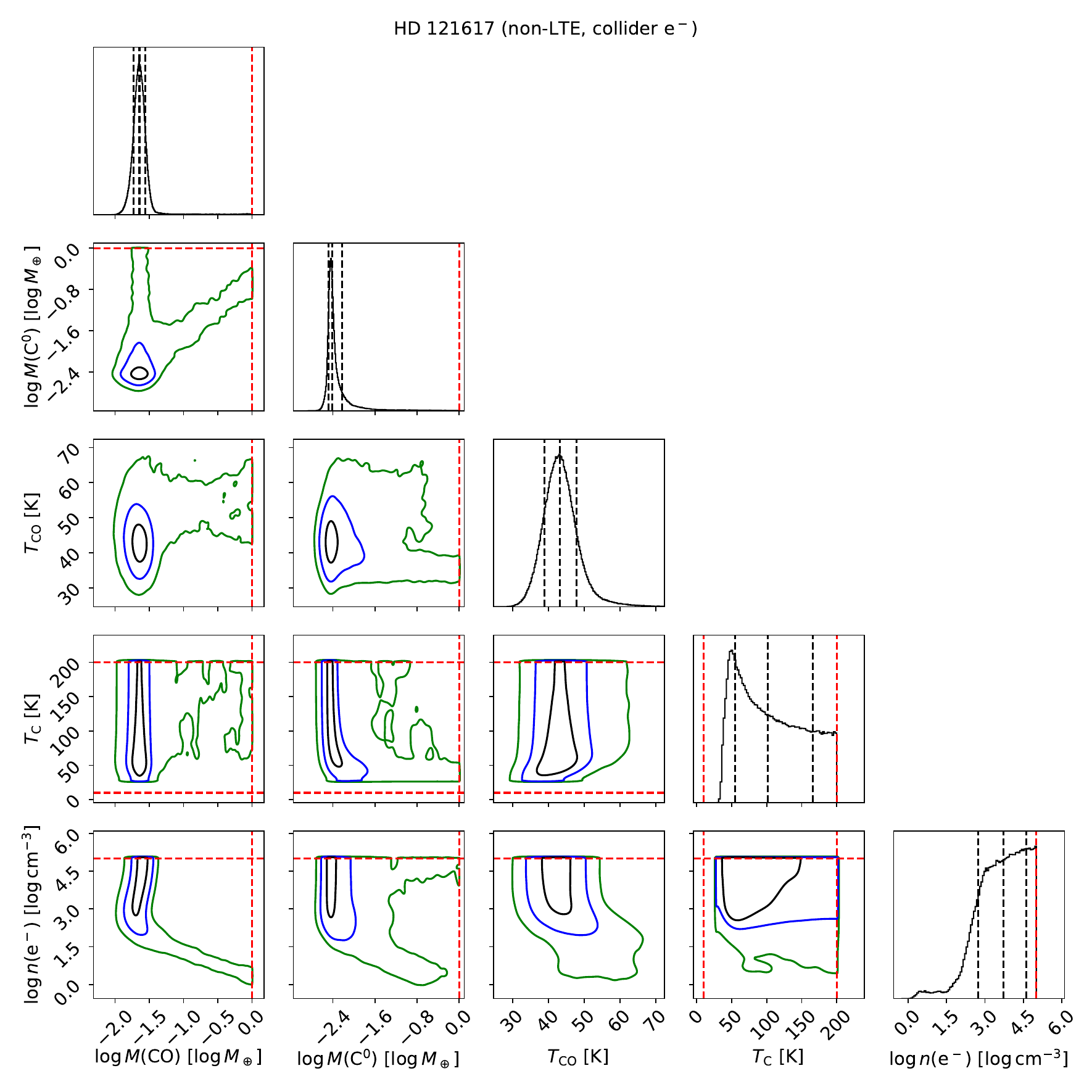}
	\caption{Same as Fig.\ \ref{fig:corner_nonLTE_e_49Ceti}, but for the disk around HD~121617.\label{fig:corner_nonLTE_e_HD121617}}
\end{figure*}

\begin{figure*}[h]
	\plotone{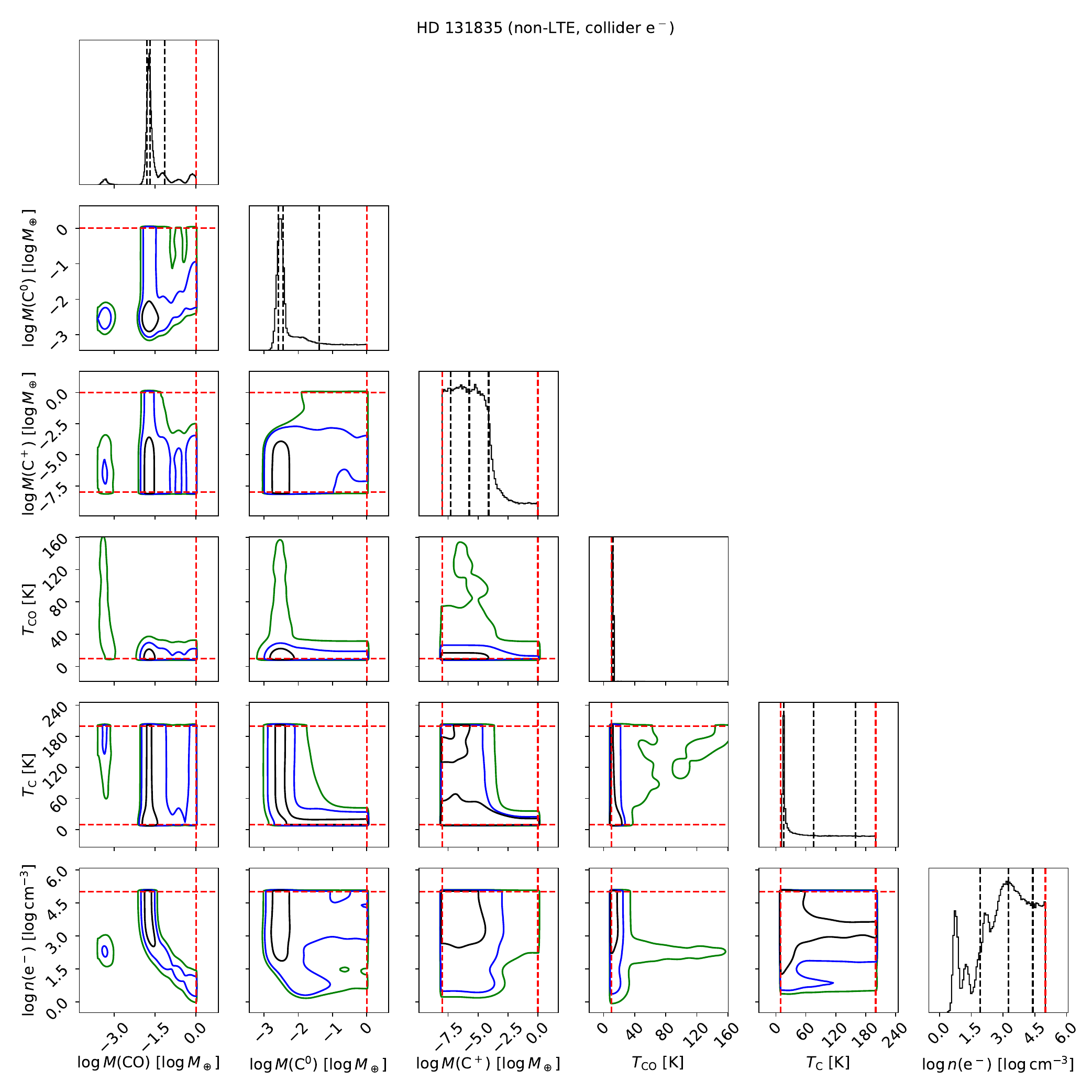}
	\caption{Same as Fig.\ \ref{fig:corner_nonLTE_e_49Ceti}, but for the disk around HD~131835.\label{fig:corner_nonLTE_e_HD131835}}
\end{figure*}

\begin{figure*}[h]
	\plotone{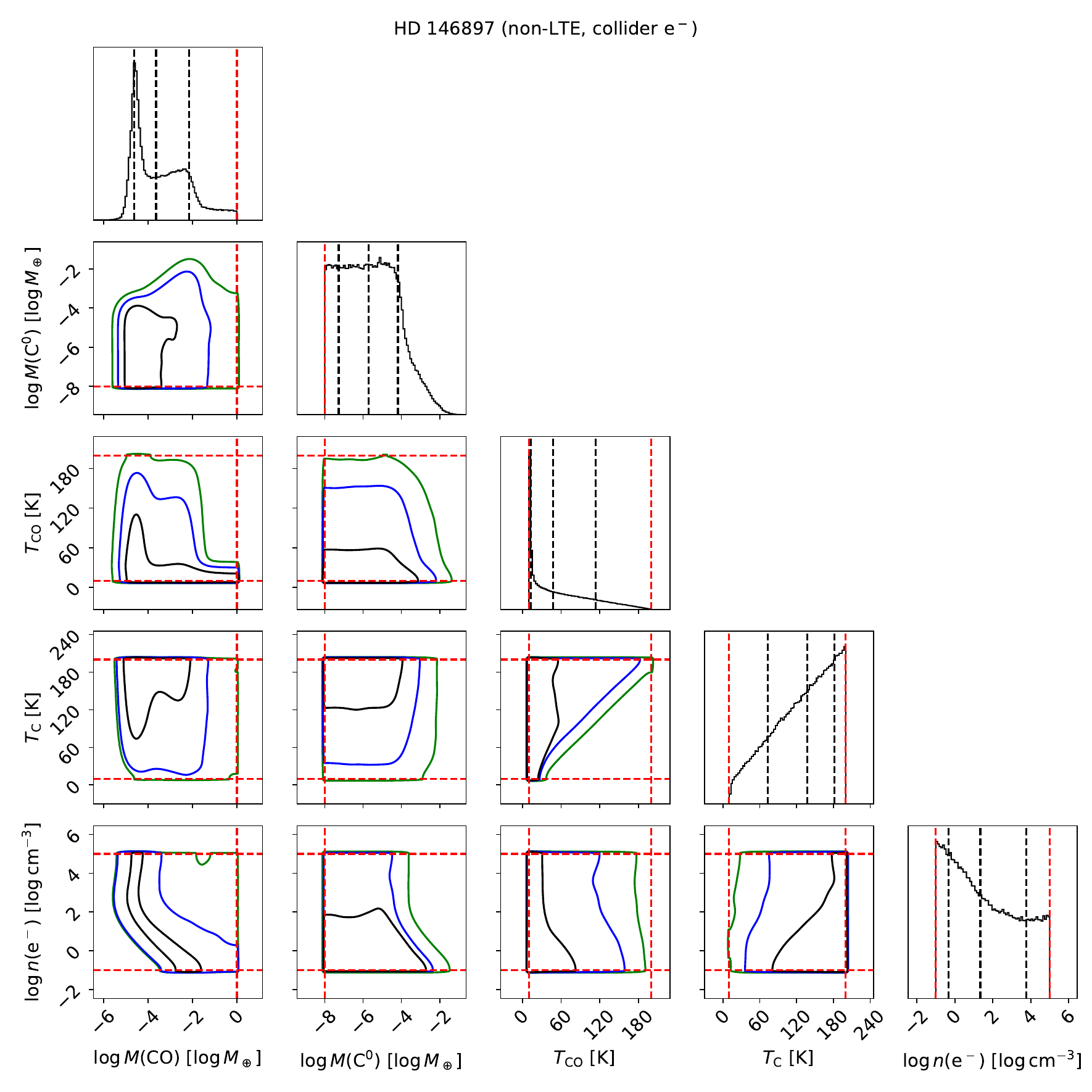}
	\caption{Same as Fig.\ \ref{fig:corner_nonLTE_e_49Ceti}, but for the disk around HD~146897.\label{fig:corner_nonLTE_e_HD146897}}
\end{figure*}

\begin{figure*}[h]
	\plotone{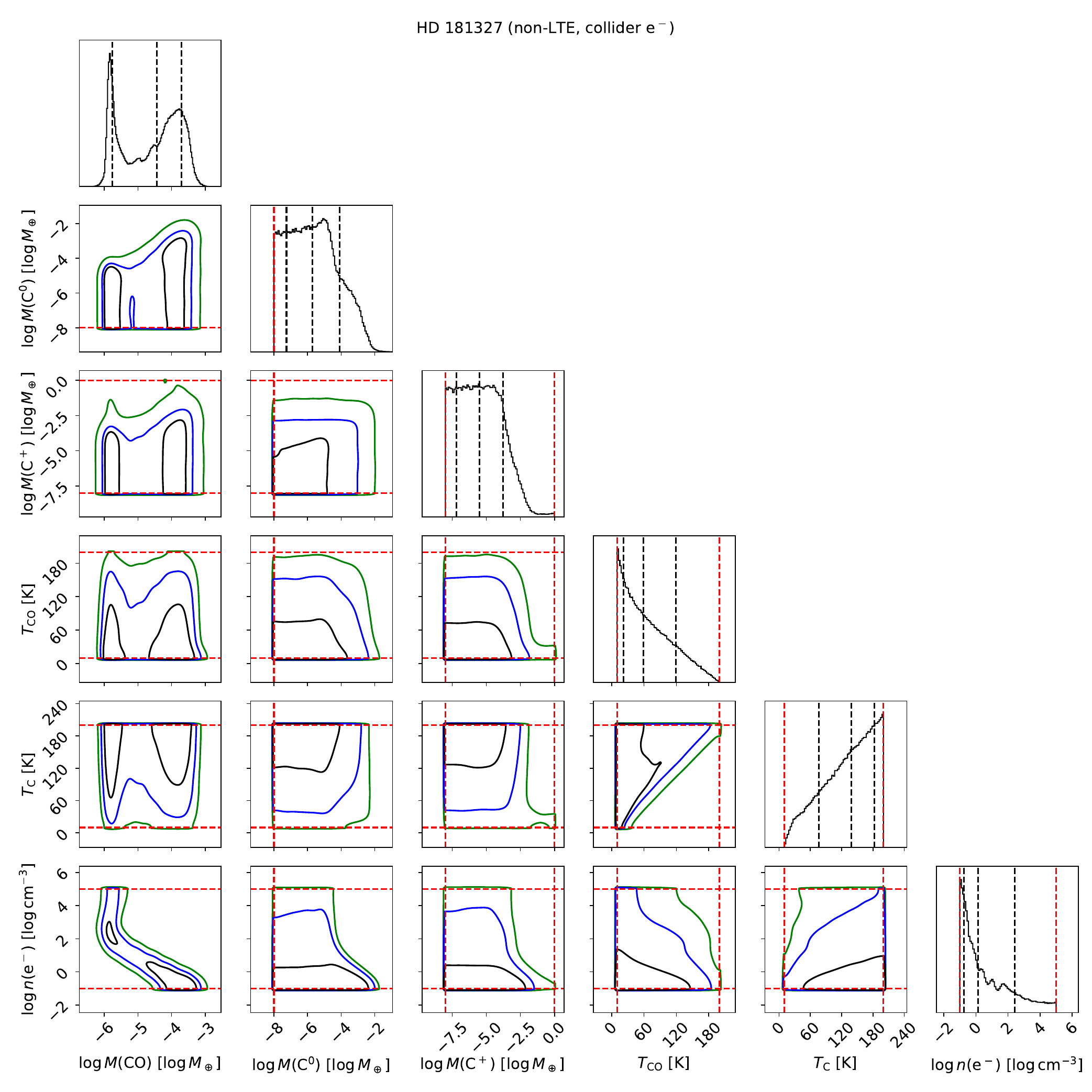}
	\caption{Same as Fig.\ \ref{fig:corner_nonLTE_e_49Ceti}, but for the disk around HD~181327.\label{fig:corner_nonLTE_e_HD181327}}
\end{figure*}

\begin{figure*}[h]
	\plotone{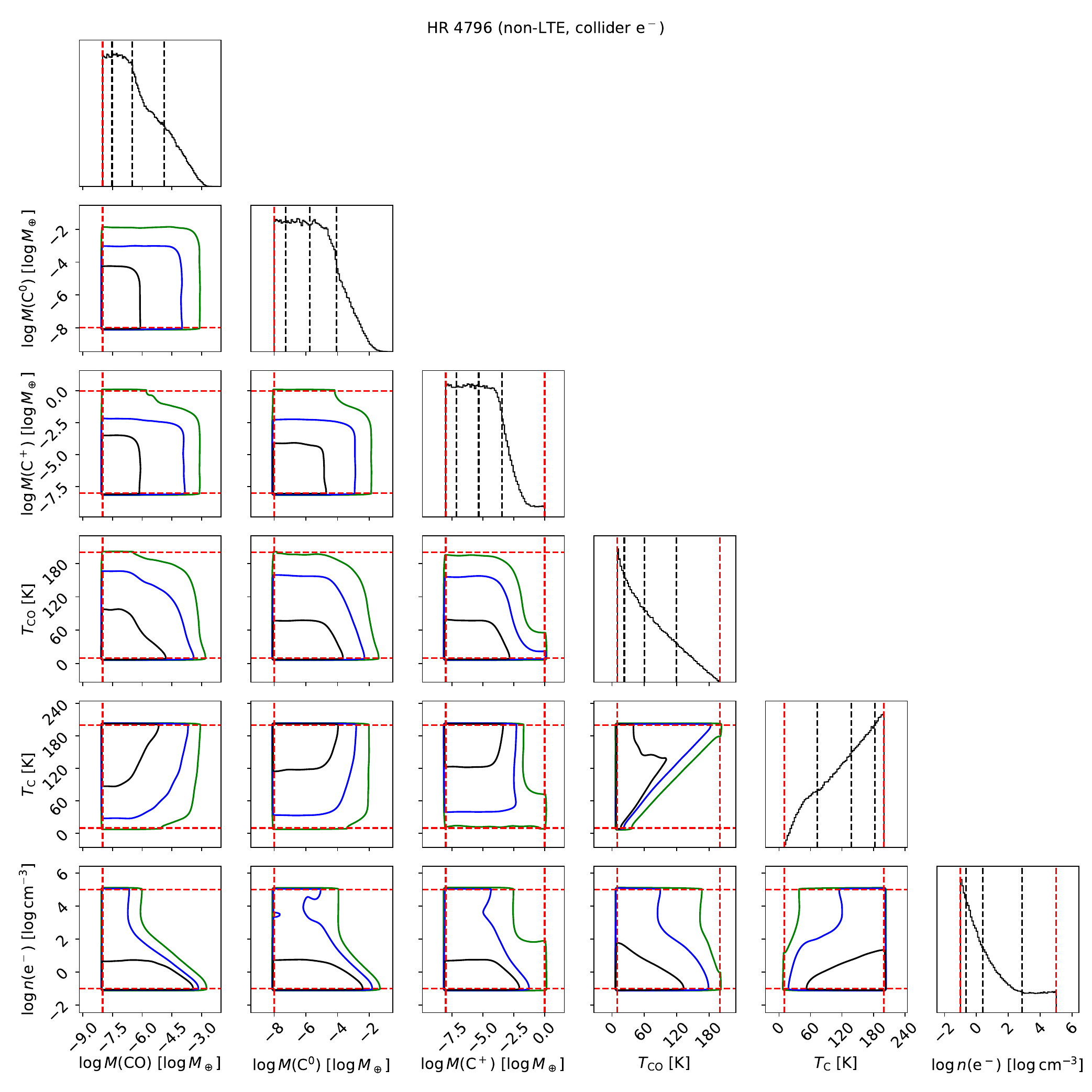}
	\caption{Same as Fig.\ \ref{fig:corner_nonLTE_e_49Ceti}, but for the disk around HR~4796.\label{fig:corner_nonLTE_e_HR4796}}
\end{figure*}

In Figures \ref{fig:corner_nonLTE_e_49Ceti} to \ref{fig:corner_nonLTE_e_HR4796} we show the corner plots of the non-LTE fits with electron colliders.

\bibliography{bibliography}{}
\bibliographystyle{aasjournal}



\end{document}